\documentclass[prr,reprint,notitlepage,noeprint]{revtex4-2}

\usepackage[english]{babel}

\usepackage{amsmath,amssymb,amsfonts,bbm,xcolor,graphicx,comment,txfonts,nicefrac}
\definecolor{darkblue}{rgb}{0.,0.,0.5}
\usepackage[colorlinks,linkcolor=darkblue,citecolor=darkblue,urlcolor=darkblue]{hyperref}

\newcommand{\sop}{string order parameter}
\newcommand{\dsop}{dual string order parameter}
\newcommand{\Dsop}{Dual string order parameter}

\newcommand{\hc}{\mathrm{H.c.}}
\newcommand{\id}{\mathbbm{1}}

\newcommand{\N}{\mathbb{N}}
\newcommand{\Z}{\mathbb{Z}}

\newcommand{\R}{\mathbb{R}}

\newcommand{\e}{e}
\newcommand{\diff}{d}
\newcommand{\imag}{i}
\newcommand{\ket}[1]{\lvert #1 \rangle}
\newcommand{\bra}[1]{\langle #1 \rvert}
\newcommand{\braket}[1]{\langle #1 \rangle}
\newcommand{\BZ}{\mathrm{BZ}}
\newcommand{\unitvec}[1]{\hat{\mathbf{#1}}}

\makeatletter
\newcommand*{\transpose}{%
  {\mathpalette\@transpose{}}%
}
\newcommand*{\@transpose}[2]{%
  \raisebox{\depth}{$\m@th#1\intercal$}%
}
\makeatother

\newcommand{\abs}[1]{\left\lvert #1 \right\rvert}

\newcommand{\tr}{\mathop{\mathrm{tr}}}
\newcommand{\sgn}{\mathop{\mathrm{sgn}}}
\newcommand{\acosh}{\mathop{\mathrm{acosh}}}
\newcommand{\asinh}{\mathop{\mathrm{asinh}}}

\newcommand{\pf}{\mathop{\mathrm{pf}}}
\renewcommand{\Re}{\mathop{\mathrm{Re}}}
\renewcommand{\Im}{\mathop{\mathrm{Im}}}

\makeatletter
\DeclareFontFamily{OMX}{MnSymbolE}{}
\DeclareSymbolFont{MnLargeSymbols}{OMX}{MnSymbolE}{m}{n}
\SetSymbolFont{MnLargeSymbols}{bold}{OMX}{MnSymbolE}{b}{n}
\DeclareFontShape{OMX}{MnSymbolE}{m}{n}{
  <-6>  MnSymbolE5
  <6-7>  MnSymbolE6
  <7-8>  MnSymbolE7
  <8-9>  MnSymbolE8
  <9-10> MnSymbolE9
  <10-12> MnSymbolE10
  <12->   MnSymbolE12
}{}
\DeclareFontShape{OMX}{MnSymbolE}{b}{n}{
  <-6>  MnSymbolE-Bold5
  <6-7>  MnSymbolE-Bold6
  <7-8>  MnSymbolE-Bold7
  <8-9>  MnSymbolE-Bold8
  <9-10> MnSymbolE-Bold9
  <10-12> MnSymbolE-Bold10
  <12->   MnSymbolE-Bold12
}{}

\let\llangle\@undefined
\let\rrangle\@undefined
\DeclareMathDelimiter{\llangle}{\mathopen}%
{MnLargeSymbols}{'164}{MnLargeSymbols}{'164}
\DeclareMathDelimiter{\rrangle}{\mathclose}%
{MnLargeSymbols}{'171}{MnLargeSymbols}{'171}
\makeatother

\begin{document}
\renewcommand\labelenumi{\arabic{enumi}.}
\renewcommand\theenumi{\thesection.\arabic{enumi}}

\title{Quantum quenches in driven-dissipative quadratic fermionic systems with
  parity-time symmetry}

\author{Elias Starchl}

\author{Lukas M. Sieberer}

\email{lukas.sieberer@uibk.ac.at}

\affiliation{Institute for Theoretical Physics, University of Innsbruck, 6020
  Innsbruck, Austria}


\begin{abstract}
We study the quench dynamics of noninteracting fermionic quantum many-body
  systems that are subjected to Markovian drive and dissipation and are
  described by a quadratic Liouvillian which has parity-time (PT) symmetry. In
  recent work, we have shown that such systems relax locally to a maximum
  entropy ensemble that we have dubbed the PT-symmetric generalized Gibbs
  ensemble (PTGGE), in analogy to the generalized Gibbs ensemble that describes
  the steady state of isolated integrable quantum many-body systems after a
  quench. Here, using driven-dissipative versions of the Su-Schrieffer-Heeger
  (SSH) model and the Kitaev chain as paradigmatic model systems, we corroborate
  and substantially expand upon our previous results. In particular, we confirm
  the validity of a dissipative quasiparticle picture at finite dissipation by
  demonstrating light cone spreading of correlations and the linear growth and
  saturation to the PTGGE prediction of the quasiparticle-pair contribution to
  the subsystem entropy in the PT-symmetric phase. Further, we introduce the
  concept of directional pumping phases, which is related to the non-Hermitian
  topology of the Liouvillian and based upon qualitatively different dynamics of
  the \dsop{} and the subsystem fermion parity in the SSH model and the Kitaev
  chain, respectively: Depending on the postquench parameters, there can be
  pumping of string order and fermion parity through both ends of a subsystem
  corresponding to a finite segment of the one-dimensional lattice, through only
  one end, or there can be no pumping at all. We show that transitions between
  dynamical pumping phases give rise to a new and independent type of dynamical
  critical behavior of the rates of directional pumping, which are determined by
  the soft modes of the PTGGE.
\end{abstract}

\maketitle

\section{Introduction}
\label{sec:introduction}

Quantum quenches are a paradigmatic scenario for inducing and studying
far-from-equilibrium dynamics in isolated quantum many-body systems. In a
quantum quench, an initial state, often chosen to be the ground state of a
prequench Hamiltonian, is evolved in time with a postquench
Hamiltonian~\cite{Polkovnikov2011, Eisert2015}. Of particular interest are
quenches across phase boundaries, which can lead to a variety of intriguing
phenomena, especially in systems with nontrivial topology~\cite{Chung2013,
  Gong2017a, Yang2018, Vajna2015, Pastori2020, Heyl2018, Budich2016}. But the
two basic questions that underlie the interest in nonequilibrium dynamics
induced by quantum quenches read as follows: Does a given system equilibrate
after a quench, i.e., do expectation values of local observables reach a steady
state?  And what is the nature of this steady state? It is well established that
integrable quantum many-body systems, which are characterized by an extensive
set of integrals of motion, relax locally to a statistical ensemble determined
by the principle of maximum entropy~\cite{Jaynes1957}---the generalized Gibbs
ensemble
(GGE)~\cite{Rigol2007,Rossini2007,Calabrese2011,Calabrese2012I,Calabrese2012II,Essler2016,Vidmar2016}. The
GGE can be regarded as being universal in the sense that its general structure
is model-independent, whereas specific model-dependent details enter the GGE
only through the form of the integrals of motion and as Lagrangian multipliers,
the values of which are determined by the state in which the system is prepared
before the quench. In this way, the GGE conserves an extensive amount of
information about the initial state. Relaxation to the GGE is accompanied by a
number of universal characteristic features, such as light cone propagation of
correlations~\cite{Essler2016,Calabrese2006}, linear growth and volume-law
saturation of the subsystem
entropy~\cite{Calabrese2005,Alba2017,Alba2018,Calabrese2020} and the
equilibration of local order
parameters~\cite{Essler2016,Vidmar2016,Calabrese2011,Calabrese2012I,Calabrese2012II}.

In stark contrast, open many-body quantum systems, which are subjected to
Markovian drive and dissipation, generally approach a highly model-dependent
steady state that is determined by the interplay between internal Hamiltonian
dynamics and the coupling to external reservoirs~\cite{Breuer2002,Diehl2008,
  Verstraete2009, Eisert2010, Sieberer2013, Sieberer2016a, Maghrebi2016a,
  Jin2016, Rota2017, Halati2022}. Through these couplings, the system can
exchange energy and particles with the reservoirs, breaking conservation laws
the system would have in isolation. In particular, for integrable systems, this
generically means that local integrals of motions of an isolated system are not
conserved anymore if the system is open. As a result, after a quench, all memory
of the initial state fades away with time, and the steady state takes the form
of a GGE only in the limit of weak coupling $\gamma$ between the system and its
environment~\cite{Lange2017,Lange2018,Lenarcic2018,Reiter2021}. Yet, as we have
shown in Ref.~\cite{Starchl2022}, the universal principles governing generalized
thermalization after a quantum quench in an isolated integrable system do
apply---in suitably generalized form---to specific driven-dissipative systems
even for finite system-bath coupling $\gamma$; and further, key signatures, such
as a driven-dissipative quasiparticle picture~\cite{Alba2022, Carollo2022,
  Alba2021} and equilibration of local observables, still accompany relaxation
to a maximum entropy ensemble. This unexpected robustness of generalized
thermalization is ensured by parity-time (PT) symmetry of the Liouvillian that
generates the dynamics of these driven-dissipative systems. To highlight the
fundamental roles that are played by PT symmetry and the principle of maximum
entropy, we have dubbed the ensemble such systems locally relax to the
PT-symmetric GGE (PTGGE)~\cite{Starchl2022}.

Originally, PT symmetry has been studied as a framework for a non-Hermitian
generalization of quantum mechanics. In conventional quantum mechanics, physical
observables are represented by Hermitian operators, mainly because these
operators have real spectra. However, as shown in seminal work by Bender et
al.~\cite{Bender1998,*Bender2008}, symmetry under the combination of spatial
inversion or parity and time reversal can also lead to real spectra in
non-Hermitian operators. In particular, an eigenvector of a PT-symmetric
non-Hermitian operator is associated with a real eigenvalue if also the
eigenvector itself is PT-symmetric, i.e., invariant under the PT
transformation. Eigenvectors that are not invariant under the PT transformation
are said to spontaneously break PT symmetry and are associated with complex
eigenvalues. Typically, if a PT-symmetric operator depends on a parameter
$\gamma$ that measures the degree of non-Hermiticity such that the operator is
Hermitian for $\gamma = 0$, all eigenvectors are PT-symmetric and the spectrum
is entirely real if $\gamma$ is sufficiently small; this situation is referred
to as the PT-symmetric phase. At intermediate values of $\gamma$, in the
PT-mixed phase, PT-symmetric and PT-breaking eigenvectors coexist. And for large
values of $\gamma$, typically all eigenvectors spontaneously break PT
symmetry. Non-Hermitian generalizations of quantum mechanics, as envisioned
originally in Refs.~\cite{Bender1998,*Bender2008}, are restricted to the
PT-symmetric phase with a real spectrum. More recently, PT-symmetric
non-Hermitian versions of paradigmatic models of condensed matter theory such as
the Kitaev chain~\cite{Kitaev2001} have received great interest due to their
unconventional topological properties, in particular, in connection with the
spontaneous breaking of PT symmetry that leads to the occurrence of complex
eigenvalues~\cite{Wang2015, Yuce2016, San-Jose2016, Zeng2016, Menke2017,
  Klett2017, Kawabata2018, Li2020, Agarwal2021}. Here, however, we are
interested in PT symmetry of a special type of non-Hermitian operator: the
Liouvillian that generates the dynamics of an open quantum many-body
system~\cite{Prosen2012, *Prosen2012a, Medvedyeva2016, VanCaspel2018,
  Minganti2019, Shibata2020, Huber2020, Huybrechts2020, Arkhipov2020,
  Curtis2021, Roccati2021, Roccati2022, Claeys2022, Nakanishi2022}. The coupling
to external reservoirs generically induces exponential decay, which is reflected
in the spectrum of the Liouvillian being complex. On the one hand, this implies
that not only the PT-symmetric but also the PT-mixed and PT-broken phases can
describe quantum dynamics of driven-dissipative systems; on the other hand, this
puts into question the very existence of a PT-symmetric phase, since all
eigenvalues of a generic Liouvillian become real only in the limit of vanishing
dissipation---we choose here the convention that the real and imaginary parts of
an eigenvalue of a Liouvillian determine, respectively, the oscillation
frequency and decay rate of the corresponding eigenmode. However, after a shift
by a constant decay rate, the single-particle spectrum of a quadratic
Liouvillian, which describes a noninteracting open quantum many-body system, can
become entirely real---a property that is known as passive PT
symmetry~\cite{Ornigotti2014, Joglekar2018}.

\begin{figure}
\includegraphics[width=\linewidth]{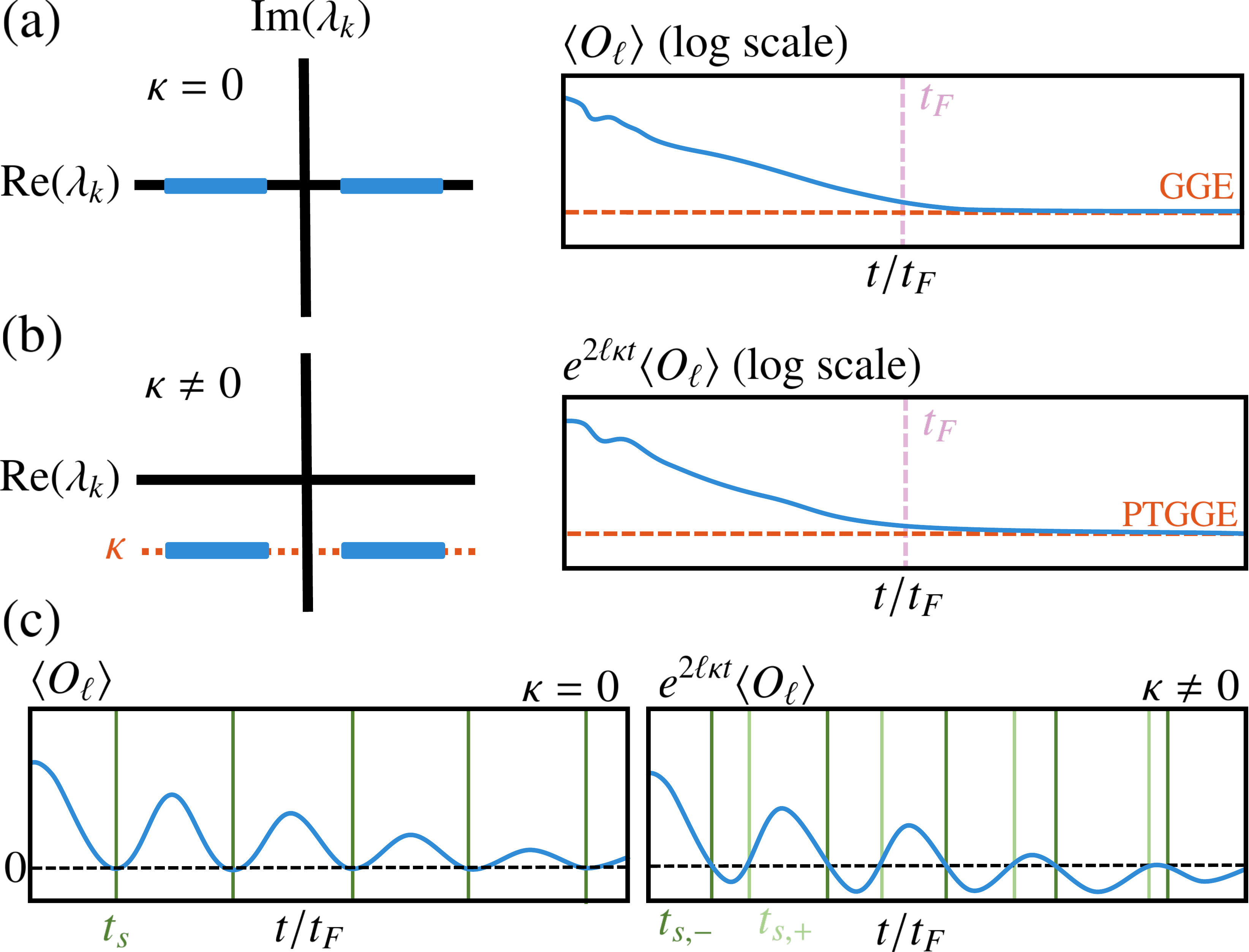}
\caption{(a) Schematic representation of (left) the single-particle spectrum
  $\lambda_k$ of an isolated system (blue lines) with $\kappa = 0$ and (right)
  relaxation of an observable $\braket{O_{\ell}}$ acting on $\ell$ sites to the
  GGE (red, dashed line) on a time scale $t_F$ (purple, dashed line). (b) For a
  driven-dissipative system with PT symmetry, (left) the spectrum acquires a
  global shift by $- \imag \kappa$, and (right) relaxation to the PTGGE (red,
  dashed line) is revealed by rescaling $\langle O_{\ell} \rangle$ with
  $\e^{2 \ell \kappa t}$. (c) Quenches from the trivial to the topological phase
  are accompanied by oscillatory dynamics of a topological disorder parameter
  are shown for (left) an isolated system, with zeros of
  $\langle O_{\ell} \rangle$ at multiples of a single soft-mode period $t_s$
  (green vertical lines), and (right) a driven-dissipative system, where
  directional pumping is characterized by two soft-mode periods $t_{s,-}$ (dark
  green) and $t_{s,+}$ (light green).}
\label{fig:overview}
\end{figure}

Passive PT symmetry of quadratic Liouvillians is the key feature that underlies
local relaxation to the PTGGE of the translationally invariant
driven-dissipative fermionic lattice systems that we study below. In the
PT-symmetric phase, the single-particle eigenvalues of such Liouvillians form
two bands that are given by $\lambda_{\pm, k} = - \imag \kappa \pm \omega_k$,
where $k$ is the quasimomentum. The momentum-independent decay rate $\kappa$
results in temporally uniform overall exponential relaxation. Independently from
that, dephasing of modes with different frequencies $\omega_k \neq \omega_{k'}$
leads to local relaxation to the PTGGE---this is completely analogous to
isolated two-band models with single-particle dispersion $\pm \varepsilon_k$,
where generalized thermalization to the GGE is induced by dephasing of purely
oscillatory modes with $\varepsilon_k \neq \varepsilon_{k'}$~\cite{Essler2016,
  Barthel2008}. However, note that the PTGGE is intrinsically time-dependent due
to the exponential decay at the rate $\kappa$. Relaxation of an isolated system
to the GGE and of a driven-dissipative system to the PTGGE, and the underlying
single-particle spectra, are illustrated in Figs.~\ref{fig:overview}(a) and~(b),
respectively. After a quench in an isolated system, a suitably chosen observable
$\langle O_{\ell} \rangle$ that acts on $\ell$ contiguous lattice sites
equilibrates to the value predicted by the GGE after the Fermi time
$t_F \sim \ell$~\cite{Calabrese2012I}. In contrast, after a quench to the
PT-symmetric phase of a driven-dissipative system, the global shift of the
spectrum by $- \imag \kappa$ results in exponential decay,
$\langle O_{\ell} \rangle \sim \e^{- 2 \ell \kappa t}$. Relaxation to the PTGGE
through dephasing can thus be revealed by considering the \emph{rescaled
  expectation value} $\e^{2 \ell \kappa t} \langle O_{\ell} \rangle$. For
stronger coupling to the environment, PT symmetry is broken spontaneously and
the imaginary part of $\lambda_{\pm, k}$ becomes dispersive. The long-time
dynamics is then determined not by dephasing of all single-particle eigenmodes
but rather by the single slowest-decaying mode that corresponds to the smallest
value of $- \Im(\lambda_{\pm, k})$. Therefore, the spontaneous breaking of
passive PT symmetry defines a sharp dynamical transition that delimits
relaxation to the PTGGE and thus the validity of the principle of maximum
entropy in driven-dissipative systems. An important caveat is that this argument
for local relaxation to the PTGGE in the PT-symmetric phase is based solely on
the form of the spectrum of the Liouvillian. Indeed, relaxation to the PTGGE
applies as long as the overall decay with constant rate $\kappa$ causes the
system to heat up to infinite temperature. Instead, for a nontrivial steady
state, local relaxation to the PTGGE occurs only transiently up to a
sharply-defined crossover time $t_\times$~\cite{Starchl2022}.

In this work, we perform a detailed study of the quench dynamics and relaxation
to the PTGGE of driven-dissipative fermionic many-body systems with PT symmetry,
corroborating and substantially expanding upon our earlier results that we have
presented in Ref.~\cite{Starchl2022}. We consider two models, driven-dissipative
versions of the Su-Schrieffer-Heeger (SSH) model~\cite{Su1979, *Heeger1988} and
the Kitaev chain~\cite{Kitaev2001}, which are representative for broad classes
of quadratic fermionic systems. Studying the time evolution of the dual string
order parameter and the subsystem fermion parity, which serve as
\emph{topological disorder parameters} for the SSH model and the Kitaev chain,
respectively, leads us to introduce \emph{directional pumping phases} that are
characterized by qualitatively different dynamics of the topological disorder
parameters as illustrated in Fig.~\ref{fig:overview}(c). Using the example of a
Kitaev chain with long-range hopping and pairing, we show that directional
pumping phases and the dynamical critical behavior at transitions between these
phases are in general independent from the phases and the associated criticality
that are defined in terms of PT symmetry and gap closings of the postquench
Liouvillian.

To make the differences between our previous~\cite{Starchl2022} and current work
more explicit, let us briefly summarize the additional results presented in this
work that have not been presented before. We extend our previous results in two
major ways. First, we present a detailed study of the quench dynamics of the
driven-dissipative SSH model, which has been mentioned only briefly in
Ref.~\cite{Starchl2022}. In particular, we derive the corresponding expression
for the PTGGE and study the evolution of the subsystem entropy for this
model. We complement our previous results for the Kitaev chain by studying the
spreading of correlations. Second, we provide an extensive analysis of the
dynamics and topological zero crossings of topological disorder parameters in
both models. The most important new concepts and results in this context are
directional pumping phases with dynamical critical behavior and a notion of
universality at the phase boundaries. We further demonstrate modifications of
this critical behavior induced by changes in the long-wavelength description due
to long-range couplings. Finally, we compare the dynamical critical behavior of
topological disorder parameters to the critical behavior of the connected
density autocorrelation function.

This paper is organized as follows: In Sec.~\ref{sec:key-results}, we summarize
our key results. The models we study, which are driven-dissipative versions of
the SSH and Kitaev chains, are introduced in Sec.~\ref{sec:models}. We then
establish the PTGGE as the maximum entropy ensemble for these models in
Sec.~\ref{sec:PTGGE}. The spreading of correlations after a quench is studied in
Sec.~\ref{sec:spread-corrs}, which is followed in
Sec.~\ref{sec:subsystem-entropy} by a discussion of the time evolution of the
subsystem entropy. We study the dynamics of the \dsop{} and the subsystem parity
in Sec.~\ref{sec:dynamics-local-observ}, where we also introduce directional
pumping phases and characterize the associated critical behavior. In
Sec.~\ref{sec:long-range-kitaev-quench}, we investigate directional pumping
phases and phase transitions in a Kitaev chain with long-range hopping and
pairing. Open research questions are presented in Sec.~\ref{sec:conclusion}, and
technical details are described in the
Appendices~\ref{sec:time-evolution-covariance-matrix}--\ref{sec:appendix-density}.

\section{Key results}
\label{sec:key-results}

We consider the following quench protocol: The system is initialized in the
ground state $\ket{\psi_0}$ of the Hamiltonian $H_0$ that describes an isolated
SSH model or Kitaev chain, where we focus on quenches starting from the
topologically trivial phases of these models. At $t = 0$, a parameter of the
Hamiltonian is changed abruptly; simultaneously, the system is coupled to
Markovian reservoirs. The ensuing dynamics is generated by a quantum Liouvillian
$\mathcal{L}$ in Lindblad form, such that the state of the system at time $t$ is
given by $\rho(t) = \e^{- \imag \mathcal{L} t} \rho_0$, where
$\rho_0 = \ket{\psi_0}\bra{\psi_0}$ is the initial state. Our main results can
be summarized as follows:

\textit{After quenches to the PT-symmetric phase, driven-dissipative free
  fermionic models relax to a maximum entropy ensemble, the parity-time
  symmetric generalized Gibbs ensemble (PTGGE).} We have introduced the PTGGE as
the maximum entropy ensemble that describes relaxation of a driven-dissipative
Kitaev chain and have briefly discussed quench dynamics of an SSH model with
incoherent loss and gain in Ref.~\cite{Starchl2022}. Here, we present a detailed
derivation of the PTGGE for the SSH model. As noted in the Introduction,
dephasing of modes with different oscillation frequencies, but with a decay rate
that is guaranteed to be equal for all modes by PT symmetry, is the fundamental
process that underlies relaxation to the PTGGE in free fermionic systems. Due to
the common decay rate of all modes, the PTGGE is inherently time-dependent. We
choose the SSH model and the Kitaev chain as representatives of the two
fundamental classes of quadratic fermionic systems: The isolated SSH model has a
$\mathrm{U}(1)$ symmetry associated with the conservation of the number of
particles, whereas due the presence of pairing terms in the isolated Kitaev
chain only the fermion parity is conserved and the symmetry group is reduced to
$\Z_2$. The driven-dissipative generalizations of these models combine their
quadratic Hamiltonians with linear Lindblad operators, which enable the systems
to exchange particles with external reservoirs and, therefore, always break
particle number conservation. However, as detailed in Sec.~\ref{sec:models}, we
can still choose a particular form of dissipation so as to preserve a weak
$\mathrm{U}(1)$ symmetry in the driven-dissipative SSH model~\cite{Buca2012}.
Then, as in the isolated SSH model, no anomalous correlations are generated in
the course of the dynamics. Therefore, the driven-dissipative versions of the
SSH model and the Kitaev chain we consider in this work can be regarded as
natural open-system generalizations of topological insulators and
superconductors.

We illustrate relaxation to the PTGGE by considering the dynamics of
\emph{topological disorder parameters,} which take finite expectation values in
the ground state in the trivial phase and vanish in the topological phase. Topological disorder parameters for the SSH model and the Kitaev chain
are the \dsop{}~\cite{Hida1992} and the subsystem fermion
parity~\cite{Starchl2022}, respectively. The corresponding operators act
nontrivially on $\ell$ contiguous lattice sites. As illustrated schematically in
Fig.~\ref{fig:overview}, relaxation of the topological disorder parameters to
the PTGGE happens on a characteristic time scale given by the Fermi time
$t_F \sim \ell$~\cite{Calabrese2012I}. We further provide analytical conjectures
for the evolution of the topological disorder parameters, which we find to be in
excellent agreement with our exact numerical results. These conjectures
generalize analytical results for the quench dynamics of the isolated transverse
field Ising model in the space-time scaling limit $\ell, t \to \infty$ with
$\ell/t$ fixed~\cite{Calabrese2011,Calabrese2012I,Calabrese2012II}, and show
clearly how the dynamics are affected by drive and dissipation: Apart from the
abovementioned uniform exponential decay, both the dispersion relation and the
statistics of the eigenmodes of the adjoint Liouvillian, which generates the
time evolution of operator expectation values, are modified, leading to
pronounced quantitative differences even after rescaling of expectation values
to compensate the exponential decay.

For the models we consider, a description of the late-time dynamics in terms of
the PTGGE applies up to arbitrarily long times for balanced loss and gain, which
leads to a steady state at infinite temperature. For a small imbalance, the
PTGGE still provides an accurate description on intermediate time scales, up to
a crossover time scale $t_{\times}$ that scales logarithmically with the
difference between loss and gain rates. However, some observables such as the
\dsop{} are not affected at all by an imbalance between loss and gain.

\textit{Correlations show ballistic light cone spreading in the PT-symmetric
  phase. In contrast, after quenches to the PT-mixed and PT-broken phases,
  correlations spread diffusively.} The single-particle spectra for isolated and
PT-symmetric driven-dissipative systems shown in Figs.~\ref{fig:overview}(a)
and~(b), respectively, suggest that PT-symmetric quadratic Liouvillians admit a
notion of quasiparticles that propagate coherently with a velocity
$v_k = \diff \omega_k/\diff k$, as is also the case for quasiparticle
excitations of an isolated system, but have a finite lifetime $\sim 1/\kappa$.
Based on this dissipative quasiparticle picture, we can expect many
characteristic features of the dynamics of isolated systems to carry over to
PT-symmetric driven-dissipative systems in the PT-symmetric phase. In
particular, we find that the spreading of correlations after quenches to the
PT-symmetric phase is described by a clear light cone structure. Interestingly,
the speed at which correlations propagate is increased as compared to isolated
systems; however, the finite lifetime of quasiparticles, which leads to an
overall exponential decay of correlations, indicates that, in fact, it is not
the case that in open systems more information is transported in a shorter time.

Light cone spreading of correlations is restricted to the PT-symmetric
phase. But also after quenches to the PT-mixed and PT-broken phases there is a
pronounced peak of correlations that propagates through the system. However, the
peak position evolves diffusively rather than ballistically as in the
PT-symmetric phase.

\textit{The growth and saturation of the subsystem entropy obeys the
  quasiparticle picture, adapted to driven-dissipative systems.} For isolated
systems, the quasiparticle picture leads to quantitative predictions for the
full time evolution of the entropy of finite subsystems~\cite{Calabrese2005,
  Alba2017, Alba2018, Calabrese2020}. One can regard the initial ground state as
a source of pairs of entangled quasiparticles with opposite momenta, which
propagate through the system with different velocities of at most
$v_{\mathrm{max}}$. If one of the two quasiparticles that form a pair is located
within the subsystem while the other one is outside of the subsystem, then this
pair contributes to the entanglement between the subsystem and its complement
and, therefore, to the subsystem entropy. Based on this picture, knowledge of
the quasiparticle velocity and the stationary value of the subsystem entropy
that is reached for $t \to \infty$ is sufficient to determine the full time
evolution of the subsystem entropy in the space-time scaling limit
$\ell, t \to \infty$, where $\ell$ is the size of the subsystem. The
quasiparticle picture has been extended to open systems, where the requirement
of ballistically propagating quasiparticles restricts its applicability to the
limit of weak dissipation $\gamma \to 0$~\cite{Alba2022, Carollo2022}, and an
additional contribution to the subsystem entropy due to the mixedness of the
state has to be accounted for. If, however, the system under consideration has
PT symmetry, then, as explained above, ballistically propagating quasiparticles
exist within the entire PT-symmetric phase. Based on this observation, in
Ref.~\cite{Starchl2022}, we have proposed an analytical conjecture for the
quasiparticle-pair contribution to the subsystem entropy of a PT-symmetric
Kitaev chain in the space-time scaling limit and for \emph{finite} dissipation
strength $\gamma$. Here, we provide further evidence for broad validity of our
conjecture by applying it to the SSH model with incoherent loss and gain, where
we again find excellent agreement with numerical data.

\textit{Quantum quenches in driven-dissipative systems can give rise to the
  unique phenomenon of directional pumping of topological disorder
  parameters. The time scales of directional pumping are determined by the soft
  modes of the PTGGE.} As detailed in Sec.~\ref{sec:dynamics-local-observ}, in
the isolated SSH model and the Kitaev chain, crossing the boundary between the
trivial and the topological phase in the course of a quench is reflected in
pumping of topological disorder parameters. That is, the \dsop{} and the fermion
parity of a subsystem of size $\ell$ exhibit oscillatory decay, crossing zero at
multiplies of a time scale $t_s$ as illustrated in the left panel in
Fig.~\ref{fig:overview}(c). The period of zero crossings $t_s$ is determined by
the momenta at which the GGE Hamiltonian vanishes---soft modes of the
GGE~\cite{Calabrese2012I, Essler2016}. In contrast, for quenches within the
trivial phase, the disorder parameters show nonoscillatory decay. Since the
Hamiltonians of the SSH model and the Kitaev chain commute with the respective
topological disorder parameters, processes that change the \dsop{} and the
fermion parity occur not in the bulk of the subsystem but at the interfaces
between the subsystem and its complement. Interestingly, for quenches in
driven-dissipative systems, the rates at which topological disorder parameters
are pumped through the left and right ends of a subsystem are
different. Therefore, as illustrated in the right panel in Fig.~\ref{fig:overview}(c),
there are two distinct time scales $t_{s, \pm}$ for zero crossings. These time
scales are determined by soft modes of the PTGGE. As shown in
Ref.~\cite{Starchl2022} for the driven-dissipative Kitaev chain, necessary
conditions for the phenomenon of directional pumping to occur are open-system
dynamics leading to mixed states, and the breaking of inversion symmetry by the
coupling to reservoirs.

\textit{A dynamical phase diagram can be defined in terms of directional
  pumping. The resulting phase boundaries and the dynamical critical behavior at
  these phase boundaries are, in general, different and independent from the
  corresponding properties defined in terms of gap closings and PT symmetry of
  the postquench Liouvillian}. PT symmetry of the Liouvillian leads to the
distinction between PT-symmetric, PT-breaking, and PT-mixed phases. For the
models we consider here, the transition from the PT-symmetric to the PT-mixed
phase occurs when the gap between the bands
$\lambda_{\pm, k} = - \imag \kappa \pm \omega_k$ closes. Note that this
corresponds to a purely dynamical phase transition that marks a qualitative
change of only the coherent dynamics: As we demonstrate using the example of the
density autocorrelation function, upon approaching the phase boundary from the
PT-symmetric phase, a characteristic period of oscillations of local correlation
functions diverges; in the PT-mixed phase, the decay of the density
autocorrelation function is overdamped. Further, this transition is purely
dynamical in the sense that it does not affect the steady state. Indeed, for
this part of our analysis, we focus on balanced loss and gain, such that the
steady state is always at infinite temperature.

Here, we show that a different and independent characterization of dynamical
phases can be given in terms of directional pumping: As explained above, the
pumping of topological disorder parameters for quenches from the trivial to the
topological phase in isolated systems becomes directional in open systems. In
particular, the topological phases of the isolated SSH model and Kitaev chain
are continuously connected to phases with pumping at different rates through
both ends of a subsystem. Surprisingly, we find transitions to phases with
pumping through only the left or the right end of a subsystem. As the transition
to such a phase is approached, one of the time scales $t_{s, \pm}$ of zero
crossings diverges.

The Kitaev chain with long-range hopping and pairing represents an especially
interesting model to study directional pumping phases and the associated
critical behavior. In particular, we find that long-range couplings can modify
the critical exponents that govern the divergence of the time scales
$t_{s, \pm}$, whereas the exponents that describe the divergence of the period
of oscillations of the density autocorrelation function at the gap closing at
the transition from the PT-symmetric to the PT-mixed phase remain
unchanged. Moreover, in the presence of long-range couplings, directional
pumping phase boundaries do not always coincide with phase boundaries that are
determined by gap closings, which implies that a divergence of one of the time
scales $t_{s, \pm}$ does not require a gap closing in the spectrum of the
Liouvillian. These findings establish directional pumping phases, phase
transitions, and the associated critical behavior as new and independent
concepts that are unique to quantum quenches in driven-open systems.

\section{Models}
\label{sec:models}

For the quench protocol outlined above, the postquench time evolution of the
system density matrix $\rho$ is described by a Markovian quantum master equation
in Lindblad form~\cite{Lindblad1976, *Gorini1976},
\begin{equation}
  \label{eq:master-equation}
  \imag \frac{\diff \rho}{\diff t} = \mathcal{L} \rho = (\mathcal{H} + \imag \mathcal{D})\rho,
\end{equation}
where the Liouvillian $\mathcal{L}$ incorporates both unitary dynamics generated
by the Hamiltonian superoperator $\mathcal{H}$ and Markovian drive and
dissipation described by the dissipator $\mathcal{D}$. These superoperators are
defined through their action on the density matrix $\rho$,
\begin{align}
  \mathcal{H} \rho & = [H, \rho], \label{eq:H-action} \\
  \mathcal{D}\rho & = \sum_{l=1}^L \left( 2L_l \rho L_l^{\dagger} - \{L_l^{\dagger}
                    L_l, \rho\} \right), \label{eq:D-action}
\end{align}
where $L_l$ are the quantum jump operators, describing the coupling between
system and environment. In this work, we focus on noninteracting fermionic
lattice models that are described by a Hamiltonian $H$ and jump operators $L_l$
that are quadratic and linear in fermionic operators, respectively, leading to a
quadratic Liouvillian $\mathcal{L}$. For the Hamiltonian $H$, we study two
models: the SSH model and the Kitaev chain, which are paradigmatic examples for
a one-dimensional topological insulator and superconductor, respectively. We
specify the Hamiltonians and jump operators in the following, and give a
detailed description of the symmetries, spectrum, and time evolution for SSH
model. For a detailed account of the driven-dissipative Kitaev chain, we refer
to Ref.~\cite{Starchl2022}.

\subsection{Driven-dissipative SSH model}
\label{sec:SSH-driven-dissipative}

The SSH model with $L$ unit cells is described by the following many-body
Hamiltonian~\cite{Su1979, *Heeger1988}:
\begin{equation}
  \label{eq:H-SSH}
  H = \sum_{l = 1}^L \left( J_1 c_{A, l}^{\dagger} c_{B, l} + J_2 c_{A, l +
      1}^{\dagger} c_{B, l} + \hc \right),
\end{equation}
where $c_{s,l}$ and $c_{s,l}^{\dagger}$ are, respectively, annihilation and
creation operators for fermions on sublattice $s\in\{A,B\}$ at lattice site
$l$. Further, $J_1, J_2 \in \R_{> 0}$ are the hopping amplitudes within and
between unit cells. Unless stated otherwise, we assume periodic boundary
conditions (PBC) with $c_{s,L+1} = c_{s,1}$; open boundary conditions (OBC) are
implemented by setting $c_{s,L+1} = 0$. We often find it convenient to
parameterize $J_1$ and $J_2$ in terms of total and relative hopping amplitudes,
\begin{equation}
  J = \frac{1}{2} \left( J_1 + J_2 \right), \qquad \Delta J = \frac{1}{2} \left(
    J_1 - J_2 \right).
\end{equation}
As detailed below, the SSH model belongs to the Altland-Zirnbauer class
BDI~\cite{Altland1997}. Therefore, the topology of the SSH model is
characterized by an integer-valued invariant, the winding number $W$, and the
ground state of the SSH model is topologically trivial and nontrivial for
$\Delta J > 0$ and $\Delta J < 0$, respectively~\cite{Hasan2010, Qi2011,
  Chiu2016}. In this work, we study quench dynamics whereby the initial state
$\ket{\psi(t)}$ at $t = 0$ is chosen to be the ground state $\ket{\psi_0}$ for
prequench parameters $J_{1, 0} > 0$ and $J_{2, 0} = 0$, corresponding to the
topologically trivial phase. The ground state $\ket{\psi_0}$ is Gaussian, i.e.,
the density matrix $\rho_0 = \ket{\psi_0} \bra{\psi_0}$ can be written as the
exponential of a quadratic form in fermionic operators; further, since the
Hamiltonian Eq.~\eqref{eq:H-SSH} is quadratic, also the time-evolved state
$\rho(t) = \e^{- \imag H t} \rho_0 \e^{\imag H t}$ is
Gaussian~\cite{Surace2022}. Therefore, after a quench in the isolated SSH model,
the state of the system is Gaussian at all times, and thus fully determined by
the covariance matrix,
\begin{equation}
  \label{eq:G}
  G_{l, l'}^{s, s'}(t) = \braket{ [c_{s, l}, c_{s', l'}^{\dagger}](t) },
\end{equation}
for $s,s' \in \{A,B\}$ and $l,l' \in \{1,\dots, L\}$, and where
$\braket{\cdots (t)} = \tr( \cdots \rho(t))$. In particular, anomalous
correlations vanish,
$\langle (c_{s, l} c_{s', l'})(t) \rangle = \langle (c_{s, l}^{\dagger} c_{s',
  l'}^{\dagger})(t) \rangle = 0$.
Turning now to a driven-dissipative generalization of the SSH model, we note
first that the dynamics generated by a quadratic Liouvillian $\mathcal{L}$
preserves the Gaussianity of the time-evolved mixed state
$\rho(t) = \e^{- \imag \mathcal{L} t} \rho_0$~\cite{Surace2022,
  Barthel2022}. The vanishing of anomalous correlations, which for the isolated
SSH model is a consequence of particle number conservation, can be ensured in an
open SSH model by imposing the weak $\mathrm{U}(1)$ symmetry that is defined by
the relation $\mathcal{L}(S \rho S^{\dagger}) = S \mathcal{L}(\rho) S^{\dagger}$
for $S = \e^{\imag \theta N}$ where the particle number operator is
$N = \sum_{s \in \{A, B\}} \sum_{l = 1}^L c_{s, l}^{\dagger} c_{s, l}$ and
$\theta \in [0, 2 \pi)$~\cite{Buca2012}. This weak symmetry condition guarantees
that no correlations are established between Hilbert space sectors with
different numbers of particles, and, therefore, anomalous correlations
vanish. For the Hamiltonian part of the time evolution in
Eq.~\eqref{eq:master-equation}, the weak symmetry is again a consequence of
particle number conservation of the SSH Hamiltonian Eq.~\eqref{eq:H-SSH},
$[H, N] = 0$. Note that the weak $\mathrm{U}(1)$ symmetry does not affect the
integrability of the model, which is guaranteed by the Liouvillian being
quadratic. The weak $\mathrm{U}(1)$ symmetry only ensures the vanishing of
anomalous correlations, which leads to a simplification of the dynamics. A
specific choice of dissipation that respects the weak $\mathrm{U}(1)$ is given
by incoherent loss and gain as described by the dissipator
$\mathcal{D} = \mathcal{D}_l + \mathcal{D}_g$ with
\begin{equation}
  \begin{split}
    \mathcal{D}_l \rho & = \sum_{s \in \{A, B\}} \sum_{l = 1}^L
    \left( 2 L_{l, s, l} \rho L_{l, s, l}^{\dagger} - \{
      L_{l, s, l}^{\dagger} L_{l, s, l}, \rho \} \right), \\
    \mathcal{D}_g \rho & = \sum_{s \in \{A, B\}} \sum_{l = 1}^L
    \left( 2 L_{g, s, l} \rho L_{g, s, l}^{\dagger} - \{
      L_{g, s, l}^{\dagger} L_{g, s, l}, \rho \} \right).
  \end{split}
\end{equation}
The jump operators corresponding to local loss and gain are given by
\begin{equation}
  \label{eq:L-SSH}
  L_{l, s, l} =  \sqrt{\gamma_{l, s}} c_{s, l}, \qquad L_{g, s, l} = \sqrt{\gamma_{g, s}} c_{s, l}^{\dagger},
\end{equation}
where we consider loss and gain rates $\gamma_{l,s}$ and $\gamma_{g,s}$,
respectively, that depend on the sublattice index. We parameterize these
rates in terms of their mean and difference,
\begin{equation}
  \gamma_{l/g} = \frac{\gamma_{l/g, A} + \gamma_{l/g, B}}{2}, \qquad \Delta
  \gamma_{l/g} = \frac{\gamma_{l/g, A} - \gamma_{l/g, B}}{2}.
\end{equation}
Below it will prove convenient to express the loss and gain rates in terms of
the four independent parameters $\gamma$, $\delta$, $\Delta \gamma$, and
$\Delta$, which are defined by
\begin{equation}
  \label{eq:parameters}
  \begin{aligned}
    \gamma &= \frac{\gamma_l + \gamma_g}{2},
    & \delta & = \gamma_l - \gamma_g, \\
    \Delta \gamma & = \frac{\Delta\gamma_l + \Delta\gamma_g}{2}, & \Delta & =
    \Delta\gamma_l - \Delta\gamma_g.
  \end{aligned}
\end{equation}
To determine the dynamics of the covariance matrix Eq.~\eqref{eq:G}, let us
first represent the jump operators in the general form
\begin{equation}
  \label{eq:L-l-L-g-Dirac-generic}
  L_{l,l} = \sum_{l' = 1}^{2L} B_{l,l,l'} c_{l'}, \qquad L_{g,l} =
  \sum_{l'=1}^{2L} B_{g,l,l'} c_{l'}^{\dagger},
\end{equation}
where we collect sublattice and lattice indices in a single index such that
$c_{A,l} = c_{2l-1}$ and $c_{B,l} = c_{2l}$. With the $L\times 2L$ matrices
$B_l$ and $B_g$, we can then introduce the bath matrices
\begin{equation}
  \label{eq:bath-matrices}
  M_l = B_l^{\dagger} B_l, \qquad M_g = B_g^{\transpose} B_g^{*}.
\end{equation}
Finally, the Liouvillian dynamics of a Gaussian state is described by the
equation of motion for the covariance matrix which reads~\cite{Song2019}
\begin{equation}
  \label{eq:G-eom}
  \frac{\diff G}{\diff t} = - \imag
  \left( Z G - G Z^{\dagger} \right) + 2 \left( M_l -
    M_g \right),
\end{equation}
where the generator of the dynamics $Z$ can be interpreted as a non-Hermitian
Hamiltonian and is defined by
\begin{equation}
  \label{eq:Z-matrix-SSH}
  Z = H - \imag \left( M_l + M_g \right).
\end{equation}
The derivation of Eq.~\eqref{eq:G-eom} is presented in
Appendix~\ref{sec:time-evolution-covariance-matrix}.

Before we proceed with the analysis of the driven-dissipative SSH model, let us
briefly comment on the relation between the formalism of Lindblad master
equations on which our work is based, and the framework of non-Hermitian
Hamiltonians~\cite{Ashida2020}. Formally, the matrix $Z$ defined in
Eq.~\eqref{eq:Z-matrix-SSH} is equivalent to non-Hermitian SSH model of
Refs.~\cite{Lieu2018,Klett2017,Chang2020} up to a shift by
$- \imag 2 \gamma \id$. Such a shift does not affect the topological properties
that are considered in these references. However, the shift guarantees that the
eigenvalues of $Z$ are negative semidefinite, which is a necessary requirement
for the dynamics of the covariance matrix generated by $Z$ to be stable and thus
physically meaningful. More generally, and in contrast to non-Hermitian
Hamiltonians, the master equation~\eqref{eq:master-equation}, from which the
evolution equation~\eqref{eq:G-eom} follows, provides a full description of the
time evolution of an open quantum system. In particular, the time evolution
generated by the Liouvillian $\mathcal{L}$ is completely positive and trace
preserving. This guarantees that the time-evolved density matrix $\rho(t)$
represents, at all times, a physical state, from which the expectation value of
an observable $O$ can be evaluated as $\braket{O(t)} = \tr(O \rho(t))$. In
Appendix~\ref{sec:biorthogonal-representation} we discuss how the left and right
eigenvectors of $Z$ enter $\braket{O(t)}$. For non-Hermitian Hamiltonians, these
eigenvectors are used to define biorthogonal expectation values, and are of key
importance for the characterization of topology~\cite{Bergholtz2021}.

In the following, we discuss some fundamental properties of the
driven-dissipative SSH model. In particular, we examine symmetries of the
isolated and driven-dissipative system, the corresponding spectrum and mode
structure, non-Hermitian topology, and finally the dynamics of the covariance
matrix.

\subsubsection{Symmetries of the isolated and driven-dissipative SSH model}
\label{sec:symmetries}

Symmetries of driven-dissipative systems are of fundamental importance not only
for their topological classification but also---as highlighted, in particular,
in our work~\cite{Starchl2022}---for their quench dynamics. In our discussion of
symmetries of the driven-dissipative SSH model, we will exploit translational
invariance and work in momentum space. To that end, we first rewrite the
Hamiltonian in Eq.~\eqref{eq:H-SSH} in terms of spinors
$C_l = \left(c_{A,l},c_{B,l}\right)^{\transpose}$ as
\begin{equation}
  H = \frac{1}{2} \sum_{l,l'=1}^L C_l^{\dagger} h_{l-l'} C_{l'},
\end{equation}
with the $2 \times 2$ matrix $h_l = \mathbf{h}_l \cdot \boldsymbol{\sigma}$
where $\boldsymbol{\sigma} = (\sigma_x,\sigma_y,\sigma_z)^{\transpose}$ is a
vector of Pauli matrices. The momentum-space representation of the SSH model is
given by the Bloch Hamiltonian $h_k$ that is defined as
\begin{equation}
  \label{eq:hk-SSH}
  h_k = \sum_{l=1}^L \e^{-\imag k l} h_l = \mathbf{h}_k \cdot \boldsymbol{\sigma},
\end{equation}
where
\begin{equation}
  \label{eq:hk-vector-SSH}
  \mathbf{h}_k = \left(J_1 + J_2 \cos(k), J_2 \sin(k), 0\right)^{\transpose}.
\end{equation}
Similarly, also the bath matrices in Eq.~\eqref{eq:bath-matrices} can be
expressed in terms of translationally invariant $2\times 2$ blocks:
\begin{equation}
  \label{eq:bath-matrices-blocks}
  m_{l,l} = \delta_{l, 0}
  \begin{pmatrix}
    \gamma_{l,A} & 0\\0 &\gamma_{l,B}
  \end{pmatrix},
  \qquad
  m_{g,l} = \delta_{l,0} 
  \begin{pmatrix}
    \gamma_{g,A} & 0\\0 &\gamma_{g,B}
  \end{pmatrix}.
\end{equation}
Combining the block representations of the Hamiltonian and the bath matrices, we
can write the matrix $Z$ given in Eq.~\eqref{eq:Z-matrix-SSH} in the form
\begin{equation}
  z_l = h_l - \imag \left( m_{l,l}+m_{g,l} \right).
\end{equation}
Then, as in Eq.~\eqref{eq:hk-SSH}, we obtain the non-Hermitian Bloch Hamiltonian
$z_k$ that generates the time evolution of the covariance matrix:
\begin{equation}
  \label{eq:zk}
  z_k = z_\id \id + \mathbf{z}_k \cdot \boldsymbol{\sigma},
\end{equation}
with 
\begin{equation}
  \label{eq:z-k-components}
  z_\id = -\imag 2 \gamma, \qquad \mathbf{z}_k = \mathbf{h}_k +\imag \mathbf{b},
  \qquad \mathbf{b} = -2 \Delta\gamma  \unitvec{e}_z,
\end{equation}
where $\unitvec{e}_z = \left( 0, 0, 1 \right)^{\transpose}$ is a unit vector
along the $z$ axis.

\paragraph{Isolated system.}

We first discuss symmetries of the isolated SSH model that is described by the
Bloch Hamiltonian in Eq.~\eqref{eq:hk-SSH}. As stated above, the SSH model
belongs to the Altland-Zirnbauer class BDI and has particle-hole symmetry (PHS),
time reversal symmetry (TRS), and chiral symmetry (CS). Due to the Hermiticity
of the Bloch Hamiltonian, $h_k = h_k^{\dagger}$, each of these symmetries can be
expressed in two equivalent ways:
\begin{equation}
  \label{eq:hk-symmetries}
  \begin{split}
    \text{PHS:} \qquad & h_k = -\sigma_z h_{-k}^{\transpose} \sigma_z = -\sigma_z
                         h_{-k}^{*} \sigma_z, \\
    \text{TRS:} \qquad & h_k = h_{-k}^{\transpose} = h_{-k}^{*}, \\
    \text{CS:} \qquad & h_k = -\sigma_z h_k \sigma_z = -\sigma_z h_k^{\dagger} \sigma_z,
  \end{split}
\end{equation}
which can be confirmed by using
$-\sigma_z \boldsymbol{\sigma} \sigma_z =
(\sigma_x,-\sigma_y,-\sigma_z)^{\transpose}$
and $\mathbf{h}_{-k} = (h_{x,k},-h_{y,k},h_{z,k})^{\transpose}$. Another symmetry
that will play a key role in the following is inversion symmetry (IS). In real
space and when expressed as a transformation of the fermionic operators
$c_{s, l}$, inversion amounts to an exchange of the sublattices and a reflection
across the middle of the chain, $c_{A/B, l} \mapsto c_{B/A, L + 1 - l}$. The
combination of IS with TRS yields PT symmetry (PTS). In momentum space, IS and
PTS are described by
\begin{equation}
  \label{eq:hk-IS-PTS}
  \begin{split}
    \text{IS:} \qquad & h_k = \sigma_x h_{-k} \sigma_x = \sigma_x h_{-k}^{\dagger}
    \sigma_x, \\
    \text{PTS:} \qquad & h_k = \sigma_x h_k^{\transpose} \sigma_x = \sigma_x
    h_k^{*} \sigma_x.
  \end{split}
\end{equation}

\paragraph{Driven-dissipative system.}

For the driven-dissipative SSH model, we can regard the non-Hermitian matrix
$z_k \neq z_k^{\dagger} $ in Eq.~\eqref{eq:zk} as the generalization of the Bloch
Hamiltonian to open systems. Of the two equivalent versions of the symmetries of
the Bloch Hamiltonian $h_k$ stated in Eq.~\eqref{eq:hk-symmetries}, only one of each
applies to $z_k$:
\begin{equation}
  \label{eq:zk-symmetries}
  \begin{split}
    \text{PHS$^{\dagger}$:} \qquad & z_k = -\sigma_z z_{-k}^{*} \sigma_z, \\
  \text{TRS$^{\dagger}$:} \qquad & z_k = z_{-k}^{\transpose}, \\
  \text{CS:} \qquad & z_k = -\sigma_z z_{k}^{\dagger} \sigma_z.
  \end{split}
\end{equation}
Therefore, in the nomenclature of Ref.~\cite{Kawabata2019}, the
driven-dissipative SSH model belongs to the class
$\mathrm{BDI}^{\dagger}$. Further, inversion symmetry in the form given in
Eq.~\eqref{eq:hk-IS-PTS} is broken by the dissipative contributions to
$z_k$. However, an inversion symmetry IS$^{\dagger}$ still applies to the
traceless part $z_k' = z_k - z_\id \id$ with $z_{\id}$ given in
Eq.~\eqref{eq:z-k-components}; and by combining $\text{IS}^{\dagger}$ with
$\text{TRS}^{\dagger}$ we obtain PTS of $z_k'$:
\begin{equation}
  \label{eq:zk-IS-PTS}
  \begin{split}
    \text{IS$^{\dagger}$:} \qquad & z_k' = \sigma_x z_{-k}^{\prime
      \dagger}\sigma_x, \\
    \text{PTS:} \qquad & z_k' = \sigma_x z_{k}^{\prime \ast}\sigma_x.
  \end{split}
\end{equation}
This form of a PT symmetry which applies after a shift that renders the
generator of the dynamics traceless is called passive
PT-symmety~\cite{Ornigotti2014,Joglekar2018,Roccati2022}.

Let us now discuss the implication of PTS for the mode structure of the
Liouvillian~\cite{Bender1998,*Bender2008}. The eigenvalues and eigenvectors of
the shifted matrix $z_k'$ are denoted by
$\lambda_{\pm,k}' = \pm\sqrt{\mathbf{z}_k\cdot \mathbf{z}_k}$ and
$\ket{\psi_{\pm,k}}$, respectively. Since $z_k'$ and $z_k$ are related by a
shift $z_{\id} \id$, they have the same eigenvectors, and their eigenvalues are
related by $\lambda_{\pm, k} = \lambda_{\pm, k}' + z_{\id}$ with
$z_{\id} = - \imag 2 \gamma$. The PTS condition in Eq.~\eqref{eq:zk-IS-PTS} then
leads to
\begin{equation}
  z_k' \sigma_x \ket{\psi_{\pm,k}}^{*} = \lambda_{\pm,k}^{\prime \ast} \sigma_x \ket{\psi_{\pm,k}}^{*}.
\end{equation}
This implies that for any eigenvalue $\lambda_{\pm, k}'$ of $z_k'$ correpsonding
to an eigenvector $\ket{\psi_{\pm, k}}$, there is also an eigenvalue
$\lambda_{\pm, k}^{\prime *}$ with eigenvector
$\sigma_x \ket{\psi_{\pm, k}}^{*}$. Since for a given value of $k$ there are
only two eigenvalues which are related by $\lambda_{+, k}' = - \lambda_{-, k}'$,
there are two possibilities: (i) PT-symmetric eigenmodes with
$\lambda_{\pm, k}' = \lambda_{\pm, k}^{\prime *} \in \R$ and
$\sigma_x \ket{\psi_{\pm,k}}^{*} = \ket{\psi_{\pm,k}}$. For the eigenvalues
$\lambda_{\pm,k}$ of the full matrix $z_k$ this implies that
$\Im(\lambda_{\pm,k}) = - 2 \gamma$ and
$\Re(\lambda_{+,k}) = -\Re(\lambda_{-,k})$. (ii) Alternatively, there exist
PT-breaking eigenmodes with
$\lambda_{\pm, k}' = - \lambda_{\pm, k}^{\prime *} \in \imag \R$ and
eigenvectors obeying $\sigma_x \ket{\psi_{\pm,k}}^{*} = \ket{\psi_{\mp,k}}$
such that $\Re(\lambda_{\pm,k})=0$ and
$\Im(\lambda_{+,k}) + 2 \gamma =- \left( \Im(\lambda_{-,k}) + 2\gamma \right)$.
That is, the eigenvalues that are associated with PT-breaking eigenmodes are
related by reflection across the line $- \imag 2 \gamma$.

\begin{figure}
  \centering
  \includegraphics[width=\linewidth]{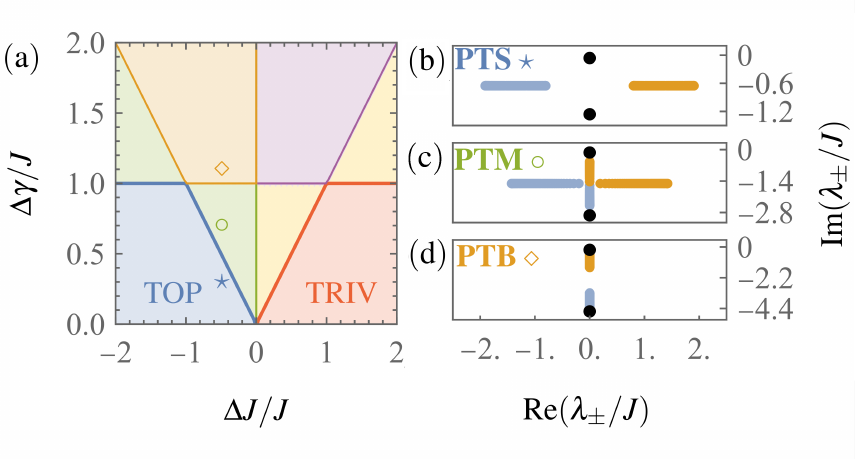}
  \caption{(a) Dynamical phase diagram of the driven-dissipative SSH model with
    topological phase for $\Delta J < 0$ and trivial phase for $\Delta J >0$.
    The model features PT-symmetric (blue, red), PT-broken (orange, purple), and
    PT-mixed phases (green, yellow). Examples of the mode structure for the
    three topological phases are shown for (b)~the PT-symmetric (PTS), (c)~the
    PT-mixed (PTM), and (d)~the PT-broken (PTB) phases. Positive (negative)
    bands are represented by orange (blue) dots and edge modes for a system with
    OBC by black dots. Star, circle and diamond markers in (a) indicate
    parameters corresponding to the spectra shown in (b)--(d).}
  \label{fig:SSH-phasediagram}
\end{figure}

Depending on the symmetry of the eigenvectors $\ket{\psi_{\pm, k}}$ under the PT
transformation, we can distinguish three different phases: (i)~In the
PT-symmetric phase, all eigenmodes are PT-symmetric; then, eigenvalues have a
constant imaginary value, while the real part is dispersive. This is shown in
Fig.~\ref{fig:SSH-phasediagram}(b). (ii)~In the PT-broken phase, all eigenmodes
are PT-breaking, and the corresponding eigenvalues are purely imaginary and
dispersive as illustrated in Fig.~\ref{fig:SSH-phasediagram}(d). (iii)~Finally,
in the PT-mixed phase, PT-symmetric and PT-breaking modes exist
simultaneously. Therefore, as shown in Fig.~\ref{fig:SSH-phasediagram}(c), there
are two sets of eigenvalues: one that is dispersive only in the real part, and
one that is dispersive only in the imaginary part. The resulting dynamical phase
diagram of the driven-dissipative SSH model is shown in
Fig.~\ref{fig:SSH-phasediagram}(a).
 
\subsubsection{Spectrum of the Liouvillian}
\label{sec:spectrum-of-the-Liouvillian}

We next consider the spectra of the isolated and driven-dissipative SSH models
with PBC. As shown in Appendix~\ref{sec:diagonalization-zk}, the Bloch
Hamiltonian $h_k$ Eq.~\eqref{eq:hk-SSH} can be diagonalized as
\begin{equation}
  \label{eq:hk-diagonal}
  U_k^{\dagger} h_k U_k = \varepsilon_k \sigma_z,
\end{equation}
where the unitary matrix $U_k$ is given by Eq.~\eqref{eq:Uk} after setting
$\gamma = \Delta \gamma = 0$, and with the single particle dispersion relation $\varepsilon_k$ of the isolated SSH model defined by the magnitude of the Bloch
vector,
\begin{equation}
  \label{eq:dispersion-isolated-SSH}
  \varepsilon_k = \abs{\mathbf{h}_k} = \sqrt{J_1^2 + J_2^2 + 2 J_1 J_2 \cos(k)}.
\end{equation}
For future reference, we note that in terms of the fermionic operators
$d_{s, k}$ defined through
\begin{equation}
  \label{eq:d-s-k}
  \begin{pmatrix}
    d_{A, k} \\ d_{B, k}
  \end{pmatrix}
  = U_k^{\dagger}
  \begin{pmatrix}
    c_{A, k} \\ c_{B, k}
  \end{pmatrix}, \qquad c_{s, k} = \frac{1}{\sqrt{L}} \sum_{l = 1}^L \e^{-\imag
    k l} c_{s, l},
\end{equation}
the Hamiltonian takes the following diagonal form:
\begin{equation}
  H = \sum_{k\in\BZ} \varepsilon_k \left(d_{A,k}^{\dagger} d_{A,k} - d_{B,k}^{\dagger} d_{B,k}\right),
\end{equation}
where the Brillouin zone is
$\BZ = \{ - \pi + \Delta k, - \pi + 2\Delta k, \dotsc, \pi \}$ with
$\Delta k = 2 \pi/L$. The ground state $\ket{\psi_0}$ is obtained by filling the
band with negative energy,
\begin{equation}
  \label{eq:ground-state}
  \ket{\psi_0} = \prod_{k\in\BZ} d_{B,k}^{\dagger} \left\vert \Omega \right\rangle,
\end{equation}
where $\left\vert \Omega \right\rangle$ is the vacuum of particles.

Applying the formalism of third quantization~\cite{Prosen2008,*Prosen2010}, one
can show that the spectrum of the Liouvillian is determined by the eigenvalues
of the matrix $z_k$ in essentially the same way as the spectrum of the
second-quantized Hamiltonian Eq.~\eqref{eq:H-SSH} is obtained by occupying
single-particle states with energies $\pm \varepsilon_k$. The eigenvalues of
$z_k$, which thus can be regarded as forming the single-particle spectrum of the
Liouvillian, form two bands and are given by
\begin{equation}
  \label{eq:lambda}
  \lambda_{\pm, k}=-\imag 2 \gamma \pm \omega_k,
\end{equation}
with the dispersion relation
\begin{equation}
  \label{eq:dispersion-open-SSH}
  \omega_k = \sqrt{\varepsilon_k^2 - 4 \Delta\gamma^2},
\end{equation}
where $\gamma$ and $\Delta \gamma$ are defined in Eq.~\eqref{eq:parameters}. The
dispersion relation $\omega_k$ determines the dynamical phase diagram of the
driven-dissipative SSH model in the $\Delta J$-$\Delta\gamma$ plane depicted in
Fig.~\ref{fig:SSH-phasediagram}(a). In particular, different phases can be
defined in terms of the gap structure of the bands
$\lambda_{\pm ,k}$~\cite{Kawabata2019}---as we explain next, this is equivalent
to the distinction of phases in terms of PT symmetry: For small values of
$\Delta \gamma$, the bands are separated by a real line gap (blue, red regions
in the figure); that is, $\omega_k > 0$ for all values of $k \in \BZ$, with a
typical complex band structure shown in Fig.~\ref{fig:SSH-phasediagram}(b). This
is the PT-symmetric phase. Upon increasing the value of $\Delta \gamma$, the
real line gap closes, and the spectrum of $z_k$ is gapless in a finite region of
the $\Delta J$-$\Delta\gamma$ plane (green, yellow). Then, for a range of
momenta, $\omega_k$ is real; for all other values of $k$,
$\omega_k = \imag \kappa_k$ is purely imaginary, with
\begin{equation}
  \label{eq:kappa-k}
  \kappa_k = \sqrt{4 \Delta \gamma^2 - \varepsilon_k^2}.
\end{equation}
A typical band structure in this gapless or PT-mixed phase is illustrated in
Fig.~\ref{fig:SSH-phasediagram}(c). The phase boundary between PT-symmetric and
PT-mixed phases is determined by a critical value $\Delta \gamma_c$. For
$\abs{\Delta J} < J$, the critical value is $\Delta \gamma_c = \abs{\Delta J}$,
and for $\abs{\Delta J} > J$, it is given by $\Delta \gamma_c = J$. At
$\abs{\Delta J} = J$, the dispersion is flat, $\omega_k = 2 J$, which enables a
direct transition between the phase with a real line gap and the phase with an
imaginary line gap or PT-broken phase (orange, purple), where $\kappa_k > 0$ for
all values of $k \in \BZ$. The band structure in this phase is exemplified in
Fig.~\ref{fig:SSH-phasediagram}(d).

\subsubsection{Non-Hermitian topology}
\label{sec:non-hermitian-topology}

We have already mentioned that the SSH model is in the Altland-Zirnbauer class
BDI. The class BDI in one spatial dimension is topologically nontrivial and
characterized by an integer-valued invariant, called the winding number $W$. For
models with a real line gap, the Altland-Zirnbauer classification can be
extended to non-Hermitian systems, where in the nomenclature of
Ref.~\cite{Kawabata2019}, the driven-dissipative SSH model belongs to the class
$\mathrm{BDI}^{\dagger}$. Accordingly, within the blue and red regions in
Fig.~\ref{fig:SSH-phasediagram}(a), which correspond to the phase with a real
line gap or, equivalently, the PT-symmetric phase, the topology of the
driven-dissipative Kitaev chain is characterized by a non-Hermitian
generalization of the winding number. To calculate the non-Hermitian winding
number, we first find the left and right eigenstates of $z_k$, which we denote
by $\ket{\psi_{\pm,k}^L}$ and $\ket{\psi_{\pm,k}^{R}}$, respectively. By
employing the representation of $z_k$ given in Eq.~\eqref{eq:zk-diag}, we find
the eigenvalue equations
\begin{equation}  
  z_k' U_k \left\vert \pm \right\rangle = \lambda_{\pm, k} U_k \left\vert \pm \right\rangle, \quad
  \left\langle \pm \right\vert U_k^{-1} z_k' = \lambda_{\pm, k} \left\langle
    \pm \right\vert U_k^{-1},
\end{equation}
where $\left\vert\pm\right\rangle$ are the eigenvectors of $\sigma_z$ with
eigenvalues $\pm 1$. The left and right eigenvectors of $z_k$ are thus given by
\begin{equation}
  \ket{\psi_{\pm, k}^L} = U_k^{- \dagger}\left\vert\pm\right\rangle,
  \qquad 
  \ket{\psi_{\pm, k}^R} = U_k \left\vert\pm\right\rangle,
\end{equation}
respectively, where $U_k^{- \dagger}$ is the short-hand notation for the inverse of the
Hermitian adjoint of $U_k$. We can now define a projector on the $-$ band:
\begin{equation}
  Q_k = \id - \ket{\psi_{-,k}^R}\bra{\psi_{-,k}^L} - \ket{\psi_{-,k}^L}\bra{\psi_{-,k}^R},
\end{equation}
and determine the non-Hermitian winding number by
\begin{equation}
  W = \frac{\imag}{4 \pi}\int_{-\pi}^\pi \diff k \tr \! \left(\Gamma Q_k^{-1}
    \frac{\diff}{\diff k} Q_k \right),
\end{equation}
where $\Gamma = \sigma_z$ is the matrix that appears in the chiral symmetry
condition Eq.~\eqref{eq:zk-symmetries}~\cite{Kawabata2019}, and we have replaced
summation of $k \in \BZ$ by an integral over $k \in [-\pi, \pi]$ in the
thermodynamic limit $L \to \infty$. With $\{Q_k,\sigma_z\} = 0$, following from
chiral symmetry, and $Q = Q^{\dagger}$ by definition, we find that $Q_k$ is
off-diagonal with
\begin{equation}
  Q_k = 
  \begin{pmatrix}
    0 & q_k \\
    q_k^{*} & 0
  \end{pmatrix},
\end{equation}
and the calculation reduces to 
\begin{equation}
  \label{eq:W-q}
  W =  -\frac{\imag}{2 \pi } \int_{-\pi}^\pi \diff k \frac{1}{q_k} \frac{\diff q_k}{\diff k}.
\end{equation}
In the PT-symmetric phase, we find
$q_k = \left. - \left( J_1 + J_2 \e^{-\imag k} \right) \middle/ \omega_k
\right.$, and, therefore,
\begin{equation}
  \frac{1}{q_k} \frac{\diff q_k}{\diff k} = -\imag \frac{J_2}{J_2 + \e^{\imag k}
    J_1} + \frac{J_1 J_2 \sin(k)}{J_1^2 + J_2^2 - 4 \Delta\gamma^2 + 2 J_1 J_2
    \cos(k)}.
\end{equation}
The second term on the right-hand side of this equation is an odd function of
$k$ and does not contribute to the symmetric integral in Eq.~\eqref{eq:W-q}; the
first term yields
\begin{equation}
  W =  \frac{1}{2 \pi} \int_{-\pi}^{\pi} \diff k  \frac{J_2}{J_2 + \e^{\imag k} J_1},
\end{equation}
which is solved by substituting $z=\e^{\imag k}$ and integrating over the unit
circle $\abs{z}=1$. Simple poles of the integrand are located at $z_1 = 0$ and
$z_2 = -J_2/J_1$. By using the residue theorem, we obtain the result
\begin{equation}
  W = 
  \begin{cases}
    0 & \text{for } \Delta J > 0,  \\
    1 & \text{for } \Delta J < 0.
  \end{cases}
\end{equation}
This is essentially the same result as for the isolated SSH chain, but extended
to the whole PT-symmetric phase. The topological PT-symmetric phase is indicated
in Fig.~\ref{fig:SSH-phasediagram}(a) by the blue area with the label ``TOP,''
and the trivial phase by the red area with the label ``TRIV.'' Interestingly,
while the definition of the non-Hermitian winding number is restricted to the
PT-symmetric phase with a real line gap, edge modes with
$\lambda_{\mathrm{edge}} = 0, \, -\imag 4 \Delta \gamma$ occur in a chain with
OBC when $\Delta J < 0$ and for arbitrarily large values of $\Delta \gamma$,
including in the PT-mixed and PT-broken phases. In
Figs.~\ref{fig:SSH-phasediagram}(b), (c), and~(d), edge modes are indicated with
black dots. The existence of these edge modes can be understood to be a
consequence of chiral symmetry~\cite{Sayyad2021}: Due to chiral symmetry, each
sublattice supports one edge mode. But on a given sublattice, the non-Hermitian
contribution to $Z$ in Eq.~\eqref{eq:Z-matrix-SSH} is constant, and its presence
does not affect the eigenvectors with support on that sublattice. Therefore, the
edge modes of $Z$ are identical to the edge modes of the Hamiltonian $H$, and
their existence is determined by the topology of $H$. This concludes our
discussion of the static properties of the SSH model, and we turn now to quench
dynamics, which are the main interest of our work.

\subsubsection{Dynamics of the driven-dissipative SSH model}
\label{sec:SSH-covariance-dynamics}

As explained at the beginning of Sec.~\ref{sec:SSH-driven-dissipative}, in
this work, we consider quench dynamics with the system initially prepared in the
ground state given in Eq.~\eqref{eq:ground-state}, for prequench parameters
$J_{1, 0} > 0$ and $J_{2, 0} = 0$ corresponding to the trivial
phase. Gaussianity of this state is preserved in the time evolution generated by
a quadratic Liouvillian, and, therefore, the state of the system is fully
determined by the covariance matrix $G$ defined in Eq.~\eqref{eq:G}, whose
dynamics are described by Eq.~\eqref{eq:G-eom}. For PBC, the covariance matrix
is a $2L\times 2L$ block Toeplitz matrix, which is built from $2\times2$ blocks
given by
\begin{equation}
  g_{l - l'} = 
  \begin{pmatrix}
    \braket{[c_{A,l},c_{A,l'}^{\dagger}]} & \braket{[c_{A,l},c_{B,l'}^{\dagger}]} \\
    \braket{[c_{B,l},c_{A,l'}^{\dagger}]} &
    \braket{[c_{B,l},c_{B,l'}^{\dagger}]}
  \end{pmatrix}.
\end{equation}
Further, for PBC and due to translational invariance of the Hamiltonian, the
Markovian baths and the chosen initial state, all matrices in
Eq.~\eqref{eq:G-eom} are block-circulant Toeplitz matrices. Therefore, the
representation of the blocks of the covariance matrix in momentum space, which is
obtained through a discrete Fourier transform,
\begin{equation}
  \label{eq:block-fourier}
  g_k = \sum_{l = 1}^L \e^{- \imag k l} g_l,  
\end{equation}
obeys the following equation of motion:
\begin{equation}
  \label{eq:gk-eom}
  \frac{\diff g_{k}}{\diff t} = -\imag \left( z_k g_k - g_k z_{k}^{\dagger} \right) + y_k,
\end{equation}
with
\begin{equation}
  \label{eq:yk}
  y_k = 2 \left( m_{l,k} - m_{g,k} \right) = 2 \left( \delta \id + \Delta
    \sigma_z \right),
\end{equation}
where $m_{l,k}$ and $m_{g,k}$ are the momentum-space representations of the
blocks of the bath matrices defined in Eq.~\eqref{eq:bath-matrices-blocks}. The
time-evolved covariance matrix in momentum space can be split into two
contributions,
\begin{equation}
  \label{eq:gk(t)}
  g_k(t)  =  g_{1,k}(t) + g_{2,k}(t),
\end{equation}
where
\begin{align}
  \label{eq:g1-g2}
  g_{1,k}(t) &=  \e^{-\imag z_k t} g_k(0) \e^{\imag z_{k}^{\dagger} t}, \\
  \label{eq:g2k(t)-explicit}
  g_{2,k}(t) &=  \int_{0}^{t} \diff t' \e^{-\imag z_k \left(t-t'\right)} y_k \e^{\imag z_{k}^{\dagger} \left(t-t'\right)}.
\end{align}
Here, $g_{1,k}(t)$ encodes information of the initial condition $g_k(0)$; the
second contribution $g_{2,k}(t)$ has no counterpart in isolated systems and
describes the approach to the steady state $\rho_{\mathrm{SS}}$ for
$t \to \infty$, where $\mathcal{L}(\rho_{\mathrm{SS}}) = 0$.

To evaluate the initial value $g_k(0)$, let us first define spinors
$C_k = \left( c_{A, k}, c_{B, k} \right)^{\transpose}$ and
$D_k = \left(d_{A,k},d_{B,k}\right)^{\transpose}$, where the operators
$c_{s, k}$ and $d_{s, k}$ are defined in Eq.~\eqref{eq:d-s-k}. Further, by
rearranging Eq.~\eqref{eq:hk-diagonal} and defining an initial unit vector
$\unitvec{n}_{0,k}=h_{0,k}/\varepsilon_{0,k}$, where $h_{0, k}$ and
$\varepsilon_{0, k}$ are, respectively, the Bloch Hamiltonian and the
single-particle dispersion relation for prequench parameters $J_{1, 0}$ and
$J_{2, 0}$, we find $\unitvec{n}_{0,k} = U_k \sigma_z U_k^{\dagger}$. Finally,
by evaluating
$\langle D_k D_k^{\dagger} \rangle_0 = \braket{\psi_0 | D_k D_k^{\dagger} |
  \psi_0} = \left(\id +\sigma_z\right) \! /2$
for the ground state $\ket{\psi_0}$ in Eq.~\eqref{eq:ground-state}, we obtain
\begin{equation}
  \label{eq:g-k-0-general}
  g_k(0) = 2 \langle C_k C_k^{\dagger} \rangle_0 - \id = 2U_k \langle D_k
  D_k^{\dagger} \rangle_0 U_k^{\dagger} - \id = \unitvec{n}_{0,k} \cdot
  \boldsymbol{\sigma}.
\end{equation} 
In particular, for our choice of prequench parameters given by $J_{1, 0} > 0$
and $J_{2, 0} = 0$, we find
\begin{equation}
  \label{eq:gk-0}
  g_k(0) = \sigma_x.
\end{equation}
With this result for the initial value, and after some algebraic
simplifications, a general form for the time-dependent covariance matrix can be
found~\cite{Starchl2022}. We first split Eqs.~\eqref{eq:g1-g2}
and~\eqref{eq:g2k(t)-explicit} into two contributions that are proportional to
the identity and traceless, respectively,
\begin{equation}
  \label{eq:g12k(t)}
  \begin{split}
    g_{1,k}(t) & = g_{1,\id,k}(t) \id + \mathbf{g}_{1,k}(t) \cdot \boldsymbol{\sigma}, \\
    g_{2,k}(t) & = g_{2,\id,k}(t) \id + \mathbf{g}_{2,k}(t) \cdot
    \boldsymbol{\sigma}.
  \end{split}
\end{equation}
In the PT-symmetric phase, the components of $g_{1,k}(t)$ read
\begin{equation}
  \label{eq:g1-id-k}
  g_{1,\id,k}(t)  = -\frac{\e^{ -4 \gamma t}}{\omega_k^2} \left(1 - \cos(2
    \omega_k t) \right) \unitvec{n}_{0,k} \cdot
  \left(\mathbf{h}_k\times\mathbf{b}\right),
\end{equation}
and
\begin{multline}
  \label{eq:g1-sigma-k}
  \mathbf{g}_{1,k}(t) = \e^{- 4 \gamma t} \left[ \unitvec{n}_{0,k} -
    \frac{\varepsilon_k^2}{\omega_k^2}\left(1-\cos(2\omega_k
      t)\right)\mathbf{h}_{\perp,k} \right. \\
  \left. + \frac{\varepsilon_k}{\omega_k}\sin(2 \omega_k t)
    \mathbf{h}_{o,k}\right],
\end{multline}
where we have introduced vectors $\mathbf{h}_{\parallel, k}$ and
$\mathbf{h}_{\perp, k}$ which are parallel and perpendicular to
$\unitvec{n}_k = \mathbf{h}_k/\varepsilon_k$, respectively, and the out-of-plane
vector $\mathbf{h}_{o, k}$ that is orthogonal to both $\unitvec{n}_{0, k}$ and
$\unitvec{n}_k$:
\begin{equation}
  \mathbf{h}_{\parallel,k} = \left(\unitvec{n}_{0,k} \cdot \unitvec{n}_k\right)\unitvec{n}_k, \quad
  \mathbf{h}_{\perp,k}  = \unitvec{n}_{0,k} - \mathbf{h}_{\parallel,k}, \quad
  \mathbf{h}_{o,k}  = - \unitvec{n}_{0,k} \times \unitvec{n}_k.
\end{equation}
Here, for $\unitvec{n}_{0,k} \nparallel \unitvec{n}_k$ the dynamics of
$\e^{4 \gamma t} \mathbf{g}_{1,k}(t)$ describe an ellipse with center
$\unitvec{n}_{0,k} -
\left(\varepsilon_k^2/\omega_k^2\right)\mathbf{h}_{\perp,k}$,
semi-major axis $\varepsilon_k^2/\omega_k^2$ pointing along
$\mathbf{h}_{\perp,k}$, and semi-minor axis $\varepsilon_k/\omega_k$ along
$\mathbf{h}_{o,k}$; for an isolated system with $\gamma = \Delta \gamma = 0$,
this reduces to the well-known precession of $\mathbf{g}_{1, k}(t)$ around
$\mathbf{h}_k$.

The components of $g_{1, k}(t)$ given in Eqs.~\eqref{eq:g1-id-k}
and~\eqref{eq:g1-sigma-k} decay exponentially at a rate $4 \gamma$. Therefore,
as indicated above, the steady state, which is determined by the limit
$t \to \infty$ of $g_k(t)$, is described by $g_{2,k}(t)$
Eq.~\eqref{eq:g2k(t)-explicit}. In turn, $g_{2,k}(t)$ is proportional to $y_k$,
which vanishes for balanced loss and gain, $\delta = \Delta = 0$. The vanishing
of all correlations indicates that balanced loss and gain lead to a steady state
at infinite temperature, $\rho_{\mathrm{SS}} = \rho_\infty = \id/2^{2 L}$. In
contrast, for $\delta, \Delta \neq 0$, also $g_{2, k}(t) \neq 0$ and the steady
state is nontrivial. The integral in the expression for $g_{2, k}(t)$ in
Eq.~\eqref{eq:g2k(t)-explicit} can be solved by elementary means, but we omit
the lengthy result.

Finally, for future reference, we briefly discuss the consequences of PHS for
$g_k(t)$. In particular, PHS of the initial state and PHS$^{\dagger}$ of $z_k$
imply that
\begin{equation}
  g_{1,k} = - \sigma_z g_{1,-k}^\transpose \sigma_z,
\end{equation}
where we omit the time argument to shorten the notation. In contrast, $y_k$ in
Eq.~\eqref{eq:yk} breaks PHS, but has TRS and commutes with $\sigma_z$. Combined
with PHS$^{\dagger}$ of $z_k$ this leads to
\begin{equation}
  g_{2,k} =  \sigma_z g_{2,-k}^\transpose \sigma_z.
\end{equation}
By inversion of Eq.~\eqref{eq:block-fourier}, one then immediately finds
\begin{equation}
  \label{eq:PHS-g1l-g2l}
  g_{1,l} = - \sigma_z g_{1,-l}^\transpose \sigma_z, \quad g_{2,l} =  \sigma_z
  g_{2,-l}^\transpose \sigma_z.
\end{equation}

\subsection{Driven-dissipative Kitaev chain}
\label{sec:driv-diss-kitaev}

In Ref.~\cite{Starchl2022}, we have presented a detailed study of the quench
dynamics of a driven-dissipative Kitaev chain with short-range hopping and
pairing. We summarize key properties of this model in the following. These
properties will form the basis for our discussion of new results that concern
the spreading of correlations and the effects of long-range hopping and pairing
in Secs.~\ref{sec:spread-corrs} and~\ref{sec:long-range-kitaev-quench},
respectively.

The Hamiltonian of a Kitaev chain~\cite{Kitaev2001} of length $L$, with hopping
matrix element $J$, pairing amplitude $\Delta$, and chemical potential $\mu$,
reads
\begin{equation}
  \label{eq:H-Kitaev}
  H = \sum_{l = 1}^L \left( -J c^{\dagger}_l c_{l + 1} + \Delta c_l c_{l + 1} + \hc
  \right) - \mu \sum_{l = 1}^L \left( c^{\dagger}_l c_l -\frac{1}{2} \right),
\end{equation}
where the fermionic annihilation and creation operators at lattice site $l$ are
$c_l$ and $c_l^{\dagger}$, respectively, with canonical anticommutation
relations $\{c_l,c _{l'}^{\dagger}\} = \delta_{l,l'}$ and
$\{c_l, c_{l'}\} = \{c_l^{\dagger}, c_{l'}^{\dagger}\} = 0$. Depending on the
observable under study, we consider both periodic boundary conditions with
$c_{L + 1} = c_1$ and open boundary conditions with $c_{L + 1} = 0$, which will
be indicated accordingly. We assume that $J$ and $\Delta$ are positive and real,
such that the Kitaev chain has TRS and belongs to the Altland-Zirnbauer class
BDI. In the following, we keep $J$ and $\Delta$ as distinct parameters in most
expressions. However, our main results are obtained for $J = \Delta$. For this
choice, the ground state of the Kitaev chain is topologically nontrivial for
$\abs{\mu} < 2 J$. In Sec.~\ref{sec:long-range-kitaev-quench}, we will also
study a generalization of the Kitaev chain that incorporates long-range hopping
and pairing.

We now subject the Kitaev chain to Markovian drive and dissipation in the form
of local particle loss and gain as described by the jump
operators~\cite{VanCaspel2019, Lieu2020, Sayyad2021}
\begin{equation}
  \label{eq:L}
  L_l = \sqrt{\gamma_l} c_l + \sqrt{\gamma_g} c_l^{\dagger},
\end{equation}
with loss and gain rates $\gamma_l$ and $\gamma_g$, respectively. A convenient
parameterization of the coupling to Markovian reservoirs is obtained by
introducing the mean and relative rates,
\begin{equation}
  \label{eq:parameters-kitaev}
  \gamma = \frac{\gamma_l + \gamma_g}{2}, \qquad \delta = \gamma_l - \gamma_g.
\end{equation}
The mean rate $\gamma$ determines the overall strength of dissipation; and the
relative rate $\delta$ can be interpreted as an effective inverse temperature in
the sense that for $\delta = 0$, the steady state $\rho_{\mathrm{SS}}$ is at
infinite temperature, $\rho_{\mathrm{SS}} = \rho_\infty = \id/2^L$, while for
$\delta \to \infty$, the jump operators Eq.~\eqref{eq:L} describe particle loss,
leading to a pure steady state, $\rho_{\mathrm{SS}} =
\ket{\Omega}\bra{\Omega}$~\cite{Starchl2022}.

In this work, we consider quenches originating from the trivial phase of the
isolated chain with $\mu_0 \to -\infty$. The initial state is then given by the
vacuum of fermions $\ket{\Omega}$. As in the case of the SSH model discussed
above, Gaussianity of the state is preserved under the evolution generated by
the quadratic Hamiltonian Eq.~\eqref{eq:H-Kitaev} and the linear jump operators
Eq.~\eqref{eq:L}, and the state is fully determined by two-point
correlations. However, for the Kitaev chain, there are also nonvanishing
anomalous correlations,
$\langle c_l c_{l'} \rangle, \langle c_l^{\dagger} c_{l'}^{\dagger} \rangle \neq
0$,
and, therefore, it is convenient to describe correlations by $2L$ real Majorana
fermions instead of $L$ complex Dirac fermions. The transformation to Majorana
operators $w_l$ reads
\begin{equation}
  \label{eq:majoranas}
  w_{2l-1} = c_l + c_l^{\dagger}, \qquad w_{2l} = \imag \left( c_l - c_l^{\dagger} \right).
\end{equation}
These operators obey the anticommutation relation
$\{w_l,w_{l'}\} = 2 \delta_{l,l'}$. The covariance matrix can then be defined
using Majorana operators as
\begin{equation}
  \label{eq:gamma-kitaev}
  \Gamma_{l,l'}(t) = \frac{\imag}{2} \braket{[w_l,w_{l'}](t)},
\end{equation}
where $\braket{\cdots(t)} = \tr(\cdots \rho(t))$. Quench dynamics can thus be
described by the time evolution of $\Gamma(t)$. For details we refer to
Ref.~\cite{Starchl2022}. Finally, analogous to the phases of the SSH model
discussed in Sec.~\ref{sec:symmetries}, the driven-dissipative Kitaev chain
features PT-symmetric, PT-breaking, and PT-mixed phases. The phase boundaries
can be defined in terms of the spectrum of the Liouvillian, determined by
$\lambda_{\pm,k} = - \imag 2 \gamma \pm \omega_k$ with the dispersion relation
\begin{equation}
  \label{eq:dispersion-open-Kitaev}
  \omega_k = \sqrt{\varepsilon_k^2 - 4 \gamma_l\gamma_g},
\end{equation}
where
\begin{equation}
  \varepsilon_k = \sqrt{\left(2 J \cos(k) + \mu \right)^2 + 4 \Delta^2 \sin(k)^2}.
\end{equation}
Of particular interest for this work is the PT-symmetric phase where
$\omega_k \in \R_{>0}$ for all $k \in \BZ$, realized for
\begin{equation}
  \label{eq:kitaev-pts-boundary}
  2 \sqrt{\gamma_l \gamma_g} < \left\vert 2J - \abs{\mu} \right\vert,
\end{equation}
while the PT-breaking phase, with $\omega_k = \imag \kappa_k$ and
$\kappa_k \in \imag \R_{>0}$ for all $k \in \BZ$, is determined by
\begin{equation}
  \label{eq:kitaev-ptb-boundary}
  2 \sqrt{\gamma_l \gamma_g} >  2J + \abs{\mu}.
\end{equation}
For values of $2 \sqrt{\gamma_l \gamma_g}$ between the boundaries given in
Eqs.~\eqref{eq:kitaev-pts-boundary} and~\eqref{eq:kitaev-ptb-boundary}, the
system is in the PT-mixed phase.

\section{PT-symmetric generalized Gibbs ensemble}
\label{sec:PTGGE}

After the technical preliminaries of the previous section, let us now focus on
the main purpose of this work, which is to study the quench dynamics and
relaxation of noninteracting driven-dissipative fermionic models. We start by
deriving the maximum entropy ensemble~\cite{Jaynes1957} for PT-symmetric free
fermionic systems in the PT-symmetric phase, the PT-symmetric generalized Gibbs
ensemble (PTGGE), using two different approaches: First, we present a derivation
based on the dephasing of the covariance matrix, which has the benefit of being
more practicable for analytical and numerical purposes; then, we show how the
PTGGE can be derived from the quadratic eigenmodes of the adjoint
Liouvillian. This, in comparison, is a more physically insightful approach,
offering a better understanding in terms of the evolution of Liouvillian
eigenmodes with modified dynamics and statistics, and connects neatly to the
structure found in isolated systems.

We have discussed the derivation of the PTGGE for the driven-dissipative Kitaev
chain in depth in Ref.~\cite{Starchl2022}. Therefore, in
Sec.~\ref{sec:SSH-PTGGE}, we present a detailed derivation of the PTGGE for the
SSH model with incoherent loss and gain, and we provide only a brief summary of
the corresponding results for the Kitaev chain in Sec.~\ref{sec:Kitaev-PTGGE}.

\subsection{Driven-dissipative SSH model}
\label{sec:SSH-PTGGE}

As stated above, we commence the derivation of the PTGGE for the SSH model in
terms of the covariance matrix. This derivation can be divided into three steps:
The first step, which we have taken in Sec.~\ref{sec:SSH-covariance-dynamics},
is to find an analytical result for the evolution of the covariance
matrix. Then, we obtain the long-time asymptotic behavior, determined by
dephasing of momentum modes. Finally, we relate the dephased covariance matrix
to the density matrix of the system. The last two steps are presented in
Sec.~\ref{sec:PTGGE-SSH-covariance-matrix} below. After this we proceed with the
derivation by finding quadratic eigenmodes of the Liouvillian in
Sec.~\ref{sec:PTGGE-SSH}.

\subsubsection{Derivation of the PTGGE from dephasing of the covariance matrix}
\label{sec:PTGGE-SSH-covariance-matrix}

In isolated integrable models that can be mapped to noninteracting fermions,
generalized thermalization to a maximum entropy ensemble following a quantum
quench happens through dephasing of momentum modes that oscillate at different
frequencies $\varepsilon_k \neq \varepsilon_{k'}$ for
$k\neq k'$~\cite{Essler2016,Barthel2008}. Consequently, local observables take
on stationary expectation values predicted by the generalized Gibbs ensemble. In
driven-dissipative integrable systems, the same mechanism is responsible for
relaxation to a maximum entropy ensemble inside the PT-symmetric
phase~\cite{Starchl2022}, yet there are fundamental differences which we will
illustrate in the following. To that end, let us study a general block $g_l(t)$
of elements of the covariance matrix, determined by the momentum-space
representation $g_k(t)$ given in Eq.~\eqref{eq:gk(t)} through inversion of
Eq.~\eqref{eq:block-fourier},
\begin{equation}
  \label{eq:g-l}
  g_l(t) =\int_{-\pi}^{\pi} \frac{\diff k}{2\pi} \, \e^{\imag k l} \left(
    g_{1,k}(t) + g_{2,k}(t) \right). 
\end{equation}
For the moment, let us consider balanced loss and gain $\delta=\Delta = 0$ such
that $g_{2,k}(t) =0$; explicit expressions for $g_{1,k}(t)$ are given in
Eqs.~\eqref{eq:g1-id-k} and~\eqref{eq:g1-sigma-k}. Note that $g_{1,k}(t)$
depends on time through an overall factor $\e^{-4 \gamma t}$, and through
oscillating factors $\sin(2 \omega_k t)$ and $\cos(2 \omega_k t)$. According to
the Riemann-Lebesgue lemma, to obtain the behavior of $g_l(t)$ for
$t \to \infty$, we have to drop these oscillating
terms~\cite{Essler2016}. Stated in terms of $g_{1,k}(t)$, in the limit
$t \to \infty$, we may write
\begin{equation}
  g_{1,k}(t) \sim g_{d,1,k} (t) = \e^{-4\gamma t} g_{d,1,k}',
\end{equation}
where the subscript ``$d$'' denotes the dephased value obtained by omitting
oscillatory contributions, and where $g_{d,1,k}'$ is time-independent and
explicitly given by
\begin{equation}
  g_{d,1,k}' = g_{d,1,\id,k}' \id + \mathbf{g}_{d,1,k}' \cdot \boldsymbol{\sigma},
\end{equation} 
where
\begin{equation}
  g_{d,1,\id,k}' = - \frac{1}{\omega_k^2}\unitvec{n}_{0,k} \cdot
  \left( \mathbf{h}_k \times \mathbf{b} \right) = \frac{2\varepsilon_k \Delta\gamma}{\omega_k^2} \sin(\Delta\phi_k),
\end{equation}
and
\begin{equation}
  \mathbf{g}_{d,1,k}' = \unitvec{n}_{0,k} -
  \frac{\varepsilon_k^2}{\omega_k^2}\mathbf{h}_{\perp,k} = \cos(\Delta \phi_k)
  \unitvec{n}_{k} + \frac{4\Delta\gamma^2}{\omega_k^2}\sin(\Delta\phi_k)
  \unitvec{e}_z \times \unitvec{n}_{k},
\end{equation}
with the angle $\Delta \phi_k$ defined by
\begin{equation}
  \label{eq:cos-sin-delta-phi}
  \cos\left(\Delta \phi_k\right)  = \unitvec{n}_{0,k} \cdot \unitvec{n}_k,
  \qquad \sin(\Delta \phi_k) = (\unitvec{n}_{0,k} \times \unitvec{n}_k) \cdot
  \unitvec{e}_z.
\end{equation}
The asymptotic behavior of $g_l(t)$ is thus given by
\begin{equation}
  \label{eq:g-l-asymptotic}
  g_l(t) \sim \e^{- 4 \gamma t} \int_{-\pi}^{\pi} \frac{\diff k}{2 \pi} \,
  \e^{\imag k l} g_{d, 1, k}'.
\end{equation}
Since for a Gaussian state any expectation value can be expressed in terms of
the covariance matrix by using Wick's theorem, this result shows that after
appropriate rescaling to compensate overall exponential decay, local observables
relax to stationary values that are determined by the dephased covariance
matrix. In particular, the expectation value of a product $O_{\ell}$ of $\ell$
fermionic operators becomes stationary after rescaling with a factor of
$\e^{2 \ell \gamma t}$. We note that this applies only when
$\langle O_{\ell} \rangle_{\mathrm{SS}} = \tr(O_{\ell} \rho_{\mathrm{SS}}) =
\tr(O_{\ell})/2^{2 L} = 0$.
Otherwise, one should consider the tracless part
$O_{\ell} - \tr(O_{\ell})/2^{2 L}$, the expectation value of which measures
actual correlations and vanishes in the steady state at infinite
temperature. Further, similarly to isolated systems, in the equilibrated values
of rescaled local observables, memory of the initial state is preserved through
the angle defined in Eq.~\eqref{eq:cos-sin-delta-phi}.

As a consequence of Gaussianity, not only arbitrary expectation values but also
the full state of the system is determined by the covariance matrix. In
particular, the density matrix that describes the PTGGE corresponding to the
dephased covariance matrix,
\begin{equation}
  \label{eq:g-PTGGE-g-d}
  g_{\mathrm{PTGGE}, k}(t) = g_{d, 1, k}(t),
\end{equation}
is given by~\cite{Peschel2009}
\begin{equation}
  \label{eq:rho-ptgge}
  \rho_{\mathrm{PTGGE}}(t) = \frac{1}{Z(t)} \e^{- 2 \sum_{k \in \BZ}
    C_k^{\dagger} \mathrm{arctanh}\left( g_{\mathrm{PTGGE},k}(t) \right)C_k},
\end{equation}
where $Z(t)$ is a normalization such that $\tr(\rho_{\mathrm{PTGGE}}(t)) = 1$.
We note that while our derivation of $\rho_{\mathrm{PTGGE}}(t)$ is based on
explicit results for the covariance matrix for the driven-dissipative SSH model,
our considerations clearly generalize to other open fermionic systems that are
described by number-conserving quadratic fermionic Hamiltonians and are
subjected to incoherent loss and gain.

Even though the same mechanism of dephasing underlies both relaxation of
isolated systems to the GGE and of PT-symmetric systems to the PTGGE, there are
several important differences: (i)~The PTGGE is intrinsically time-dependent due
to the exponential decay of correlations at a rate determined by the imaginary
part of the eigenvalues $\lambda_{\pm, k}$ of $z_k$. Crucially, in the
PT-symmetric phase, the decay rate is identical for all momentum
modes. Relaxation of local observables to the PTGGE is then visualized best by
factoring out the overall exponential decay. (ii)~The oscillation frequencies in
driven-dissipative models are given by the Liouvillian dispersion $\omega_k$ and
not by the bare Hamiltonian dispersion relation $\varepsilon_k$. This affects
the characteristic time scale of relaxation to the PTGGE. (iii)~Until now, we
have considered balanced loss and gain rates with $\delta = \Delta = 0$.  For
finite values of $\delta$ and $\Delta$, also $g_{2,k}(t)\neq 0$ in
Eq.~\eqref{eq:g-l} is nonzero, and this contribution to the covariance matrix
has no counterpart in isolated systems. While in driven-dissipative systems
information about the initial state is incorporated in $g_{1,k}(t)$, the
contribution $g_{2,k}(t)$ describes the approach to the steady state, which is
nontrivial for $\delta, \Delta \neq 0$. Hence, for finite $\delta$ and $\Delta$,
the state of the system will deviate from the PTGGE for $t \to \infty$. However,
as we explain in the following, for sufficiently small values of $\delta$ and
$\Delta$, one can clearly observe transient relaxation to the PTGGE.

\begin{figure}
  \centering
  \includegraphics[width=\linewidth]{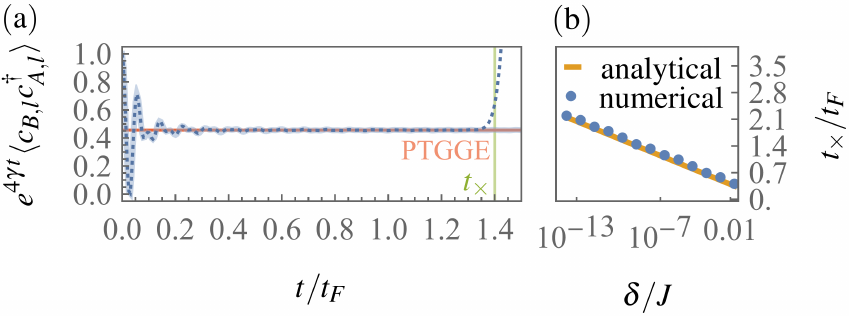}
  \caption{Relaxation of local correlations for a PT-symmetric quench with
    $\Delta J = -0.5 J$, $\Delta \gamma = \gamma = 0.2 J$, and with small
    imbalance between loss and gain. (a) The dashed line corresponds to an
    imbalance of $\Delta = \delta/2 = 10^{-9} J$, while the solid line shows the
    evolution for balanced loss and gain, $\Delta = \delta = 0$. For finite
    imbalance, the rescaled correlation function relaxes to the PTGGE prediction
    on intermediate time scales up to $t_\times$, indicated by a green vertical
    line. (b) The analytical prediction from
    Eq.~\eqref{eq:t-cross-intracell-hopping} is in excellent agreement with the
    numerically determined values for a range of values of $\delta = 2 \Delta$.
    Numerically, we define $t_{\times}$ as the time at which the deviation from
    the PTGGE value equals one.}
  \label{fig:crossover-time}
\end{figure}

Similarly to $g_{1,k}(t)$, the contribution $g_{2,k}(t)$ dephases in the
long-time limit and contains terms that decay exponentially. However, as
mentioned before, the integral in Eq.~\eqref{eq:g2k(t)-explicit} also includes a
time-independent steady-state contribution,
\begin{equation}
  g_{2,k}(t)\sim g_{d,2,k}(t) + g_{\mathrm{SS},k} = \e^{-4\gamma t}
  g_{d,2,k}'(t) + g_{\mathrm{SS},k},
\end{equation}
where
\begin{equation}
  g_{d,2,k}' = -\imag \frac{\delta}{2\gamma \omega_k^2} \left[ 2\Delta\gamma \left( \mathbf{h}_k \times \unitvec{e}_z \right)\cdot\boldsymbol{\sigma} -\varepsilon_k^2 \id \right],
\end{equation}
and with the steady state contribution given by
\begin{multline}
  \label{eq:g-SS-k}
  g_{\mathrm{SS},k} = - \imag \frac{\Delta}{4\gamma^2 + \omega_k^2} \left\{
    \left[ \left(\mathbf{h}_k \times \unitvec{e}_z \right) + 2 \gamma
      \unitvec{e}_z \right] \boldsymbol{\sigma} - 2\Delta\gamma \id\right\} \\
  - \imag \frac{\delta}{4\gamma^2 + \omega_k^2} \frac{\Delta\gamma}{\gamma}
  \left\{ \left( 2\Delta\gamma + \frac{4\gamma^2 + \omega_k^2}{2 \Delta
        \gamma}\right) \id - \left[ 2 \gamma + \left( \mathbf{h}_k \times
        \unitvec{e}_z \right) \right] \boldsymbol{\sigma} \right\}.
\end{multline}
Hence, for $\delta, \Delta \neq 0$, Eq.~\eqref{eq:g-l-asymptotic} is replaced by
\begin{equation}
  \label{eq:g-l-asymptotic-crossover}
  \begin{split}
    g_l(t) & \sim \int_{-\pi}^{\pi} \frac{\diff k}{2\pi} \, \e^{\imag k l}
    \left[ \e^{-4 \gamma t}\left( g_{d,1,k}' + g_{d,2,k}'\right) +
      g_{\mathrm{SS},k} \right] \\ & = \e^{-4 \gamma t} g_{d, l}' +
    g_{\mathrm{SS}, l}.
  \end{split}
\end{equation}
Clearly, at some point the steady state contribution will dominate over the
exponentially decaying terms, and the covariance matrix will assume its steady
state form $g_l(t) \sim g_{\mathrm{SS},l}$. Yet, for sufficiently small values
of $\delta$ and $\Delta$, the PTGGE still gives an accurate description for
relaxation of local observables on intermediate time scales, up to the crossover
time $t_\times$ at which $g_{\mathrm{SS},l}$ becomes dominant. The explicit
value of $t_\times$ depends on the observable under consideration. For the
example of correlations within a unit cell as measured by
$\braket{c_{B,l} c_{A,l}^{\dagger} (t)} = G^{B, A}_{l, l}(t)/2 = g_0^{2, 1}(t)/2$,
we can estimate the crossover time as follows: Assuming that dephasing happens
on a time scale that is shorter than $t_\times$, the crossover time $t_\times$
is determined by the condition that the absolute value of the exponentially
decaying term in Eq.~\eqref{eq:g-l-asymptotic-crossover} is equal to the steady
state contribution, which leads to
\begin{equation}
  \label{eq:t-cross-intracell-hopping}
  t_\times = \frac{1}{4\gamma} \log \! \left( \abs{\left. g_{d,l}^{\prime 2,1}
        \middle/ g_{\mathrm{SS}, l}^{2 ,1} \right.}\right).
\end{equation}
Since $g_{\mathrm{SS}, l}$ is proportional to $y_k$ given in Eq.~\eqref{eq:yk},
this estimate implies that $t_{\times}$ diverges logarithmically for
$\delta, \Delta \to 0$. The time evolution of
$\braket{c_{B,l} c_{A,l}^{\dagger} (t)}$ and the logarithmic divergence of
$t_{\times}$ are illustrated in Fig.~\ref{fig:crossover-time}.

Finally, we want to contrast relaxation to the PTGGE with the asymptotic
behavior of the covariance matrix in the PT-mixed and PT-breaking phases, again
based on the dynamics of $\braket{c_{B,l} c_{A,l}^{\dagger} (t)}$, which, for
$\delta = \Delta = 0$, is given by
\begin{equation}
  \label{eq:c-B-A-PT-mixed-broken}
  \braket{c_{B,l} c_{A,l}^{\dagger} (t)} = \int_{-\pi}^{\pi} \frac{dk}{4
    \pi} \, g_{1,k}^{2 ,1}(t).
\end{equation}
In the PT-breaking phase, the dispersion is purely imaginary,
$\omega_k = \imag \kappa_k$ with $\kappa_k \in \R_{>0}$ given in
Eq.~\eqref{eq:kappa-k}. After some algebraic simplifications of $g_{1,k}(t)$,
the above integral reads~\cite{Starchl2022}
\begin{equation}
  \begin{split}
    \braket{c_{B,l} c_{A,l}^{\dagger} (t)} & = \e^{-4 \gamma t} \int_{-\pi}^{\pi}
    \frac{dk}{4 \pi} \left(1 - 2\frac{h_{y, k}^2}{\kappa_k^2}\sinh(\kappa_k
      t)^2\right) \\ & \sim - \e^{- 4 \gamma t} \int_{-\pi}^{\pi} \frac{\diff k}{8
    \pi} \frac{h_{y, k}^2}{\kappa_k^2} \e^{2 \kappa_k t}.
  \end{split}
\end{equation}
Clearly, the dominant contribution is due to momenta in the vicinity of the
maximum of $\kappa_k$. In the PT-mixed phase, the integration in
Eq.~\eqref{eq:c-B-A-PT-mixed-broken} can be split into a contribution only
consisting of PT-symmetric modes and another contribution due to PT-breaking
modes, and again the dominant contribution at late times is due to PT-breaking
modes in the vicinity of the maximum of $\kappa_k$. For example, for
$\abs{\Delta J} > J$, the dispersion $\kappa_k$ has a maximum at
$k_{\mathrm{max}} = 0$. Using standard asymptotic expansion
techniques~\cite{Bender1999a}, we obtain
\begin{equation}
  \braket{c_{B,l} c_{A,l}^{\dagger} (t)} \sim -\e^{-4 \gamma t} \e^{4 t \sqrt{\Delta\gamma^2 - J^2 }} t^{-3/2}.
\end{equation}
That is, in the PT-mixed and PT-breaking phases, there is a single momentum mode
$k_{\mathrm{max}}$ that maximizes $\kappa_k$ and dominates the dynamics, and the
continuum of modes in the vicinity of $k_{\mathrm{max}}$ leads to additional
algebraic decay. In contrast, all momenta $k \in \BZ$ contribute to the result
for the PT-symmetric phase given in Eq.~\eqref{eq:g-l-asymptotic}.

\subsubsection{Derivation of the PTGGE from the principle of maximum entropy}
\label{sec:PTGGE-SSH}

As we show next, the PTGGE can also be derived from the principle of maximum
entropy~\cite{Jaynes1957} by properly taking into account the modified
statistics of Liouvillian eigenmodes and the expectation values of their
commutators. This approach is explained best by considering first the isolated
SSH model. The dynamics of the isolated SSH model are generated by the
Hamiltonian superoperator $\mathcal{H}$, whose action is defined in
Eq.~\eqref{eq:H-action}, and with eigenmodes $d_{s,k}$ given in
Eq.~\eqref{eq:d-s-k}. Since we are concerned with Gaussian states that are by
definition fully determined by two-point functions, let us consider also
quadratic forms of operators. Any quadratic form of eigenmodes $d_{s, k}$ can be
expressed in terms of commutators and anticommutators through the decomposition
\begin{equation}
  d_{s,k} d_{s',k'}^{\dagger} = \frac{1}{2} \left( [d_{s,k},
    d_{s',k'}^{\dagger}] + \{ d_{s,k}, d_{s',k'}^{\dagger} \}\right).
\end{equation}
For fermions, statistics are encoded in anticommutators,
\begin{equation}
  \label{eq:canonical-commutation}
  \{d_{s,k},d_{s',k'}^{\dagger}\} = \delta_{s,s'} \delta_{k,k'},  \quad \{d_{s,
    k}, d_{s', k'}\} = \{d_{s, k}^{\dagger}, d_{s', k'}^{\dagger}\} = 0.
\end{equation}
By contrast, commutators describe the dynamics. In particular, commutators of
the modes $d_{s, k}$ are also eigenmodes of the Hamiltonian superoperator
$\mathcal{H}$:
\begin{equation}
  \label{eq:isolated-commutators}
  \begin{split}
    \mathcal{H} [d_{A,k},d_{A,k}^{\dagger}] & = 0, \\
    \mathcal{H} [d_{A,k},d_{B,k}^{\dagger}] & = - 2\varepsilon_k
    [d_{A,k},d_{B,k}^{\dagger}].
  \end{split}
\end{equation}
That is, mode-diagonal commutators are conserved, while mixed-index commutators
oscillate with frequency $2 \varepsilon_k$ and dephase. In quadratic fermionic
models or in integrable models that can be mapped to noninteracting fermions,
the GGE is usually stated as the maximum entropy ensemble that is compatible
with conserved mode occupation numbers
$n_{s, k} = d_{s, k}^{\dagger} d_{s, k}$~\cite{Essler2016, Vidmar2016}.
According to our discussion, we can equivalently define the GGE as the maximum
entropy ensemble that is consistent with canonical anticommutations relations
Eq.~\eqref{eq:canonical-commutation} and the conservation of mode-diagonal
commutators Eq.~\eqref{eq:isolated-commutators}. Crucially, this latter
definition generalizes to the PTGGE, however, with some important differences:
First, the adjoint Liouvillian $\mathcal{L}^{\dagger}$ that generates the
dynamics of operators in the driven-dissipative setting, and is specified below
for the SSH model, is a non-Hermitian operator. Therefore, its eigenmodes obey
modified noncanonical anticommutation relations. Second, mode-offdiagonal
commutators of eigenmodes of $\mathcal{L}^{\dagger}$ oscillate at modified
frequencies $2 \omega_k$, directly affecting the dynamics. And third,
mode-diagonal commutators are not conserved; instead, these commutators decay
exponentially. But crucially, they do not oscillate and are, therefore, not
affected by dephasing. Based on the above definition of the PTGGE, in the
following, we present a detailed derivation of the PTGGE for the
driven-dissipative SSH model. The computation consists of three steps:
(i)~Specifying the generator of operator dynamics $\mathcal{L}^{\dagger}$ and
the corresponding eigenvalue equation. (ii)~Solving the eigenvalue equation to
obtain (ii.a)~nonoscillatory, mode-diagonal and (ii.b)~oscillatory,
mode-offdiagonal commutators of eigenmodes of
$\mathcal{L}^{\dagger}$. (iii)~Constructing the PTGGE as the maximum entropy
ensemble that is compatible with the statistics of the eigenmodes of the adjoint
Liouvillian $\mathcal{L}^{\dagger}$ and the expectation values of nonoscillatory
commutators.

\paragraph*{(i) Adjoint Liouvillian.} As detailed in
Appendix~\ref{sec:time-evolution-covariance-matrix}, for a density matrix $\rho$
evolving according to a Liouvillian superoperator $\mathcal{L}$, the expectation
value of an operator $O$ follows the equation of motion given by
\begin{equation}
  \label{eq:O-eom}
  \frac{\diff}{\diff t} \left\langle O \right\rangle =  \imag \left\langle \mathcal{L}^{\dagger}
    O \right\rangle,
\end{equation}
with the adjoint Liouvillian
\begin{equation}
  \label{eq:app-L-adjoint}
  \mathcal{L}^{\dagger} = \mathcal{H} - \imag \mathcal{D}^{\dagger}.
\end{equation}
Hermitian conjugation is defined here with respect to the Hilbert-Schmidt scalar
product, leading to $\mathcal{H} = \mathcal{H}^{\dagger}$, and
\begin{equation}
  \label{eq:D-adjoint}
  \begin{split}
    \mathcal{D}^{\dagger} O & = \sum_{l=1}^L \left( 2 L_l^{\dagger} O L_l -
      \{L_l^{\dagger} L_l, O\}\right) \\ & = \sum_{l = 1}^L \left(
      [L_l^{\dagger}, O] L_l + L_l^{\dagger} [O, L_l] \right).
  \end{split}
\end{equation}
According to Eq.~\eqref{eq:O-eom}, the adjoint Liouvillian
$\mathcal{L}^{\dagger}$ generates the dynamics of operator expectation
values. Suppose now that $O$ is an eigenmode of $\mathcal{L}^{\dagger}$ in the
following sense:
\begin{equation}
  \label{eq:adjoint-eigenmodes}
  \mathcal{L}^{\dagger} O = \lambda^{*} \left( O - O_{\mathrm{SS}} \right),
\end{equation}
where $O_{\mathrm{SS}}$ is a number. Then, the equation of motion of the
expectation value $\langle O \rangle$ reduces to
\begin{equation}
  \label{eq:opertaor-eom}
  \frac{\diff}{\diff t} \left\langle O \right\rangle = \imag \lambda^{*}
  \left(\left\langle O \right\rangle-O_{\mathrm{SS}}\right).
\end{equation}
Dynamical stability requires the imaginary part of the eigenvalue $\lambda$ to
be negative, such that the expectation value $\langle O \rangle$ approaches
$O_{\mathrm{SS}}$ for $t \to \infty$.

\paragraph*{(ii) Eigenmodes of the adjoint Liouvillian.} We want to solve the
eigenvalue equation~\eqref{eq:adjoint-eigenmodes}, where we consider quadratic
eigenmodes of the adjoint Liouvillian. In particular, we seek eigenmodes in the
form of mode-diagonal and mode-offdiagonal commutators,
\begin{equation}
  \label{eq:eta-k-general-form}
  \eta_k = [d_{s,k},d_{s,k}^{\dagger}], \qquad \chi_k = [d_{A,k},d_{B,k}^{\dagger}].
\end{equation}
We note that only the commutators $\eta_k$ and $\chi_k$ and not the modes
$d_{s, k}$ themselves satisfy the eigenvalue
equation~\eqref{eq:adjoint-eigenmodes}. Nevertheless, for simplicity, we refer
to both the modes $d_{s,k}$ and the commutators $\eta_k$ and $\chi_k$ as
eigenmodes of $\mathcal{L}^{\dagger}$. In analogy to
Eq.~\eqref{eq:isolated-commutators}, we anticipate that the expectation values
of the diagonal commutators $\eta_k$ are nonoscillatory whereas the expectation
values of the offdiagonal commutators $\chi_k$ are oscillatory and, therefore,
subject to dephasing.

\paragraph*{(ii.a) Nonoscillatory eigenmodes.} To find the nonoscillatory
eigenmodes of the adjoint Liouvillian, we use a general bilinear ansatz given by
\begin{equation}
  \eta = \sum_{l,l'=1}^{2L} Q_{l,l'} [c_l,c_{l'}^{\dagger}],
\end{equation} 
where the goal is to find $Q$ such that $\eta$ satisfies the eigenvalue
equation~\eqref{eq:adjoint-eigenmodes}. This approach does not rely on
translational invariance and, therefore, we omit the momentum index for the time
being. Plugging the ansatz into the eigenvalue
equation~\eqref{eq:adjoint-eigenmodes}, we obtain two contributions on the
left-hand side that are due to the Hamiltonian superoperator $\mathcal{H}$ and the
adjoint dissipator $\mathcal{D}^{\dagger}$, respectively. The Hamiltonian part
reads
\begin{equation}
  \label{eq:H-eta}
  \mathcal{H} \eta = \sum_{l,l'=1}^{2L} \left( c_l [Q,H]_{l,l'} c_{l'}^{\dagger} -
    c_{l'}^{\dagger} [Q^{\transpose},H]_{l',l} c_l \right),
\end{equation}
and the action of the dissipator is given by
\begin{multline}
  \label{eq:D-eta}
  \mathcal{D}^{\dagger} \eta = - \sum_{l, l' = 1}^{2L}\left( c_l \{ Q, M_l + M_g
    \}_{l,l'} c_{l'}^{\dagger} + c_{l'}^{\dagger} \{M_l + M_g, Q^{\transpose}\}_{l',l}
    c_l \right) \\ +2\tr \! \left( \left( M_l - M_g \right) Q \right).
\end{multline}
We write the eigenvalue $\lambda$ in Eq.~\eqref{eq:adjoint-eigenmodes} as
$\lambda = - \imag \kappa$ with an undetermined real parameter $\kappa \in \R$,
and we identify the steady-state value $\eta_{\mathrm{SS}}$ with the last term
in Eq.~\eqref{eq:D-eta},
\begin{equation}
  \label{eq:eta-ss}
  \eta_{\mathrm{SS}} = \frac{2}{\kappa} \tr \! \left( \left( M_l - M_g \right) Q
  \right).
\end{equation}
Then, the eigenvalue equation takes the form
\begin{multline}
  \sum_{l, l' = 1}^{2L} \left[ c_l \left( QZ^{\dagger} - ZQ \right)_{l,l'}
    c_{l'}^{\dagger} - c_{l'}^{\dagger} \left( Z^{\dagger} Q^{\transpose} -
      Q^{\transpose} Z \right)_{l',l}c_l \right] \\
  = \imag \kappa \left(c_lQ_{l,l'}c_{l'}^{\dagger} - c_{l'}^{\dagger}
    Q_{l',l}^{\transpose} c_l \right).
\end{multline}
Multiplying this equation with $c_m c_{m'}^{\dagger}$ and using
\begin{equation}
  \begin{split}
    \frac{1}{2^{2L}} \tr \! \left( c_m c_{m'}^{\dagger} c_l c_{l'}^{\dagger}
    \right) & = \frac{1}{4} \left( \delta_{m,m'}\delta_{l,l'} +
      \delta_{m,l'}\delta_{l,m'} \right), \\
    \frac{1}{2^{2L}} \tr \! \left( c_m c_{m'}^{\dagger} c_l^{\dagger} c_{l'}
    \right) & = \frac{1}{4} \left( \delta_{m,m'}\delta_{l,l'} -
      \delta_{m,l}\delta_{m',l'} \right),
\end{split}
\end{equation}
we obtain two equations for $Q$ and its transpose $Q^{\transpose}$, respectively,
\begin{align}
  \label{eq:eqm-Q}
  Q Z^{\dagger} - ZQ & = \imag \kappa Q \\
  \label{eq:eqm-Q-transpose}
  Z^{\dagger} Q^{\transpose} - Q^{\transpose} Z & = \imag \kappa Q^{\transpose},
\end{align}
where $Z$ is defined in Eq.~\eqref{eq:Z-matrix-SSH}. Since $Z = Z^{\transpose}$,
the two equations above are actually equivalent. To make further progress, we
use translational invariance of the driven-dissipative SSH model, which implies
that the eigenmodes $\eta_k$ are labeled by a momentum $k$, and that $Q_k$ are
block Toeplitz matrices with $2 \times 2$ blocks
\begin{equation}
  q_{k,l-l'} = \begin{pmatrix} Q_{k,2l-1,2l'-1} & Q_{k,2l-1,2l'} \\ Q_{k,2l,2l'-1} & Q_{k,2l,2l'}
  \end{pmatrix},
\end{equation}
such that the commutator of Liouvillian eigenmodes in
Eq.~\eqref{eq:eta-k-general-form} now takes the form
\begin{equation}
  \eta_k = \sum_{l, l' = 1}^L \left(C_l^{\transpose} q_{k,l-l'} C_{l'}^{\dagger\transpose} - C_{l'}^{\dagger} q_{k,l-l'}^{\transpose} C_l\right),
\end{equation}
with $C_l = \left( c_{2l-1},c_{2l} \right)^{\transpose}$. Defining the discrete
Fourier transformations of the blocks $q_{k, l}$ and the spinors $C_l$ as
\begin{equation}
  q_{k,k'} = \sum_{l=1}^L \e^{-\imag k' l} q_{k,l}, \qquad C_k =
  \frac{1}{\sqrt{L}} \sum_{l=1}^L \e^{-\imag k l} C_l,
\end{equation}
we can recast Eq.~\eqref{eq:eqm-Q-transpose} as
\begin{equation}
  \label{eq:qk-eigenvalue}
  z_{k'}^{\dagger} q_{k,-k'}^{\transpose} - q_{k,-k'}^{\transpose} z_{k'} = \imag
  \kappa q_{k,-k'}^{\transpose}, 
\end{equation}
and write the commutator in momentum space,
\begin{equation}
  \label{eq:eta-k}
  \eta_k  = \sum_{k'\in\BZ} \left(C_{k'}^{\transpose} q_{k,-k'}
    C_{k'}^{\dagger\transpose} - C_{k'}^{\dagger} q_{k,-k'}^{\transpose}
    C_{k'}\right).
\end{equation}
In the PT-symmetric phase, $z_k$ can be diagonalized as stated in
Eq.~\eqref{eq:zk-diag}, and the spectrum of $z_k$ is given by
$\sigma(z_k) = \{-\imag 2 \gamma \pm \omega_k\}$ where $\omega_k \in \R_{>
  0}$.
Using this representation of $z_k$ in Eq.~\eqref{eq:qk-eigenvalue}, we obtain
\begin{equation}
  \omega_{k'} \left( U_{k'}^{-\dagger}\sigma_z U_{k'}^{\dagger}
    q_{k,-k'}^{\transpose} - q_{k,-k'}^{\transpose} U_{k'}\sigma_z U_{k'}^{-1}
  \right) = \imag \left( \kappa - 4\gamma \right) q_{k,-k'}^{\transpose},
\end{equation}
with the shorthand notation $U_k^{-\dagger} = (U_k^{\dagger})^{-1}$. Then,
identifying $\kappa = 4 \gamma$, we can rewrite this equation as a commutator,
\begin{equation}
  \left[ \sigma_z,U_{k'}^{\dagger} q_{k,-k'}^{\transpose} U_{k'} \right] = 0.
\end{equation}
Therefore, for $\eta_k$ to be an eigenmode of the adjoint Liouvillian with
eigenvalue $\lambda = - \imag 4 \gamma$, $q_{k,k'}$ has to satisfy the above
commutation relation. The general solution for $q_{k, k'}$ reads
\begin{equation}
  \label{eq:q-k-k-prime-general-solution}
  q_{k,-k'}^{\transpose} = U_{k'}^{-\dagger} \left(\alpha_{\id,k,k'} \id +
    \alpha_{z,k,k'} \sigma_z \right) U_{k'}^{-1},
\end{equation}
where $\alpha_{\id,k,k'}$ and $\alpha_{z,k,k'}$ are undetermined parameters. To
obtain the solution anticipated in Eq.~\eqref{eq:eta-k-general-form}, we choose
$\alpha_{\id,k,k'}= \alpha_{z,k,k'} = \alpha_{0,k} \delta_{k, k'}/2$. Then,
Eq.~\eqref{eq:eta-k} takes the form
\begin{equation}
  \eta_k  = \frac{\alpha_{0,k}}{2} \left(C_k^{\transpose} U_k^{-\transpose}
    P_{z,+} U_k^{-\ast}C_k^{\dagger\transpose} - C_k^{\dagger} U_k^{-\dagger}
    P_{z,+} U_k^{-1}C_k \right),
\end{equation}
where
\begin{equation}
  P_{z, \pm} = \frac{1}{2} \left( \id \pm \sigma_z \right).
\end{equation}
Defining now the Liouvillian eigenmodes as
\begin{equation}
  \label{eq:D-k}
  D_k = 
  \begin{pmatrix}
    d_{A,k}\\d_{B,k}
  \end{pmatrix}
  = V_k^{\dagger} C_k
\end{equation}
with 
\begin{equation}
  \label{eq:Vk}
  V_k = \sqrt{\frac{2}{\tr \! \left( U_k^{-1}U_k^{-\dagger} \right)}} U_k^{-\dagger}= \frac{1}{\sqrt{2}}
  \begin{pmatrix}
    \e^{-\imag \left(\phi_k + \psi_k \right) \! /2 } & -\e^{-\imag \left(\phi_k
        - \psi_k \right) \! /2 } \\ \e^{\imag \left(\phi_k + \psi_k \right) \!
      /2 } & \e^{\imag \left(\phi_k - \psi_k \right) \! /2 }
  \end{pmatrix},
\end{equation}
we obtain the final form of $\eta_k$:
\begin{equation}
  \label{eq:dAk-commutator}
  \eta_k = [d_{A,k},d_{A,k}^{\dagger}].
\end{equation}
The angle $\phi_k$ is defined in Eq.~\eqref{eq:phi-k} and $\psi_k$ is determined
by the relation
\begin{equation}
  \label{eq:psik}
  \varepsilon_k \e^{\imag \psi_k} = \omega_k + \imag 2 \Delta \gamma.
\end{equation}
In the transformation to the Liouvillian eigenmodes in Eq.~\eqref{eq:Vk}, we
have chosen the normalization such that $\{d_{A,k},d_{A,k}^{\dagger}\} =
1$. This leads to
\begin{equation}
  \alpha_{0,k} = \tr \! \left( U_k^{-1}U_k^{-\dagger} \right) = 2
  \cosh(\theta_k),
\end{equation}
with $\theta_k$ given in Eq.~\eqref{eq:theta-k} and related to $\psi_k$ by
\begin{equation}
  \label{eq:psi-k-theta-k}
  \tan(\psi_k) = \sinh(\theta_k).
\end{equation}
Note that the alternative choice
$\alpha_{\id,k,k'}=-\alpha_{z,k,k'} = \alpha_{0, k} \delta_{k, k'}/2$ in
Eq.~\eqref{eq:q-k-k-prime-general-solution} leads to
$\eta_k = [d_{B,k},d_{B,k}^{\dagger}]$. Further, the transformation $V_k$ given in
Eq.~\eqref{eq:Vk} determines the statistics of the Liouvillian eigenmodes. Since
$V_k$ is not unitary, the operators $d_{s, k}$ do not obey the usual canonical
anticommutation relations. The statistics of these modes are encoded in the
anticommutators collected in
\begin{equation}
  \begin{pmatrix}
    \{d_{A,k},d_{A,k'}\} & \{d_{A,k},d_{B,k'}\} \\
    \{d_{B,k},d_{A,k'}\} & \{d_{B,k},d_{B,k'}\}
  \end{pmatrix}
  = f_k \delta_{k,k'},
\end{equation}
where
\begin{equation}
  \label{eq:fk-SSH}
  f_k = V_k^{\dagger} V_k =  \id +  \sin(\psi_k) \sigma_y.
\end{equation}
Next, having specified the solution for the matrix $Q$ that leads to eigenmodes
$\eta_k$ in the form given in Eq.~\eqref{eq:dAk-commutator}, we can determine
the steady state contribution in Eq.~\eqref{eq:eta-ss}. We find
\begin{equation}
  \label{eq:dAk-commutator-ss}
  \braket{[d_{A,k},d_{A,k}^{\dagger}]}_{\mathrm{SS}}= \eta_{\mathrm{SS}} = \frac{\delta}{\gamma}.
\end{equation}
Finally, to fully determine the time evolution of the expectation value
$\braket{\eta_k}$, we have to calculate its initial value. Relating the
expectation value of the commutator to the covariance matrix by
\begin{equation}
  \braket{[d_{A,k},d_{A,k}^{\dagger}]} = \frac{\omega_k}{\varepsilon_k} \tr \! \left(U_k^{-\dagger} P_{z,+} U_k^{-1} g_k \right),
\end{equation} 
we find the initial value, determined by the covariance matrix $g_k(0)$ in
Eq.~\eqref{eq:g-k-0-general}, to be given by
\begin{equation}
  \label{eq:dAk-commutator-initial}
  \braket{[d_{A,k},d_{A,k}^{\dagger}]}_0  = \cos(\Delta \phi_k - \psi_k),
\end{equation}
where the angle $\Delta \phi_k$ is defined in Eq.~\eqref{eq:cos-sin-delta-phi}.
Therefore, by solving Eq.~\eqref{eq:opertaor-eom} for the commutator in
Eq.~\eqref{eq:dAk-commutator}, we obtain the time evolution of the expectation
value of diagonal commutators:
\begin{multline}
  \label{eq:dAk-commutator-eom}
  \braket{[d_{A,k},d_{A,k}^{\dagger}](t)} = \e^{-4 \gamma t}
  \braket{[d_{A,k},d_{A,k}^{\dagger}]}_0 \\ + \left(1-\e^{-4 \gamma t}\right)
  \braket{[d_{A,k},d_{A,k}^{\dagger}]}_{\mathrm{SS}},
\end{multline}
with the initial and steady-state values given in
Eq.~\eqref{eq:dAk-commutator-initial} and Eq.~\eqref{eq:dAk-commutator-ss},
respectively.

\paragraph*{(ii.b) Oscillatory eigenmodes.} We proceed by showing that the
offdiagonal commutators $\chi_k=[d_{A,k},d_{B,k}^{\dagger}]$ with mode operators
$d_{s, k}$ given in Eq.~\eqref{eq:D-k} are oscillatory eigenmodes of the
Liouvillian. To that end we write $\chi_k$ in the form
\begin{equation}
  \chi_k = \sum_{k'\in \BZ} \left(C_{k'}^{\transpose} p_{k,k'}^{\transpose}
    C_{k'}^{\dagger \transpose} -C_{k'}^{\dagger} p_{k,k'} C_k \right),
\end{equation}
where we define
\begin{equation}
  p_{k,k'} = \delta_{k,k'} V_{k'} \sigma_- V_{k'}^{\dagger},
\end{equation}
and $\sigma_{\pm} = (\sigma_x \pm \imag \sigma_y)/2$. The fact that the
offdiagonal commutators $\chi_k$ are eigenmodes of $\mathcal{L}^{\dagger}$ with
eigenvalue $\lambda = - 2 \left( \omega_k + \imag 2 \gamma \right)$ can be
confirmed by noting that $p_{k, k'}$ satisfies the eigenvalue equation that is
similar to Eq.~\eqref{eq:qk-eigenvalue}:
\begin{equation}
  z_{k'}^{\dagger} p_{k,k'} - p_{k,k'} z_{k'} = - \left( 2 \omega_k -\imag
    4\gamma \right) p_{k,k'}.
\end{equation}
The time evolution of expectation values of $\chi_k$ thus reads
\begin{multline}
  \label{eq:dAk-dBk-eom}
  \braket{[d_{A,k},d_{B,k}^{\dagger}](t)} = \e^{-\imag 2 \left(\omega_k - \imag 2 \gamma \right) t} \braket{[d_{A,k},d_{B,k}^{\dagger}]}_0 \\
  + \left(1 - \e^{-\imag 2(\omega_k - \imag 2 \gamma)t}\right) \braket{[d_{A,k},d_{B,k}^{\dagger}]}_{\mathrm{SS}},
\end{multline}
where the initial and steady-state values are given by
\begin{equation}
  \begin{split}
    \braket{[d_{A,k},d_{B,k}^{\dagger}]}_0 & = \imag \sin(\Delta\phi_k), \\
    \braket{[d_{A,k},d_{B,k}^{\dagger}]}_{\mathrm{SS}} & = \frac{\imag \Delta
      \cos(\psi_k) - \delta \sin(\psi_k)}{2 \left( \omega_k-2\imag \gamma
      \right)}.
  \end{split}
\end{equation}

\paragraph*{(iii) Construction of the PTGGE.} We consider now the case of
balanced loss and gain with $\delta = \Delta = 0$, such that the steady state
contributions to Eqs.~\eqref{eq:dAk-commutator-eom} and~\eqref{eq:dAk-dBk-eom}
vanish,
\begin{equation}
  \braket{[d_{A,k},d_{A,k}^{\dagger}]}_{\mathrm{SS}} =
  \braket{[d_{A,k},d_{B,k}^{\dagger}]}_{\mathrm{SS}} = 0.
\end{equation}
Further, due to the oscillatory exponent in~\eqref{eq:dAk-dBk-eom}, the
contribution of offdiagonal commutators to local observables vanishes at late
times due to dephasing between offdiagonal commutators with different momenta
\begin{equation}
  \braket{[d_{A,k},d_{B,k}^{\dagger}](t)} \to 0.
\end{equation}
Therefore, the state of the system at late times is determined by the
expectation values of diagonal commutators given in
Eq.~\eqref{eq:dAk-commutator-eom} and the statistics of the modes $d_{s, k}$
that follow from the nonunitary transformation $V_k$ Eq.~\eqref{eq:Vk} and are
encoded in the matrix $f_k$ in Eq.~\eqref{eq:fk-SSH}. The PTGGE can be defined
as the maximum entropy ensemble that is compatible with these requirements. We
can also collect the same information in the dephased covariance given in
Eq.~\eqref{eq:g-PTGGE-g-d}, which can be related to the eigenmodes by
\begin{equation}
  \label{eq:g-PTGGE-V-zeta}
  g_{\mathrm{PTGGE},k}(t) = V_k^{-\dagger} \zeta_k (t) V_k^{-1}.
\end{equation}
where $\zeta_k (t) = \e^{- 4 \gamma t} \zeta_k'$ with
\begin{equation}
  \label{eq:zeta-SSH}
  \begin{split}
    \zeta_k' & = 
    \begin{pmatrix}
      \braket{[d_{A,k},d_{A,k}^{\dagger}]}_0 & 0 \\
      0 & \braket{[d_{B,k},d_{B,k}^{\dagger}]}_0 
    \end{pmatrix}  \\
    & = 
    \begin{pmatrix}
      \cos(\Delta\phi_k - \psi_k) & 0 \\
      0 & -\cos(\Delta\phi_k + \psi_k)
    \end{pmatrix}.
  \end{split}
\end{equation}
Then for the given two-point function $g_{\mathrm{PTGGE}, k}(t)$, the entropy is
maximized for the Gaussian state that is uniquely determined through
Eq.~\eqref{eq:rho-ptgge}. In terms of the eigenmodes of the adjoint Liouvillian,
the PTGGE can be written as
\begin{equation}
  \label{eq:rho-PTGGE-SSH}
  \rho_{\mathrm{PTGGE}}(t)= \frac{1}{Z_{\mathrm{PTGGE}}(t)} \e^{-2
    \sum_{k\in\BZ} D_k^{\dagger} f_k^{-1}
    \mathrm{arctanh}\left(\zeta_k(t)f_k^{-1}\right)D_k},
\end{equation}
which shows most transparently both the dependence of the PTGGE on the initial
conditions through $\zeta_k(t)$ and the effect of the noncanonical
anticommutation relations of the Liouvillian eigenmodes described by $f_k$.

\subsection{Driven-dissipative Kitaev chain}
\label{sec:Kitaev-PTGGE}

Having presented the derivation of the PTGGE for the SSH model, we now turn to
the driven-dissipative Kitaev chain. Since the latter has been discussed in
detail in Ref.~\cite{Starchl2022}, we restrict ourselves here to highlighting
the aspects in which the derivation for the Kitaev chain differs from the one
for the SSH model. In particular, for the Kitaev chain, we define bilinear forms
of eigenmodes as
\begin{equation}
  \eta_k = [d_k, d_k^{\dagger}], \qquad \chi_k=[d_k,d_{-k}],
\end{equation}
which can be regarded as normal and anomalous commutators, respectively, and
where the Liouvillian eigenmodes $d_k$ are related to the original fermionic
operators $c_k = \frac{1}{\sqrt{L}} \sum_{l = 1}^L \e^{-\imag k l} c_l$ via
\begin{equation}
  \label{eq:Kitaev-liouvillian-eigenmodes}
  D_k =
  \begin{pmatrix} d_k \\ d_{-k}^{\dagger} \end{pmatrix} = V_k^{\dagger} \begin{pmatrix} c_k \\ c_{-k}^{\dagger} \end{pmatrix}, 
  \quad V_k =
  \begin{pmatrix}
    \cos \! \left( \frac{\theta_k + \phi_k}{2} \right) & \imag \sin \! \left(
      \frac{\theta_k - \phi_k}{2} \right) \\ \imag \sin \! \left( \frac{\theta_k
        + \phi_k}{2} \right) & \cos \! \left( \frac{\theta_k - \phi_k}{2}
    \right)
  \end{pmatrix},
\end{equation}
with angles $\theta_k$ and $\phi_k$ that are defined through the relations
\begin{equation}
  \begin{split}
    \varepsilon_k \e^{\imag \theta_k } & = -2 J \cos(k) - \mu + \imag 2 \Delta \sin(k),\\
    \varepsilon_k \e^{\imag \phi_k } & = \omega_k + \imag 2
    \sqrt{\gamma_l\gamma_g}.
  \end{split}
\end{equation} 
The Liouvillian eigenmodes $d_k$ obey noncanonical anticommutation relations,
\begin{equation}
  \begin{pmatrix}
    \{ d_k, d_{k'}^{\dagger} \} & \{ d_k, d_{-k'} \} \\ \{ d_{-k}^{\dagger},
    d_{k'}^{\dagger} \} & \{ d_{-k}^{\dagger}, d_{-k'} \}
  \end{pmatrix}
  = f_k \delta_{k, k'},  
\end{equation}
where $f_k = V_k^{\dagger} V_k = \id + \sin(\phi_k) \sigma_y$. In analogy to the
mode-diagonal commutators for the SSH model, the normal commutators $\eta_k$ are
nonoscillatory,
\begin{equation}
  \label{eq:normal-commutator-eom}
  \braket{[d_{k},d_{k}^{\dagger}](t)} = \e^{-4 \gamma t}
  \braket{[d_{k},d_{k}^{\dagger}]}_0 + \left(1-\e^{-4 \gamma t}\right)
  \braket{[d_{k},d_{k}^{\dagger}]}_{\mathrm{SS}},
\end{equation}
where $\braket{[d_{k},d_{k}^{\dagger}]}_0$ is determined by the initial
conditions and $\braket{[d_{k},d_{k}^{\dagger}]}_{\mathrm{SS}}$ contains the
steady state contribution. The nonoscillatory normal commutators decay with an
overall decay rate $\gamma$ defined in Eq.~\eqref{eq:parameters-kitaev}, and are
not affected by dephasing. In contrast, and analogously to the mode-offdiagonal
commutators of the SSH model, the anomalous commutators $\chi_k$ are
oscillatory,
\begin{multline}
  \label{eq:anomalous-commutator-eom}
  \braket{[d_{k},d_{-k}](t)} = \e^{-\imag 2 \left(\omega_k - \imag 2 \gamma \right) t} \braket{[d_{k},d_{-k}]}_0 \\
  + \left(1 - \e^{-\imag 2(\omega_k - \imag 2 \gamma)t}\right)
  \braket{[d_{k},d_{-k}]}_{\mathrm{SS}},
\end{multline}
with initial values $\braket{[d_{k},d_{-k}]}_0$ and steady-state contributions
$\braket{[d_{k},d_{-k}]}_{\mathrm{SS}}$. The anomalous commutators oscillate
with frequency $\omega_k$ and are thus affected by dephasing. As above, the
PTGGE describes the late-time relaxation dynamics for vanishing steady-state
contributions
$\braket{[d_{k},d_{k}^{\dagger}]}_{\mathrm{SS}} =
\braket{[d_{k},d_{-k}]}_{\mathrm{SS}} = 0$,
accomplished by balanced loss and gain rates $\delta = \gamma_l - \gamma_g = 0$,
and after dephasing of the contributions due to anomalous commutators. The explicit form of
the PTGGE is given by
\begin{equation}
  \label{eq:PTGGE-Kitaev}
  \rho_{\mathrm{PTGGE}}(t) = \frac{1}{Z_{\mathrm{PTGGE}}(t)} \e^{-2\sum_{k \geq
      0} D_k^{\dagger} f_k^{-1} \mathrm{arctanh} \left(\zeta_k(t)f_k^{-1} \right)
    D_k },
\end{equation}
where $\zeta_k (t) = \e^{-4 \gamma t} \zeta_k'$ with
\begin{equation}
  \label{eq:zeta-k-prime}
  \zeta_k'  =
  \begin{pmatrix}
    \langle [d_k, d_k^{\dagger}] \rangle_0 & 0 \\ 0 & \langle [d_{-k}^{\dagger}, d_{-k}] \rangle_0
  \end{pmatrix}.
\end{equation}
The key difference between Eqs.~\eqref{eq:PTGGE-Kitaev}
and~\eqref{eq:rho-PTGGE-SSH} for the Kitaev chain and the SSH model,
respectively, is that in general anomalous correlations such as
$\langle c_l c_{l'} \rangle$ do not vanish for the Kitaev chain; in contrast,
the weak $\mathrm{U}(1)$ symmetry of the driven-dissipative SSH model ensures
that anomalous correlations like $\langle c_{s, l} c_{s', l'} \rangle$ vanish at
all times.

\section{Spreading of correlations}
\label{sec:spread-corrs}

\begin{figure}
  \centering
  \includegraphics[width=\linewidth]{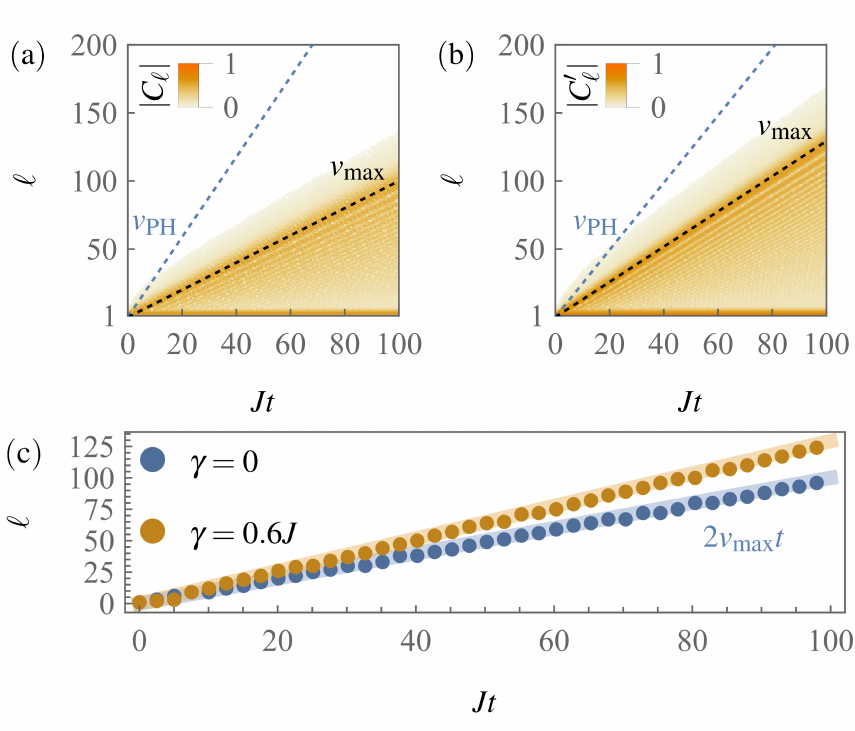}
  \caption{Propagation of correlations after quenches to the topological
    PT-symmetric phase ($\mu = -0.5J$) for (a) the isolated system ($\gamma=0$)
    and (b) the driven dissipative system ($\gamma = 0.6J$). The black and blue
    dashed lines are described by $\ell = 2 v_{\mathrm{max}} t$ and
    $\ell = 2 v_{\mathrm{PH}} t$, respectively. (c) Direct comparison between
    the numerically determined peak positions (dots) and
    $\ell = 2 v_{\mathrm{max}} t$ (solid lines).}
  \label{fig:light-cones-PT}
\end{figure}

Now that we have derived the ensemble that describes the late-time dynamics for
quenches to the PT-symmetric phase, we turn to a detailed study of relaxation to
the PTGGE. In the following sections, both for the SSH model and the Kitaev
chain, we focus on balanced loss and gain with $\delta = \Delta = 0$, such that
the steady state is at infinite temperature, and a description of the dynamics
in terms of a PTGGE applies on arbitrarily long time scales.

A well-established property of isolated integrable systems that exhibit
generalized thermalization after a quantum quench is the ballistic propagation
of correlations through the system~\cite{Essler2016,Calabrese2006}. This is also
referred to as light cone spreading of correlations, since correlations outside
of the light cone that is defined by the group velocity $v_{\mathrm{max}}$ are
suppressed exponentially~\cite{Lieb1972}. As we demonstrate in the following,
light cone propagation of correlations is not unique to isolated
systems~\cite{Bernier2018, YagoMalo2018}. We illustrate this behavior for the
driven-dissipative Kitaev chain, where we consider the evolution of the normal
commutators $C_{l - l'}(t) = \langle [c_l, c_{l'}^{\dagger}](t) \rangle$ for
$\delta = \Delta = 0$.  In the PT-symmetric phase, we find ballistic propagation
of quasiparticles which, however, have a finite lifetime, and a maximum
velocity adjusted to the modified dispersion relation $\omega_k$; in contrast,
in the PT-mixed and PT-broken phases, we observe diffusive spreading of
correlations.

\subsection{PT-symmetric phase}

\begin{figure}
  \centering
  \includegraphics[width=1\linewidth]{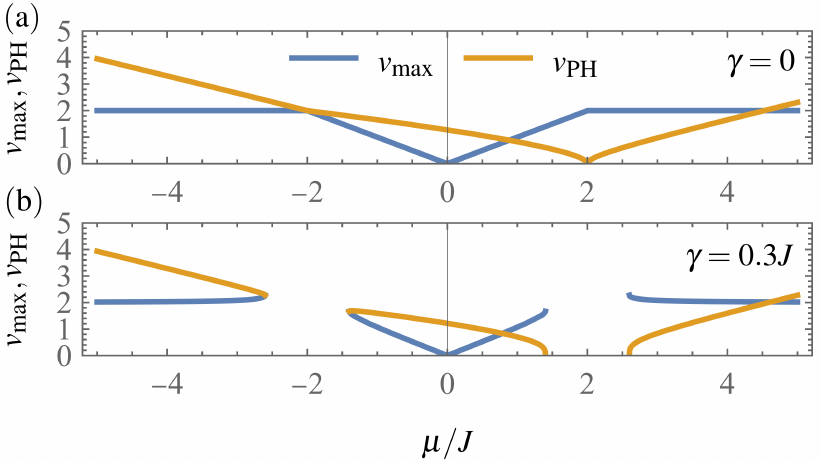}
  \caption{Maximum of the quasiparticle velocity $v_{\mathrm{max}}$ and the
    phase velocity $v_\mathrm{PH}$ as a function of the chemical potential $\mu$
    for (a) the isolated system ($\gamma = 0$) and (b) the driven-dissipative
    system ($\gamma = 0.3J$). For $\mu<0$, at the phase boundary,
    $v_{\mathrm{max}}=v_{\mathrm{PH}} = \sqrt{2 J \abs{\mu}}$, while for
    $\mu>0$, we find $v_{\mathrm{max}}=\sqrt{2 J \abs{\mu}}$ and
    $v_{\mathrm{PH}} = 0$.}
  \label{fig:vmax-vph}
\end{figure}

In Figs.~\ref{fig:light-cones-PT}(a) and (b), we show the absolute value of
rescaled normal commutators $C'_{\ell} (t) = \e^{4 \gamma t} C_{\ell}(t) $ for
the isolated and the driven-dissipative Kitaev chain in the PT-symmetric phase,
respectively. A clear light cone structure is visible in both figures. The peak
of correlations, defining the boundary of the light cone and found at position
$\ell= 2 v_{\mathrm{max}} t$, is well described by the ballistic propagation of
quasiparticles with finite lifetime $\sim 1/\gamma$ and velocity
\begin{equation}
  \label{eq:v-max}
  v_{\mathrm{max}}= \max_{k\in\BZ} \! \left( \abs{v_k} \right) = \max_{k\in\BZ}
  \! \left( \abs{\diff \omega_k/\diff k} \right).
\end{equation}
We conclude that the spreading of correlations with finite velocity is
maintained also in the driven-dissipative model throughout the whole
PT-symmetric phase. For a finite rate of dissipation $\gamma$, the light cone
velocity with dispersion $\omega_k$ is increased compared to the velocity
corresponding to the isolated system with dispersion $\varepsilon_k$. This is
further verified in Fig.~\ref{fig:light-cones-PT}(c), where the position of the
light cone boundary is traced numerically and compared to the analytical
prediction. However, the increased speed at which correlations propagate is
offset by the exponential decay of correlations at rate $4 \gamma$. Further, the
phase velocity defined as~\cite{Cevolani2018}
\begin{equation}
  v_{\mathrm{PH}} =  \abs{\omega_k/k}_{k = k_{\mathrm{max}}},
\end{equation}
where $k_{\mathrm{max}}$ is the momentum for which $v_k = v_{\mathrm{max}}$,
decreases with increasing reservoir coupling. While $v_{\mathrm{max}}$ defines
the boundary beyond which correlators become exponentially suppressed, the phase
velocity describes the propagation of local peaks within the light cone. At the
boundaries of the PT-symmetric phase, the phase velocity and maximum particle
velocity line up with $v_{\mathrm{max}}=v_{\mathrm{PH}}=\sqrt{2 J \abs{\mu}}$
for $\mu < 0$, while for $\mu >0$ we still have
$v_{\mathrm{max}}=\sqrt{2 J \abs{\mu}}$ but the phase velocity
$v_{\mathrm{PH}}=0$. This behavior is depicted in Fig.~\ref{fig:vmax-vph}, where
$v_{\mathrm{max}}$ and $v_{\mathrm{PH}}$ are plotted as a function of $\mu$. For
the value $\mu = - 0.5 J$ chosen in Figs.~\ref{fig:light-cones-PT}(a) and (b),
we obtain $v_{\mathrm{PH}} > v_{\mathrm{max}}$. However, as can be seen in Fig.~\ref{fig:vmax-vph}, also the opposite order $v_{\mathrm{PH}} <
v_{\mathrm{max}}$ is realized for different values of $\mu$.

\subsection{PT-mixed and PT-broken phases}

\begin{figure}
  \centering
  \includegraphics[width=\linewidth]{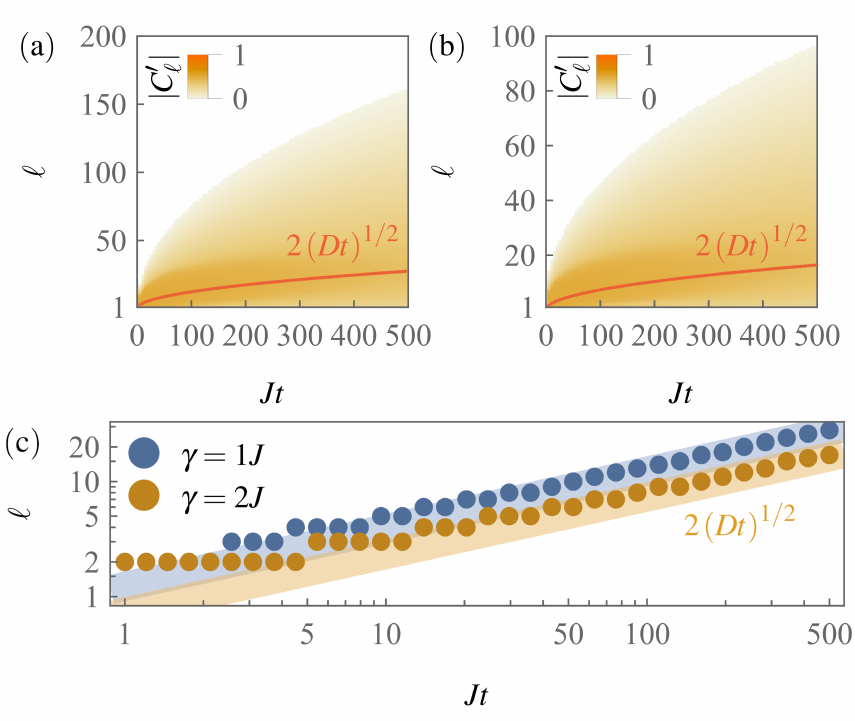}
  \caption{Propagation of correlations after quenches to (a) the PT-mixed phase
    ($\mu = -0.5 J$, $\gamma=1J$) and (b) the PT-broken phase ($\mu = -0.5$,
    $\gamma = 2J$). The red dots indicate the position of the peak of
    correlations, spreading diffusively as $\ell = 2 \left(Dt\right)^{1/2}$. (c)
    Direct comparison between the numerically determined peak positions (dots)
    and diffusive propagation in log-log scale.}
  \label{fig:light-cones-MB}
\end{figure}

In the PT-mixed and PT-broken phases, where
$\gamma>\gamma_c=\left\vert J - \abs{\mu}/2 \right\vert$, the spreading of
correlations is dominated by the single slowest-decaying mode with decay rate
\begin{equation}
  \label{eq:gamma-s}
  2 \gamma_s = \min_{k \in \BZ} \! \left( - \Im(\lambda_{+, k}) \right).
\end{equation}
Accordingly, we define rescaled normal commutators as
$C_l'(t) = \e^{4\gamma_{s} t} C_l(t)$. In Fig.~\ref{fig:light-cones-MB}(a)
and~(b), the absolute value of rescaled normal commutators is shown for the
PT-mixed and PT-broken phase, respectively. It is worthwhile to first discuss
the qualitative differences of correlations in the phases with PT-symmetric and
PT-breaking modes. In the PT-symmetric phase, at any time, the normal
commutators show multiple oscillations inside the light cone, with a peak at the
boundary of the light cone and ensuing suppression of correlations outside the
light cone. This creates the sharp boundaries in
Fig.~\ref{fig:light-cones-PT}. In contrast, in the presence of PT-breaking
modes, normal commutators show a single peak without oscillations and long
decaying tails, giving the unstructured appearance inside the boundary in
Fig.~\ref{fig:light-cones-MB}. Crucially, a correlation boundary can still be
defined through this single peak. The correlation boundary, however, spreads
diffusively according to $\ell = 2 (Dt)^{1/2}$, where the diffusion constant is
given by
\begin{equation}
  \label{eq:D}
  D = \left. - \frac{1}{2} \frac{\diff^2 \kappa_k}{\diff k^2}
  \right\rvert_{k = k_{\mathrm{max}}} =  \frac{J \abs{\mu}}{\sqrt{4 \gamma_l \gamma_g
      - \left( 2 J - \abs{\mu} \right)^2}},
\end{equation}
with $\kappa_k$ defined in Eq.~\eqref{eq:kappa-k} and $k_{\mathrm{max}}$ the momentum maximizing $\kappa_k$. The diffusive evolution of
the peak position is illustrated further in Fig.~\ref{fig:light-cones-MB}(c).

Let us briefly comment on the spreading of correlations in the
driven-dissipative SSH model. Based on the analytical form of the covariance
matrix derived in Sec.~\ref{sec:SSH-covariance-dynamics}, which is structurally
similar to the covariance matrix of the Kitaev chain, we can expect that the
dynamics of correlations is qualitatively the same in both models. This
expectation is confirmed by numerical results, which we do not show here.

\section{Time evolution of the subsystem entropy}
\label{sec:subsystem-entropy}

The linear growth and volume-law saturation of the von Neumann entropy of a
finite subsystem is a key signature of thermalization in isolated
systems\cite{Calabrese2005, Alba2017, Alba2018, Calabrese2020}. As we show in
the following, the exact same phenomenology can be observed after quenches to
the PT-symmetric phase of the driven-dissipative SSH model---if we consider the
contribution to the entropy that measures the spreading of correlations due to
the propagation of pairs of entangled quasiparticles, and after appropriate
rescaling to compensate for exponential decay. The corresponding analysis for
the driven-dissipative Kitaev chain is provided in Ref.~\cite{Starchl2022}.

We consider a subsystem of the SSH chain that contains $\ell$ contiguous unit
cells. The corresponding reduced density matrix is obtained by tracing out the
remaining $L - \ell$ unit cells, $\rho_{\ell} = \tr_{L-\ell}(\rho)$, and the von
Neumann subsystem entropy is defined by
$S_{\mathrm{vN},\ell}= - \tr(\rho_{\ell} \ln(\rho_{\ell}))$. For a Gaussian
state, the subsystem entropy can be calculated as~\cite{Peschel2009}
\begin{equation}
  S_{\mathrm{vN},\ell} = \sum_{l=1}^{\ell} S(\xi_l),
\end{equation}
where
\begin{equation}
  S(\xi) = - \frac{1 + \xi}{2} \ln\left(\frac{1 + \xi}{2}\right) - \frac{1 - \xi}{2} \ln\left(\frac{1 - \xi}{2}\right),
\end{equation} 
and where $\pm \xi_l$ with $0\leq \xi_l \leq 1$ and $l\in\{1,\dots,\ell\}$ are
the eigenvalues of the reduced covariance matrix $G_{\ell}$ defined by
\begin{equation}
  \label{eq:G-ell}
  G_{\ell} = \left( G^{s, s'}_{l, l'} \right)^{s, s' \in \{ A, B \}}_{l, l' \in
    \{1, \dotsc, \ell\}},
\end{equation}
with $G^{s, s'}_{l, l'}$ given in Eq.~\eqref{eq:G}. The set of eigenvalues
$\{ \pm \xi_l \}$ of $G_{\ell}$ forms the single-particle entanglement
spectrum~\cite{Peschel2009}. In pure states of isolated systems, the subsystem
entropy measures the entanglement between the subsystem and its
compliment. After a quench, the subsystem entropy grows linearly in $t$ before
it saturates to a volume-law value, $S_{\mathrm{vN}, \ell} \propto \ell$.  This
behavior can be explained through a quasiparticle picture, which does not only
provide a qualitative interpretation but also quantitative predictions for the
full time evolution after a quench in the space-time scaling
limit~\cite{Calabrese2005,Alba2017, Alba2018, Calabrese2020}. In this picture,
the initial state acts as a source of pairs of entangled quasiparticles with
opposite momenta. After creation, these quasiparticles move ballistically with a
velocity of at most $v_{\mathrm{max}}$ through the system. All pairs whose
members are separated by the boundary between the subsystem and its compliment
contribute to the subsystem entropy. Consequently, the subsystem entropy starts
to saturate when all maximum-velocity pairs generated in the subsystem have left
the subsystem.

\begin{figure}
  \centering
  \includegraphics[width=\linewidth]{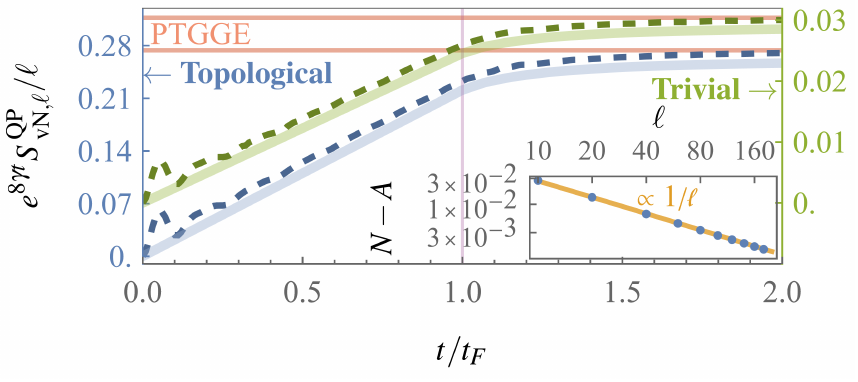}
  \caption{Quasiparticle-pair contribution to the subsystem entropy in the
    driven-dissipative SSH model after quenches to the trivial (green,
    $\Delta J = 0.5 J$) and topological (blue, $\Delta J = -0.5 J$) PT-symmetric
    phases for $\gamma = \Delta\gamma = 0.1 J$, $\delta = \Delta = 0$, and
    $\ell = 20$. Dashed lines show numerical data, and solid lines correspond to
    the analytical conjecture Eq.~\eqref{eq:quasiparticle-conjecture} for the
    space-time scaling limit. The approach of the numerical results to the
    conjecture with increasing subsystem size is shown in the inset: For the
    topological quench at $t = 2 t_F$, the difference between the numerical data
    and Eq.~\eqref{eq:quasiparticle-conjecture} (blue dots) vanishes as
    $1 / \ell$ (orange line).}
  \label{fig:entropy-PT-sym}
\end{figure}

In open systems, the time-evolved state is no longer pure and the subsystem
entropy involves two contributions~\cite{Maity2020, Alba2021,
  Carollo2022, Alba2022}: $S_{\mathrm{vN},\ell}^{\mathrm{QP}}$, which describes
correlations due to quasiparticle pairs as in an isolated system; and
$S_{\mathrm{vN}}^{\mathrm{stat}} = S_{\mathrm{vN},L}$, the statistical entropy
due to the mixedness of the state. The quasiparticle-pair contribution is thus
given by the difference
\begin{equation}
  \label{eq:quasiparticle-contribution}
  S_{\mathrm{vN},\ell}^{\mathrm{QP}} =S_{\mathrm{vN},\ell} - \frac{\ell}{L} S_{\mathrm{vN}}^{\mathrm{stat}}.
\end{equation}
In the following, we consider quenches to the PT-symmetric phase of the
driven-dissipative SSH model. Building upon findings of Refs.~\cite{Carollo2022,
  Alba2022}, in Ref.~\cite{Starchl2022}, we have proposed an analytical
conjecture for the evolution of the quasiparticle-pair contribution
$ S_{\mathrm{vN},\ell}^{\mathrm{QP}}$ in the space-time scaling limit:
\begin{equation}
  \label{eq:quasiparticle-conjecture}
  S_{\mathrm{vN},\ell}^{\mathrm{QP}}(t) \sim \int_0^{\pi} \min(2\abs{v_k}t,\ell)
  \tr\left( S \! \left( \zeta_k(t) f_k^{-1}\right) - S(g_k(t))_d \right),
\end{equation}
where for the SSH model $\zeta_k(t)$ and $f_k$ are defined in
Eqs.~\eqref{eq:zeta-SSH} and~\eqref{eq:fk-SSH}, respectively, and $g_k(t)$ is
given in Eq.~\eqref{eq:gk(t)} with the subscript ``\textit{d,}'' which stands
for dephasing, indicating that only nonoscillatory components contribute. For
long times, $\gamma t\gg 1$, we can expand the quasiparticle-pair entropy in the
exponentially decaying factors $\zeta_k(t),g_k(t) \sim \e^{-4 \gamma t}$. To
lowest nontrivial order, we find $S(\xi)\sim \ln(2) - \xi^2/2$. Then, the
constant contributions in Eq.~\eqref{eq:quasiparticle-conjecture} cancel, and we
find $S_{\mathrm{vN},\ell}^{\mathrm{QP}}(t) \sim \e^{-8 \gamma t}$. Therefore,
relaxation to the PTGGE can be revealed by considering the rescaled
quasiparticle-pair entropy
$\e^{8 \gamma t} S_{\mathrm{vN},\ell}^{\mathrm{QP}}(t)$, which is shown in
Fig.~\ref{fig:entropy-PT-sym}. The rescaled quasiparticle-pair entropy grows
linearly up to the Fermi time~\cite{Calabrese2012I},
\begin{equation}
  \label{eq:tF}
  t_F = \ell/(2 v_{\mathrm{max}}),
\end{equation}
where saturation to the value predicted by the PTGGE sets in. As shown in the
inset, the difference between the numerical results and the analytical
conjecture Eq.~\eqref{eq:quasiparticle-contribution} vanishes as $1/\ell$. The
numerical method we have used to obtain this data is described in
Ref.~\cite{Starchl2022}.

\section{Time evolution of string order and subsystem fermion parity}
\label{sec:dynamics-local-observ}

Generalized thermalization after a quantum quench in isolated systems is defined
as relaxation of the averages of local observables to stationary values that are
determined by the GGE. Similarly, based on the arguments given in
Sec.~\ref{sec:PTGGE}, after quenches to the PT-symmetric phases of the
driven-dissipative Kitaev and SSH chains, we expect to observe local relaxation
to the PTGGE. To confirm this expectation, we proceed to study the time
evolution of the \dsop{} for the SSH model and the subsystem parity for the
Kitaev chain, which are defined below in Secs.~\ref{sec:dual-string-order}
and~\ref{sec:subsystem-parity}, respectively. These observables are of special
interest due to their connection to topology: On the one hand, in the ground
state of an isolated system, the \dsop{} and the subsystem parity function as
\emph{topological disorder parameters.} That is, they are finite in the trivial
phase and vanish in the topological phase. On the other hand, after quenches
originating from the trivial phase, the decay of dual string order and subsystem
parity carries a robust signature of the topology of the postquench Hamiltonian:
For quenches across the topological phase boundary, the topological disorder
parameters exhibit periodically recurring zero crossings; in contrast, they do
not cross zero for quenches within the trivial phase. Physically, these zero
crossings correspond to pumping of spin order and fermion parity between a
finite subsystem and its complement. As we discuss in the following, in the
presence of drive and dissipation, this phenomenology acquires qualitative
modifications that are unique to open systems.

\subsection{\Dsop{}}
\label{sec:dual-string-order}

The concept of string order has originally been introduced and is mostly
discussed in the context of spin chains~\cite{DenNijs1989, Tasaki1991}. For the
SSH model Eq.~\eqref{eq:H-SSH}, an equivalent formulation in terms of a
dimerized spin-\nicefrac{1}{2} chain can be obtained through the Jordan-Wigner
transformation~\cite{Jordan1928}, which yields
\begin{equation}
  \label{eq:H-dimerized-spin-chain}
  H = 2 \sum_{l = 1}^{2 L} \left( J - \left( -1 \right)^l \Delta J
  \right) \left( S^x_l S^x_{l + 1} + S^y_l S^y_{l + 1}
  \right),
\end{equation}
where $S_l^{\mu}$ with $\mu \in \{ x, y, z \}$ are spin-\nicefrac{1}{2}
operators that act on the spin at site $l$ with $l \in \{ 1, \dotsc, 2 L
\}$.
The Jordan-Wigner transformation, which maps the set of $2 L$ fermionic
operators $c_l$ and $c_l^{\dagger}$ with $c_{A, l} = c_{2 l - 1}$ and
$c_{B, l} = c_{2 l}$ to spin operators $S_l^{\pm} = S_l^x \pm \imag S_l^y$, is
given by
\begin{equation}
  \label{eq:Jordan-Wigner}  
  S_l^- = \e^{\imag \pi \sum_{l' = 1}^{l - 1} n_{l'}} c_l, \quad S_l^+ =
  \e^{\imag \pi \sum_{l' = 1}^{l - 1} n_{l'}} c_l^{\dagger}, \quad S_l^z = n_l -
  \frac{1}{2}.
\end{equation}
where $n_l = c_l^{\dagger} c_l$. For $\Delta J < 0$, the ground state of the
dimerized spin chain is topologically ordered. Indeed, the topologically ordered
phase of the dimerized spin-\nicefrac{1}{2} chain is continuously connected to
the Haldane phase of the antiferromagnetic spin-1 Heisenberg
chain~\cite{DeLeseleuc2019}. The topological and trivial phases of the dimerized
spin chain can be distinguished through the \sop{}, which takes a finite
expectation value in the topological phase, and vanishes in the trivial
phase. Equivalently, the \emph{dual} \sop{} is nonzero in the trivial phase, and
vanishes in the topological phase~\cite{Hida1992}. The designation as dual
refers here to the self duality of the dimerized spin chain: Under a translation
by one lattice site, $S_l^{\mu} \mapsto S_{l + 1}^{\mu}$, the Hamiltonian
Eq.~\eqref{eq:H-dimerized-spin-chain} maps to itself with
$\Delta J \mapsto - \Delta J$. Consequently, also the trivial and topological
phases are exchanged, and, therefore, the \dsop{}, which is obtained by
performing the translation by one lattice site on the usual string order
parameter, serves as a topological disorder parameter. The \dsop{} for the
dimerized spin-\nicefrac{1}{2} chain is given by~\cite{Hida1992,
  Bahovadinov2019}
\begin{equation}
  \label{eq:dual-SOP-spin}
  O_{\ell}^{\mu} = \e^{\imag \pi \sum_{l=1}^{2\ell} S_l^\mu},
\end{equation}
with $\mu \in \{x,y\}$. We note that the invariance of the Hamiltonian
Eq.~\eqref{eq:H-dimerized-spin-chain} under rotations around the $z$-axis
implies that $\braket{O_{\ell}^x} = \braket{O_{\ell}^y}$. An alternative
interpretation of string order is obtained through a nonlocal unitary
transformation that maps the dimerized spin chain
Eq.~\eqref{eq:H-dimerized-spin-chain} to two coupled Ising
models~\cite{Takada1992}, which have the doubled Ising symmetry
$\Z_2 \times \Z_2$. This symmetry is broken in the topologically ordered phase,
and correlations of the Ising order parameters map to the string order
parameters $O_{\ell}^x$ and $O_{\ell}^y$.

To compute the \dsop{} for the SSH model, it is convenient to introduce $4 L$
Majorana operators $w_l$ according to Eq.~\eqref{eq:majoranas}. For
concreteness, we consider the \dsop{} with $\mu = x$, which can be written in
terms of Majorana fermions as
\begin{equation}
  \label{eq:dual-SOP}
  O_{\ell}^x = (-\imag)^{\ell} \prod_{l=1}^{\ell} w_{4l-2} w_{4l-1}.
\end{equation}
Since the quench dynamics we consider here only involve Gaussian states, we can
employ Wick's theorem~\cite{Lieb1961, Barouch1970I, *Barouch1971II,
  *Barouch1971III, Kraus2009}, according to which the expectation value of the
\dsop{} is given by the Pfaffian of a submatrix of the covariance matrix
$\Gamma$ for Majorana fermions defined in Eq.~\eqref{eq:gamma-kitaev},
\begin{equation}
  \label{eq:dual-SOP-expectation}
  \braket{O_{\ell}^x} = \pf(\Gamma_{\ell}).
\end{equation}
The submatrix $\Gamma_{\ell}$ consists of the elements of $\Gamma$ that
correspond to the Majorana operators appearing in Eq.~\eqref{eq:dual-SOP},
\begin{equation}
  \label{eq:Gamma-ell-SSH}
  \Gamma_{\ell} = \left( \Gamma_{l, l'} \right)_{l, l' \in \{ 2, 3, 6, 7,
    \dotsc, 4 \ell - 2, 4 \ell - 1 \}}.
\end{equation}
As shown in Appendix~\ref{sec:appendix-majoran-cov-SSH}, up to a factor of
$\imag$, $\Gamma_{\ell}$ is unitarily equivalent to the contribution
$G_{1, \ell}$ to the reduced covariance matrix of complex fermions defined in
Eq.~\eqref{eq:G-ell},
\begin{equation}
  \label{eq:Gamma-l-G-l-SSH}
  \Gamma_{\ell} = \imag P_{\ell}^{\dagger} G_{1,\ell} P_{\ell},
\end{equation}
where
\begin{equation}
  \label{eq:P-ell}
  P_{\ell} = \bigoplus_{l=1}^{\ell}
  \begin{pmatrix}
    1 & 0 \\ 0 &  \imag
  \end{pmatrix}.
\end{equation}
The decomposition of the reduced covariance matrix as
$G_{\ell} = G_{1, \ell} + G_{2, \ell}$ employed here is analogous to that of
$g_k$ in Eq.~\eqref{eq:gk(t)}.

\subsubsection{Relaxation of the \dsop{}}
\label{sec:SOP-relaxation}

\begin{figure}
  \centering
  \includegraphics[width=\linewidth]{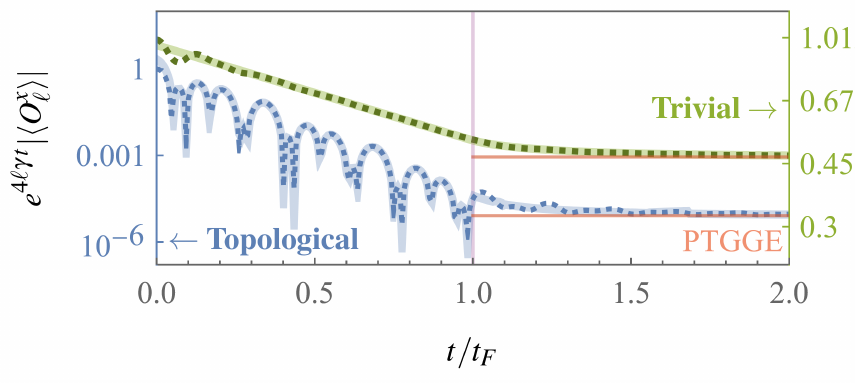}
  \caption{\Dsop{} after quenches to the trivial (green,
    $\Delta J = 0.5 J$) and topological (blue, $\Delta J = - 0.5 J$)
    PT-symmetric phases for $\gamma = \Delta\gamma = 0.3 J$,
    $\delta = \Delta = 0$, and $\ell = 20$. Dashed lines correspond to numerical
    data, and the solid lines show the analytical conjectures in
    Eqs.~\eqref{eq:subsystem-parity-space-time-scaling-trivial}
    and~\eqref{eq:subsystem-parity-space-time-scaling-topological} with
    $\alpha_+ = \alpha_- = 0.15$. We indicate the asymptotic PTGEE prediction by
    straight horizontal red lines and the the change of dynamics at the
    characteristic time scale $t = t_F$ by the purple vertical line. The system
    size $L$ is chosen sufficiently large to avoid finite-size effects.}
  \label{fig:SOP-PT-sym}
\end{figure}

To illustrate relaxation to the PTGGE, we consider the time evolution of the
\dsop{} after quenches to the trivial and topological PT-symmetric phases. As
discussed in Sec.~\ref{sec:PTGGE-SSH-covariance-matrix}, the expectation value
of a product of $\ell$ fermionic operators becomes stationary after rescaling
with a factor of $\e^{2 \ell \gamma t}$. Accordingly, in
Fig.~\ref{fig:SOP-PT-sym}, we show the evolution of the rescaled \dsop{}
$\e^{4 \ell \gamma t} \abs{\braket{O_{\ell}^x}}$. For quenches to both the
topological (blue) and the trivial (green) phase, we observe fast decay up to
the Fermi time $t_F$ Eq.~\eqref{eq:tF}, followed by much slower relaxation to
the value predicted by the PTGGE. As an aside, it is interesting to note that,
since $G_{2, \ell}$ does not contribute to $\Gamma_{\ell}$ in
Eq.~\eqref{eq:Gamma-l-G-l-SSH}, relaxation of the \dsop{} is described by the
PTGGE even if $\delta, \Delta \neq 0$. The numerical data for the quench to the
trivial phase is in excellent agreement with the following analytical form:
\begin{equation}
  \label{eq:subsystem-parity-space-time-scaling-trivial}
  \braket{O^x_{\ell}(t)} \sim O_0 \e^{-4 \ell \gamma t + \int_{0}^{\pi}
    \frac{\diff k}{2 \pi} \min( 2 \abs{v_k} t, \ell) \tr\left( \ln\left(
        \abs{\zeta_k' f_k^{-1}} \right) \right) },
\end{equation}
where $\zeta_k'$ and $f_k$ are defined in Eqs.~\eqref{eq:zeta-SSH}
and~\eqref{eq:fk-SSH}, respectively, and the prefactor $O_0$ is determined by
fitting the late-time asymptotic behavior to the PTGGE
prediction. Equation~\eqref{eq:subsystem-parity-space-time-scaling-trivial} is
an analytical conjectures we have proposed originally for the dynamics of the
subsystem parity in the driven-dissipative Kitaev chain~\cite{Starchl2022}, and
which is based on analytical results obtained by Calabrese et
al.~\cite{Calabrese2011,Calabrese2012I,Calabrese2012II} for the relaxation of
order parameter correlations in the transverse field Ising model and in the
space-time scaling limit $\ell,t \to \infty$ with $\ell/t$ fixed. As mentioned
above, the SSH model can be mapped to two coupled Ising models by a nonlocal
unitary transformation~\cite{Takada1992}, which leads us to expect that our
conjecture also applies to the \dsop. Numerically, this
expectation is confirmed in Fig.~\ref{fig:SOP-PT-sym}, where the solid lines
indicate the analytical predictions.

The analytical conjecture
Eq.~\eqref{eq:subsystem-parity-space-time-scaling-trivial} shows that as
compared to the isolated SSH model, there are two important modifications:
(i)~The dynamics are determined the Liouvillian dispersion relation $\omega_k$
rather than the Hamiltonian dispersion relation $\varepsilon_k$, leading here to
a shorter Fermi time $t_F$; (ii)~noncanonical anticommuation relations of
Liouvillian quasiparticles encoded in $f_k \neq \id$ affect the result
quantitatively.

For quenches to the topologically nontrivial phase with $\Delta J < 0$, the
\dsop{} exhibits additional oscillatory dynamics for $t < t_F$, well described
in the space-time scaling limit by
\begin{equation}
  \label{eq:subsystem-parity-space-time-scaling-topological}
  \braket{O^x_{\ell}(t)} \sim 2 \cos(\omega_{k_{s,+}}t + \alpha_+)
  \cos(\omega_{k_{s,-}}t + \alpha_-) \braket{O^x_{\ell}(t)}_{\mathrm{nonosc}},
\end{equation}
where $\braket{O^x_{\ell}(t)}_{\mathrm{nonosc}}$ is the nonoscillatory evolution
described by Eq.~\eqref{eq:subsystem-parity-space-time-scaling-trivial},
$\alpha_{\pm}$ are numerically determined phase shifts, and $k_{s,\pm}$ are the
soft modes of the PTGGE. These modes are determined by the condition that the
matrix in the exponent in Eq.~\eqref{eq:rho-ptgge} has an eigenvalue that is
equal to zero. According to Eq.~\eqref{eq:g-PTGGE-V-zeta}, this condition can be
stated as
\begin{multline}  
  \det \! \left( g_{\mathrm{PTGGE}, k}(t) \right) = \e^{- 8 \gamma t} \det
  \!  \left( V_k^{-\dagger} V_k^{-1} \right) \det \! \left( \zeta_k' \right) \\
  = - \frac{\varepsilon_k^2}{\omega_k^2} \e^{- 8 \gamma t} \cos(\Delta \phi_k -
  \psi_k) \cos(\Delta \phi_k + \psi_k) = 0,
\end{multline}
or, equivalently, $\cos(\Delta \phi_k \pm \psi_k) = 0$. As we show below,
$\cos(\Delta \phi_k - \psi_k) = 0$ has two solutions $k = k_{s, \pm}$. The
solutions to $\cos(\Delta \psi_k + \psi_k) = 0$ are then given by
$- k_{s, \pm}$: Indeed, IS of the Bloch Hamiltonian $h_k$ and IS$^{\dagger}$ of
the non-Hermitian Bloch Hamiltonian $z_k$ imply that
$\Delta \phi_k = - \Delta \phi_k$ and $\psi_k = \psi_{-k}$,
respectively~\cite{Starchl2022}, and, therefore,
$\cos \! \left( \Delta \phi_{- k_{s, \pm}} + \psi_{- k_{s, \pm}} \right) = \cos
\! \left( \Delta \phi_{k_{s, \pm}} - \psi_{k_{s, \pm}} \right) = 0$.
It is thus sufficient to consider the condition
$\cos(\Delta \phi_k - \psi_k) = 0$. For our choice of initial state with
vanishing intercell hopping, $J_{2, 0} =0$, this condition leads to
\begin{equation}
  \label{eq:soft-mode-eqs-SSH}
  J_1 + J_2 \cos(k) = - \sgn(\sin(k)) 2 \Delta \gamma, \quad \omega_k = J_2
  \abs{\sin(k)}.
\end{equation}
The two solutions are given by
\begin{equation}
  \label{eq:SSH-soft-modes}
  k_{s,\pm} = \mp \sgn(J_1) \arccos \! \left(- \left( J_1 \mp \sgn(J_1) 2
      \Delta\gamma \right) \middle/ J_2 \right).
\end{equation}
We have chosen the designation of each of the solutions of
Eq.~\eqref{eq:soft-mode-eqs-SSH} as either $k_{s, +}$ or $k_{s, -}$ such that
$k_{s, -}$ and $k_{s, +}$ are associated with the pumping of string order
through the right and left boundary of a subsystem, respectively, as detailed
below. Note that the designation is reversed when $J_1 = J + \Delta J$ changes
sign. Crucially, for this choice, the frequencies $\omega_{k_{s, \pm}}$ are
continuous functions of $\Delta J$ and $\Delta \gamma$. For
$\Delta \gamma \neq 0$, the PTGGE has two distinct soft modes
$k_{s,+} \neq - k_{s,-}$. In contrast, for the GGE, which is obtained for
$\Delta \gamma \to 0$, the soft modes are locked onto each other by inversion
symmetry, $k_{s, +} = - k_{s, -}$. Associated with the soft modes are two time
scales~\footnote{We notice that the definition of the time scales $t_{s, \pm}$
  in Ref.~\cite{Starchl2022} erroneously contains a factor of $1/2$.},
\begin{equation}
  \label{eq:SSH-soft-modes-time-scales}
  t_{s,\pm} = \pi/\omega_{k_{s,\pm}} = \left. \pi \middle/ \left( J_2
      \abs{ \sin(k_{s,\pm}) } \right), \right.
\end{equation}
which determine the oscillation periods of the \dsop{}, and, in particular, the
periodicity of the recurring zero crossings of the \dsop. Note that since
$\omega_k = \omega_{-k}$, there is only a single time scale
$t_s = t_{s, +} = t_{s, -}$ for isolated systems. As shown in
Fig.~\ref{fig:SOP-PT-sym}, the periodicity of zero crossings persists up to
$t_F$, and the subsequent relaxation of the \dsop{} for $t > t_F$ approximately
follows the prediction given in
Eq.~\eqref{eq:subsystem-parity-space-time-scaling-trivial}.

The oscillatory dynamics of the \dsop{} for quenches to the topological phase
are connected to the topological zero crossings in the entanglement spectrum. We
explore this connection in more detail in the following section.

\subsubsection{Topological zero crossings}
\label{sec:SOP-zero-crossings}

\begin{figure}
  \centering
  \includegraphics[width=\linewidth]{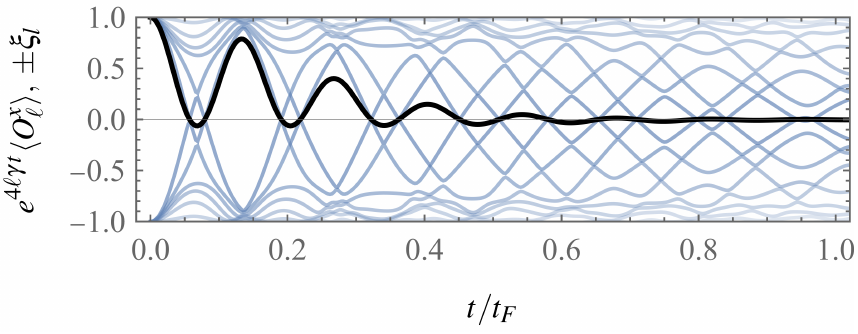}
  \caption{Simultaneous zero crossings of the \dsop{} (black
    line) and in the single-particle entanglement spectrum (blue lines) for a
    quench to the topological phase, $\Delta J=-1.5 J$ and
    $\gamma = \Delta\gamma = 0.3J$, with balanced loss and gain
    $\delta = \Delta = 0$, and for a subsystem of size $\ell=10$.}
  \label{fig:entanglement-spectrum}
\end{figure}

For quantum quenches in the isolated SSH model, zero crossings of the \dsop{}
are a robust signature of the topology of the postquench Hamiltonian. To explain
why that is the case, we observe first that according to
Eq.~\eqref{eq:dual-SOP-expectation} zero crossings of the \dsop{} occur whenever
an eigenvalue of $\Gamma_{\ell}$, or, equivalently as per
Eq.~\eqref{eq:Gamma-l-G-l-SSH}, of the reduced covariance matrix $G_{\ell}$,
crosses zero. The coincidence of zero crossings of the \dsop{} and in the
single-particle entanglement spectrum applies to all Gaussian states, and is
illustrated for a quench to the topological PT-symmetric phase in
Fig.~\ref{fig:entanglement-spectrum}. Furthermore, the dynamical entanglement
spectrum bulk-boundary correspondence for one-dimensional lattice
models~\cite{Gong2017a} that belong to the class BDI~\cite{Lu2019} states that
for quenches originating from the trivial phase, zero crossings in the
entanglement spectrum occur if and only if the postquench Hamiltonian is
topologically nontrivial. Consequently, also zero crossings of the \dsop{} occur
if and only if the postquench Hamiltonian of the isolated SSH model is
topologically nontrivial. As we discuss in more detail below, for the Kitaev
chain, the role of the \dsop{} is played by the subsystem fermion
parity~\cite{Starchl2022}.

\begin{figure}
  \centering
  \includegraphics[width=\linewidth]{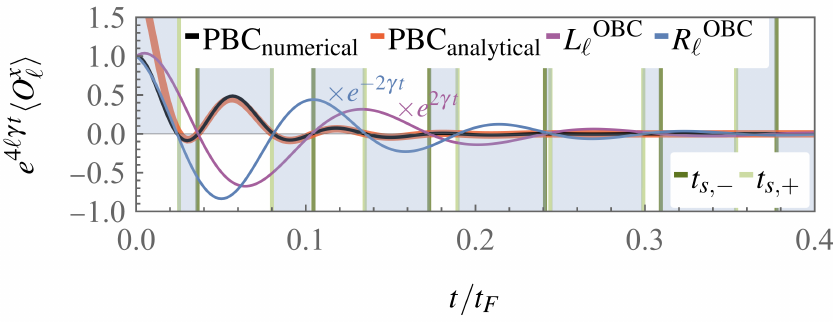}
  \caption{Directional string order pumping for a quench to the topological
    PT-symmetric phase with $\Delta J=-0.5 J$, $\gamma = \Delta\gamma = 0.2J$,
    $\delta = \Delta = 0$, and $\ell=40$. For PBC, the numerical data (black
    line) crosses zero at multiples of both $t_{s, +}$ (light green) and
    $t_{s, -}$ (dark green), with blue shading indicating the sign of the \dsop. The analytical conjecture
    Eq.~\eqref{eq:subsystem-parity-space-time-scaling-topological} agrees well
    with the numerics (red line). For OBC, zero crossings occur at multiples of
    $t_{s, -}$ or $t_{s, +}$ depending on whether the subsystem is at the left
    ($L_{\ell}$, violet line) or the right end of the chain ($R_{\ell}$, blue
    line). Additional rescaling with $\e^{\pm 2 \gamma t}$ of the data for
    $L_{\ell}$ and $R_{\ell}$ compensates for exponential decay and growth,
    respectively, due to edge modes.}
  \label{fig:SSH-crossings-PT-symmetric}
\end{figure}

Turning now to the driven-dissipative SSH model, the continuity of non-Hermitian
real-line-gap topology suggests that the relation between zero crossings of the
\dsop{} and nontrivial topology of the generator of the postquench dynamics
remains valid through the entire PT-symmetric phase. An important qualitative
modification due to drive and dissipation has already been mentioned above:
There are now two distinct time scales for zero crossings of the \dsop, which
are determined by soft modes of the PTGGE. In Ref.~\cite{Starchl2022}, for the
driven-dissipative Kitaev chain, we have provided a physical interpretation of
these two time scales in terms of directional parity pumping: The two time scales
correspond to different rates of the exchange of parity between a subsystem and
its complement through the left and right ends of the subsystem. This effect
requires the breaking of inversion symmetry and mixedness of the time-evolved
state, and is thus unique to driven-dissipative systems. As we illustrate in
Fig.~\ref{fig:SSH-crossings-PT-symmetric}, the driven-dissipative SSH model
exhibits an analogous effect of directional string order pumping. The figure
shows the dynamics of the rescaled \dsop{} for a quench to the PT-symmetric
topological phase. After a short initial period, the numerical data for PBC
(black line) are well described by the analytical prediction
Eq.~\eqref{eq:subsystem-parity-space-time-scaling-topological} (red line), and
exhibit zero crossings at multiples of both $t_{s, -}$ and $t_{s, +}$. In
contrast, the \dsop{} for subsystems $L_{\ell} = \{1,\dots,\ell\}$ at the left
end and $R_{\ell} = \{L-\ell+1,\dots,L\}$ at the right end of a chain with OBC,
crosses zero at multiples of only $t_{s, -}$ and $t_{s, +}$, respectively. These
observations confirm that string order pumping does indeed occur at different
rates through the left and right ends of a subsystem. We note that a previous
study of topological entanglement spectrum crossings in the driven-dissipative
Kitaev chain has focused on subsystems of type $L_{\ell}$ and has, therefore,
not observed directional pumping~\cite{Sayyad2021}.

\begin{figure}
  \centering
  \includegraphics{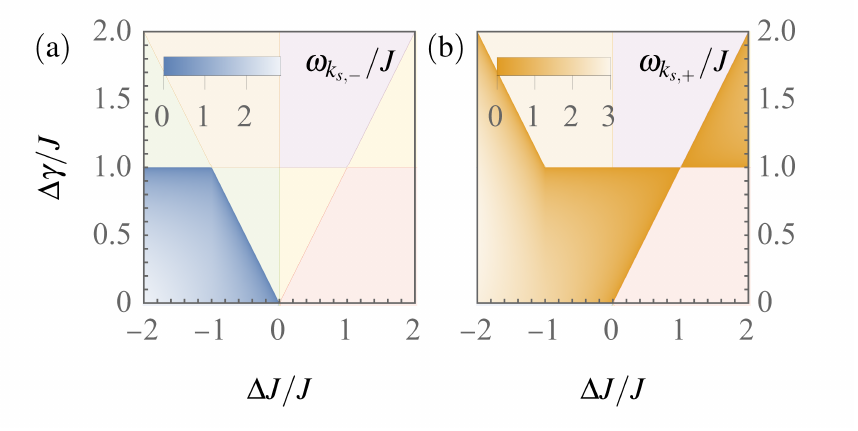}
  \caption{Directional string order pumping phase diagram of the
    driven-dissipative SSH model determined by the soft-mode frequencies (a)
    $\omega_{k_{s,-}}$ and (b) $\omega_{k_{{s, +}}}$. The phases determined by
    PT symmetry, introduced in Fig.~\ref{fig:SSH-phasediagram}, are shown in the
    background.}
  \label{fig:SSH-soft-mode-phasediagram}
\end{figure}

\begin{figure}
  \centering
  \includegraphics[width=\linewidth]{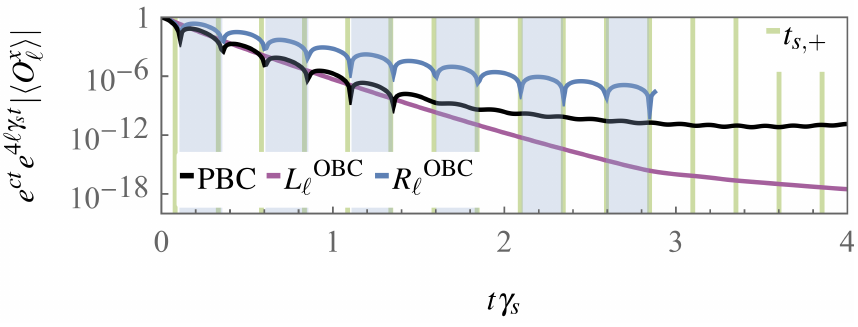}
  \caption{Directional string order pumping for quenches to the PT-mixed phase
    with $\Delta J = -5 J$, $\Delta \gamma = \gamma = 1.1 J$, and $\ell = 30$,
    for PBC (black line) and for OBC with subsystems located at the left
    ($L_{\ell}$, purple line) and the right edge of the chain ($R_{\ell}$, blue
    line). The blue shading indicates the sign of the \dsop{} for the subsystem
    $R_{\ell}$. Zero crossings of the \dsop{} for PBC and the subsystem
    $R_{\ell}$ for OBC fit well to the soft mode time scale $t_{s,+}$. We include
    an additional exponential rescaling $\e^{c t}$ with $c = 51$ to improve the
    presentation of the data.}
  \label{fig:crossings-PT-mixed}
\end{figure}

As explained above, our expectation for the occurrence of zero crossings of the
\dsop{} in the topological PT-symmetric phase is based on the continuity of
topological properties throughout this phase. Indeed, for the driven-dissipative
SSH model, two soft modes $k_{s, \pm}$ of the PTGGE given by
Eq.~\eqref{eq:SSH-soft-modes} exist within the entire topological PT-symmetric
phase. The corresponding frequencies $\omega_{k_{s, \pm}}$ are shown in
Fig.~\ref{fig:SSH-soft-mode-phasediagram}. Upon increasing $\Delta\gamma$, one
of the soft mode frequencies, $\omega_{s, -}$, vanishes at the transition from
the PT-symmetric to the PT-mixed phase. In contrast, $\omega_{k_{s, +}}$ remains
finite within the entire PT-mixed phase, and vanishes at the boundaries
separating the PT-mixed from the trivial PT-symmetric and the PT-broken
phases. This observation suggests that zero crossings of the \dsop{} can also
occur in the PT-mixed phase---an expectation that is confirmed in
Fig.~\ref{fig:crossings-PT-mixed}. The numerical analysis of zero crossings of
the \dsop{} for quenches to the PT-mixed phase is more challenging than for the
PT-symmetric phase: In the PT-symmetric phase, through a simple rescaling, the
overall exponential decay of the covariance matrix can be removed analytically
before evaluating the covariance matrix numerically~\cite{Starchl2022}. In
contrast, due to the finite bandwidth of decay rates in the PT-mixed phase,
exponential decay cannot be fully accounted for beforehand, leading at late
times to extremely small numbers below the numerical precision. The data shown
in Fig.~\ref{fig:crossings-PT-mixed} is rescaled by
$\e^{\left( 4 \ell \gamma_s + c \right) t}$, where $2 \gamma_s$ is the decay
rate of the slowest-decaying PT-breaking mode given in Eq.~\eqref{eq:gamma-s},
and we have chosen $c = 51$ to best present the numerical results. In agreement
with our expectations based on the values of $\omega_{k_{s, \pm}}$ shown in
Fig.~\ref{fig:SSH-soft-mode-phasediagram}, the \dsop{} for the subsystem
$L_{\ell}$ does not exhibit zero crossings, while for $R_{\ell}$ there are zero
crossings at multiples of $t_{s,+}$. Note that the numerical data for $R_{\ell}$
stops abruptly due to the occurrence of exponentially small numbers in the
calculation of the Pfaffian. For PBC we, further observe a transition from a
fast initial decay to a much slower relaxation behavior, in analogy to but much
smoother than the transition at $t_F$ for the quenches to the PT-symmetric phase
shown in Fig.~\ref{fig:SOP-PT-sym}.

Our findings lead us to speculate that the non-Hermitian topology of a suitably
defined restriction of the matrix $z_k$ to its PT-symmetric eigenmodes remains
intact even upon crossing the transition to the PT-mixed phase, and that a
finite gap is not essential for such a construction. In physical terms, our
results suggest to define \emph{directional pumping phases} based on the
existence of soft modes. For the driven-dissipative SSH model, as can be seen
in Fig.~\ref{fig:SSH-soft-mode-phasediagram}, the phase boundaries of
directional pumping phases are also phase boundaries in terms of the gap
structure and PT symmetry. However, as we discuss for the example of a
driven-dissipative Kitaev chain with long-range hopping and pairing in
Sec.~\ref{sec:long-range-kitaev-quench} below, this does not have to be the case.

\subsubsection{Dynamical criticality}
\label{sec:SSH-dynamical-criticality}

Having introduced the concept of directional pumping phases, we may ask what
kind of critical behavior occurs at the transitions between these phases, and
whether this behavior is universal. Usually, dynamical critical phenomena in
driven-dissipative systems are associated with symmetry-breaking second order
phase transitions in the steady state~\cite{Diehl2008, Verstraete2009,
  Eisert2010, Sieberer2013, Sieberer2016a, Maghrebi2016a, Jin2016, Rota2017,
  Halati2022}. Then, critical behavior is induced by the closing of the
dissipative gap, given by the decay rate of the slowest-decaying eigenmode of
the Liouvillian, at the critical point. In particular, the correlation length
diverges as $\xi \sim \abs{\delta_g}^{-\nu}$, where $\delta_g = g - g_c$ is the
deviation of the parameter $g$ from its critical value $g_c$, and the divergence
of the relaxation time scale $\tau \sim \abs{\delta_g}^{- \nu z}$, where $z$ is
the dynamical critical exponent, leads to the phenomenon of critical slowing
down. Contrary to that, directional pumping phase transitions mark a qualitative
change in the oscillatory dynamics of the \dsop{} and are not associated with
any changes in the steady state. Indeed, we consider here dynamics which, for
$\delta = \Delta = 0$, always lead to a steady state at infinite temperature
with a vanishing correlation length. Further, the rate of relaxation to the
steady state is finite for any value of $\gamma > 0$. Instead, directional
pumping phase transitions are characterized by divergences of the soft-mode time
scales $t_{s, \pm}$, which determine the periodicity of zero crossings of the
\dsop. The soft modes $k_{s, \pm}$ and the associated time scales $t_{s, \pm}$
are genuinely dynamical quantities and depend on both pre- and postquench
parameters. However, as we discuss in the following, the critical exponents that
govern the power-law behavior of $k_{s, \pm}$ and $t_{s, \pm}$ at the phase
boundaries do not depend on the specific choice of parameters. These exponents
are determined by an effective long-wavelength description, which indicates that
their values are indeed universal.

At the boundaries of the directional pumping phases, the soft modes $k_{s, \pm}$
approach values $k_{s, \pm, c}$. The frequencies
$\omega_{k_{s, \pm}} = J_2 \, \lvert\sin(k_{s, \pm})\rvert$ vanish at the phase
boundaries, implying that $k_{s, \pm, c} = 0, \pm \pi$. In analogy to the
correlation length exponent $\nu$ and the dynamical exponent $z$, we define
critical exponents $\nu'$ and $z'$ through the scaling behavior
\begin{equation}
  \label{eq:critical-exponents}
  \abs{k_{s, \pm} - k_{s, \pm, c}} \sim \abs{\delta_g}^{\nu'}, \qquad
  \omega_{k_{s, \pm}} \sim \abs{\delta_g}^{\nu' z'},
\end{equation}
for $\delta_g = g - g_{c, \pm} \to 0$, leading to a divergence of the soft-mode
time scales $t_{s, \pm} \sim \abs{\delta_g}^{- \nu' z'}$. For the SSH model, we
consider the parameters $g = \Delta J$ and $g = \Delta \gamma$. Let us assume
that the values $\Delta J_{c, \pm}$ and $\Delta \gamma_{c, \pm}$ correspond to a
particular point on the phase boundary of $\omega_{k_{s, \pm}}$, where
$k_{s, \pm}$ takes the value $k_{s, \pm, c}$. To obtain the behavior of
$k_{s, \pm}$ and $\omega_{k_{s, \pm}}$ in the vicinity of this critical point,
we set $\Delta J = \Delta J_{c, \pm} + \delta_{\Delta J}$ and
$\Delta \gamma = \Delta \gamma_{c, \pm} + \delta_{\Delta \gamma}$ in
Eq.~\eqref{eq:soft-mode-eqs-SSH}, and expand in $k$ around $k_{s, \pm, c}$. We
thus find
\begin{equation}
  \abs{k_{s,\pm} - k_{s, \pm, c}} \sim \omega_{k_{s,\pm}} \sim
  \abs{\delta_g}^{1/2} \quad \text{for } \delta_g \to 0,  
\end{equation}
for both $g = \Delta J$ and $g = \Delta \gamma$. Therefore, the time scales
$t_{s, \pm}$ exhibit a square-root divergence with critical exponents
\begin{equation}
  \label{eq:critical-exponents-SSH}
  \nu' = 1/2, \qquad z' = 1,
\end{equation}
irrespective of the direction from which the phase boundary is approached in the
$\Delta J$-$\Delta \gamma$ plane.

\subsection{Subsystem fermion parity}
\label{sec:subsystem-parity}

\begin{figure}
  \centering
  \includegraphics{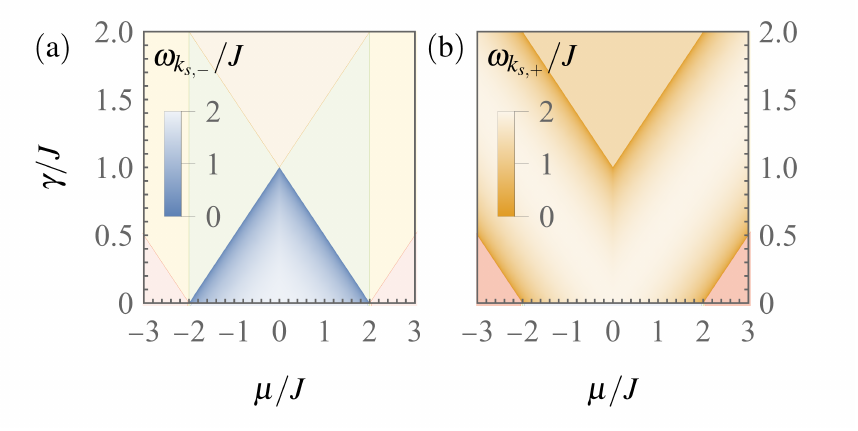}
  \caption{Directional parity pumping phases determined by the soft-mode
    frequencies (a) $\omega_{k_{s, -}}$ and (b) $\omega_{k_{s, +}}$ for the
    driven-dissipative Kitaev chain. The background colors indicate the phases
    defined by PT symmetry, with PT-symmetric phases (blue, red), PT-mixed
    phases (green, yellow) and a PT-broken phase
    (orange)~\cite{Sayyad2021,Starchl2022}. For better visual differentiation the saturation of background colors in (b) has been increased.}
  \label{fig:soft-mode-phasediagram}
\end{figure}

As mentioned above, the subsystem fermion parity serves as a topological
disorder parameter for the Kitaev chain. This can be seen by noting that a
combination of the Jordan-Wigner~\cite{Jordan1928} and
Kramers-Wannier~\cite{Kramers1941,Fisher1995} transformations maps the Kitaev
chain to the transverse field Ising model such that the trivial phase of the
Kitaev chain corresponds to the ferromagnetically ordered phase of the Ising
model. Then, order-parameter correlations of the Ising model over a distance
$\ell$ are equivalent to the fermion parity $P_{\ell}$ of a subsystem of size
$\ell$ of the Kitaev chain~\cite{Starchl2022}, which is defined by
\begin{equation}
  P_{\ell} = \e^{\imag \pi \sum_{l=1}^{\ell} c_l^{\dagger} c_l}.
\end{equation}
For a Gaussian state, the expectation value
$\braket{P_{\ell}} = \pf(\Gamma_{\ell})$ is given by the Pfaffian of the reduced
covariance matrix $\Gamma_{\ell}$ defined in
Eq.~\eqref{eq:gamma-kitaev}~\cite{Lieb1961, Barouch1970I, *Barouch1971II,
  *Barouch1971III}.

\subsubsection{Topological zero crossings}

In complete analogy to our the discussion for the SSH model in
Sec.~\ref{sec:SOP-zero-crossings}, the subsystem parity can be seen to exhibit
zero crossings for quenches of the isolated Kitaev chain from the trivial to the
topological phase. As shown in Refs.~\cite{Starchl2022}, the connection between
zero crossings and topology remains valid throughout the entire PT-symmetric
phase of the driven-dissipative Kitaev chain. The periodicities $t_{s, \pm}$ of
these zero crossings are determined by the soft modes $k_{s, \pm}$, which, for
$\mu_0 \to - \infty$, are given by the solutions to~\cite{Starchl2022}
\begin{equation}
  \label{eq:soft-mode-eqs-Kitaev}
  2 J \cos(k) + \mu = \sgn(\sin(k)) 2 \gamma, \quad \omega_k = 2 \Delta \abs{\sin(k)}.
\end{equation}
We find
\begin{equation}
  \label{eq:soft-modes-Kitaev}
  k_{s,\pm} = \pm \sgn(\mu) \arccos \! \left( - \left( \mu \mp \sgn(\mu) 2
      \gamma \right) \middle/ \left( 2 J \right) \right),
\end{equation}
and the corresponding frequencies $\omega_{k_{s, \pm}}$ determine the
directional pumping phases shown in Fig.~\ref{fig:soft-mode-phasediagram}. As in
the case of the SSH model, the frequency $\omega_{k_{s, -}}$, which describes
parity pumping through the right end of a subsystem, is nonzero only within the
topological PT-symmetric phase; in contrast, the frequency $\omega_{k_{s, +}}$,
corresponding to parity pumping through the left end of a subsystem, remains
finite within the entire PT-mixed phase. The phase boundaries defined by
directional parity pumping coincide with those that are related to
PT symmetry. However, as we show in Sec.~\ref{sec:long-range-kitaev} where we
consider a Kitaev chain with long-range hopping and pairing, this does not have
to be the case.

\subsubsection{Dynamical criticality}
\label{sec:dynamical-criticality-kitaev}

The critical exponents $\nu'$ and $z'$, which describe the behavior of the soft
modes $k_{s, \pm}$ and the frequencies $\omega_{k_{s, \pm}}$ in the vicinity of
transitions between directional pumping phases according to
Eq.~\eqref{eq:critical-exponents}, take on the same values for the Kitaev chain
as for the SSH model. To obtain this result, we insert
$\mu = \mu_c + \delta_{\mu}$ and $\gamma = \gamma_c + \delta_{\gamma}$ in
Eq.~\eqref{eq:soft-mode-eqs-Kitaev} and expand in $k$ around
$k_{s, \pm, c} = 0, \pm \pi$. We obtain
\begin{equation}
  \omega_{k_{s,\pm}} \sim \abs{k_{s,\pm} - k_{s, \pm, c}} \sim
  \abs{\delta_g}^{1/2} \quad \text{for } \delta_g \to 0,  
\end{equation}
for both $g = \mu$ and $g = \gamma$, which leads to the values of $\nu'$ and
$z'$ given in Eq.~\eqref{eq:critical-exponents-SSH}. As we show in the next
section, the values of these exponents can be modified in the presence of
long-range hopping and pairing.

\section{Driven-dissipative Kitaev chain with long-range hopping and pairing}
\label{sec:long-range-kitaev-quench}

We have introduced the concept of directional pumping phases to distinguish
parameter regions with qualitatively different dynamics of string order and
fermion parity. Transitions between directional pumping phases are characterized
by divergent time scales $t_{s, \pm}$ for string order and parity pumping, and
there is evidence for universality of the exponents that govern the critical
behavior of $t_{s, \pm}$: We have found the same exponents given in
Eq.~\eqref{eq:critical-exponents-SSH} for two models that differ by the presence
of a weak $\mathrm{U}(1)$ symmetry but belong to the same Altland-Zirnbauer
class; and these exponents can be obtained from an expansion in momenta around
critical values $k_{s, \pm, c}$, which indicates that models with the same
low-momentum or, equivalently, long-wavelength description will have the same
exponents, while microscopic details that require the entire Brillouin zone for
their description are irrelevant. However, in the examples we have considered so
far, the boundaries of directional pumping phases coincide with gap closings of
the Liouvillian single-particle spectrum in the complex plane, suggesting that
also directional pumping phase transitions are a mere manifestation of gap
closings, and that the exponents that describe the critical behavior of
$t_{s, \pm}$ are determined by those that govern the divergences of oscillation
periods of low-order correlation functions~\cite{Sayyad2021}. To better
understand the relation between directional pumping phases and the more
elementary notion of dynamical phases defined in terms of gap closings or
PT symmetry, we now consider quench dynamics in a Kitaev chain with long-range
hopping and pairing~\cite{Vodola2014, Vodola2016, VanRegemortel2016,
  Buyskikh2016, Viyuela2016, Lepori2016, Lepori2017, Lepori2017a, Dutta2017,
  Alecce2017, Bhattacharya2018a, Su2020, Maity2020a, Jager2020, Uhrich2020,
  Francica2022, Mondal2022}. Long-range couplings are known to modify critical
properties at gap closings in isolated systems. As we discuss in the following,
for a Kitaev chain with Markovian drive and dissipation, the presence of
long-range couplings can likewise lead to modifications of the critical
exponents that are associated with the pumping time scales $t_{s, \pm}$, but
does not affect the exponents that govern the divergence of the oscillation
period of the density autocorrelation function. Furthermore, even the boundaries
of the directional pumping phases of the driven-dissipative long-range Kitaev
chain do not always coincide with gap closings. These results indicate that
directional pumping phases and the associated critical behavior are indeed new
and independent concepts.

\subsection{Long-range Kitaev chain}
\label{sec:long-range-kitaev}

We consider a Kitaev chain with long-range hopping and
pairing as described by the Hamiltonian~\cite{Uhrich2020}
\begin{multline}
  \label{eq:H-Kitaev-LR}
  H = \sum_{l = 1}^L \sum_{r = 1}^{\lfloor L/2 \rfloor} \left( - J_r
    c^{\dagger}_l c_{l+r} + \Delta_r c_l c_{l+r} + \hc \right) \\ - \mu
  \sum_{l=1}^L \left( c^{\dagger}_l c_l - \frac{1}{2} \right),
\end{multline}
where both the hopping matrix element
$J_r = \left. J \middle/ \left( \mathcal{N}_\alpha r^{\alpha} \right) \right.$
and the pairing amplitude
$\Delta_r = \left. \Delta \middle/ \left( \mathcal{N}_\alpha r^{\alpha} \right)
\right.$
decay with distance as a power law with exponent $\alpha > 1$, and the Kac
normalization factors are defined as
$\mathcal{N}_\alpha = \sum_{r = 1}^{\lfloor L/2 \rfloor}
r^{-\alpha}$~\cite{Kac1963}.
As with the short-range Kitaev chain, we assume that the coupling to Markovian
reservoirs is described by the jump operators given in Eq.~\eqref{eq:L}, where
we focus on $\gamma_l = \gamma_g$ in this section. Consequently, the
single-particle dispersion relation of the Liouvillian takes the form given in
Eq.~\eqref{eq:dispersion-open-Kitaev}, but with the Hamiltonian dispersion
relation $\varepsilon_k$ given by
\begin{equation}
  \label{eq:epsilon-LRKC}
  \varepsilon_k = \sqrt{\left( 2 J_k + \mu \right)^2 + 4 \Delta_k^2},
\end{equation}
where, in the thermodynamic limit $L \to \infty$,
\begin{equation}
  \label{eq:J-k-Delta-k}
  J_k = \frac{J}{\zeta \! \left(\alpha\right)} \Re \! \left( \mathrm{Li}_{\alpha} \! \left(\e^{\imag k}\right)\right), \quad
  \Delta_k =  \frac{\Delta}{\zeta \! \left(\alpha\right)} \Im \! \left( \mathrm{Li}_\alpha \! \left(\e^{\imag k}\right) \right).
\end{equation}
Explicit expressions for the polylogarithm $\mathrm{Li}_{\alpha}(z)$ and the
Riemann zeta function $\zeta(\alpha)$ are provided in
Appendix~\ref{sec:polylog-zeta}.

\subsection{Phase diagram of the long-range Kitaev chain}
\label{sec:long-range-dispersion}

\begin{figure}
  \centering
  \includegraphics{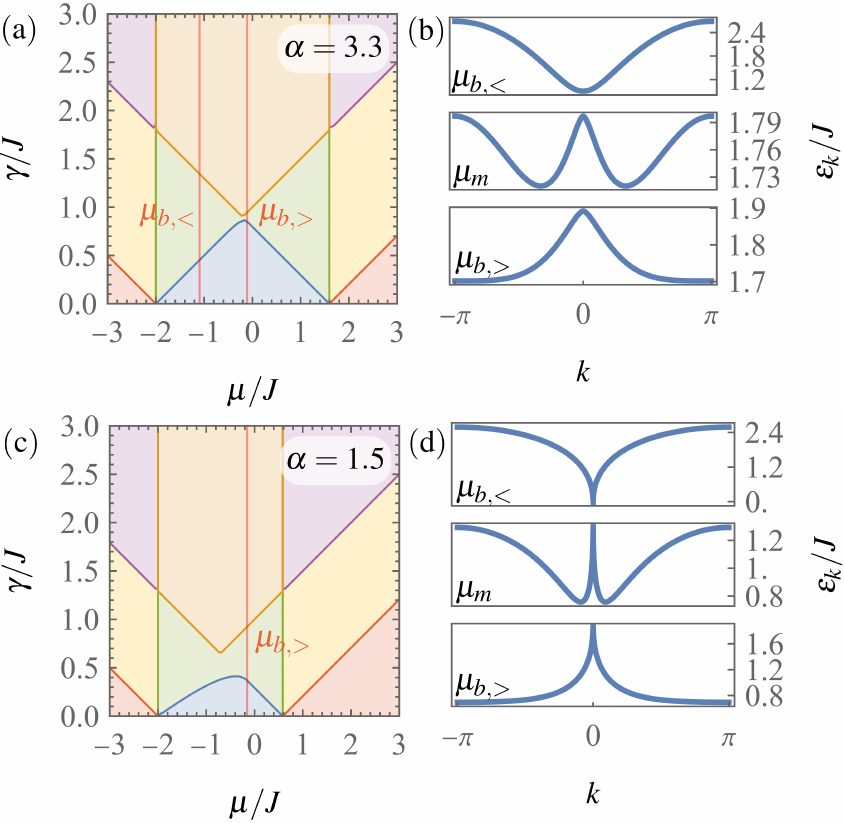}
  \caption{Phase diagrams and dispersion relations of the long-range Kitaev
    chain for (a), (b) $\alpha = 3.3$ and (c), (d) $\alpha = 1.5$. (a), (c)
    Colors indicate PT-symmetric (blue, red), PT-mixed (green, yellow) and
    PT-broken phases (orange, purple). (b), (d) Dispersion relations
    $\varepsilon_k$ with a single minimum at $k = 0$ for $\mu = \mu_{b, <}$, two
    degenerate minima at $k = \pm k_b$ for $\mu = \mu_m$, and a single minimum
    at $k = \pi$ for $\mu = \mu_{b, >}$. The bifurcation points
    $\mu_{b, \lessgtr}$ are shown in (a), (c) as red lines. For $\alpha = 1.5$,
    the first bifurcation occurs at $\mu_{b, <} = \mu_{c, <} = -2 J$.}
  \label{fig:long-range-phasediagram}
\end{figure}

Long-range couplings do not affect the symmetry properties of the Kitaev chain;
in particular, the driven-dissipative long-range Kitaev chain is
PT-symmetric. The phase diagram of the long-range Kitaev chain, determined by
the spontaneous breaking of PT symmetry, is shown in
Figs.~\ref{fig:long-range-phasediagram}(a) and (c) for $\alpha = 3.3$ and
$\alpha = 1.5$, respectively. In comparison to the phase diagram of the
short-range Kitaev chain~\cite{Starchl2022, Sayyad2021}, the key qualitative
differences are the absence (i)~of mirror symmetry with respect to axis
$\mu = 0$ and (ii)~of a direct transition between the PT-symmetric and the
PT-broken phase, and (iii)~a segment of the boundary of the topological
PT-symmetric phase being curved instead of described by straight lines.

To understand these modifications, let us first discuss how the corresponding
properties come about in the short-range Kitaev chain: (i)~The unitary
transformation $c_l \mapsto \left( -1 \right)^l c_l$ maps the Hamiltonian $H$
Eq.~\eqref{eq:H-Kitaev} and the chemical potential $\mu$ to $- H$ and $-\mu$,
respectively. Therefore, gap closings that determine phase boundaries occur
symmetrically with respect to $\mu = 0$. This symmetry of the isolated Kitaev
chain extends to the driven-dissipative model. However, in the isolated
long-range Kitaev chain, the mapping $c_l \mapsto \left( -1 \right)^l c_l$ does
not result in a simple sign change and, therefore, the phase diagram does not
have reflection symmetry with respect to $\mu = 0$. The critical values
$\mu_{c, \lessgtr}$ of the chemical potential that separate the topological
phase from the trivial phase are given by
\begin{equation}
  \mu_{c,<} = - 2 J_0 = - 2 J, \quad \mu_{c,>} = - 2 J_{\pi} = 2 J \eta(\alpha)/\zeta(\alpha).
\end{equation}
We note that also the midpoint between $\mu_{c,<}$ and $\mu_{c,>}$, given by
$\mu_m = - \left( J_{\pi} + J_0 \right)$, does not describe a mirror symmetry of
the phasediagram, but will prove to be convenient in the characterization of the
long-range Kitaev chain.

(ii)~At $\mu = 0$, the dispersion relation of the short-range Kitaev chain is
flat, $\varepsilon_k = 2 \sqrt{J^2 + \Delta^2}$. Therefore, upon increasing
$\gamma$, the Liouvillian dispersion relation
Eq.~\eqref{eq:dispersion-open-Kitaev} becomes imaginary simultaneously for all
$k \in \BZ$, leading to a direct transition from the PT-symmetric to the
PT-broken phase. In contrast, for $1 < \alpha < \infty$, the dispersion relation
Eq.~\eqref{eq:epsilon-LRKC} of the long-range Kitaev chain is never flat. The
resulting absence of a direct transition between the PT-symmetric and PT-broken
phase is clearly visible in Fig.~\ref{fig:long-range-phasediagram}.

\begin{figure}
  \centering
  \includegraphics[width=\linewidth]{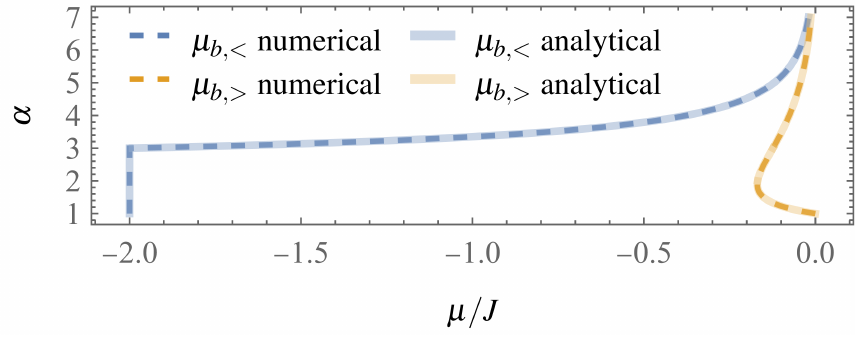}
  \caption{Numerical evaluation of the left and right bifurcation points
    $\mu_{b, <}$ (blue, dashed) and $\mu_{b, >}$ (orange, dashed), respectively,
    together with the analytical predictions in Eqs.~\eqref{eq:mu-b-left-1}
    and~\eqref{eq:mu-b-left-2} (blue, solid) and~\eqref{eq:mu-b-right} (orange,
    solid).}
  \label{fig:mbl-mbr}
\end{figure}

(iii)~Spontaneous breaking of PT symmetry, which determines the boundary of the
PT-symmetric phase, occurs when $\omega_k$ Eq.~\eqref{eq:dispersion-open-Kitaev}
where $\gamma = \gamma_l = \gamma_g$ with $\varepsilon_k$ given in
Eq.~\eqref{eq:epsilon-LRKC} becomes imaginary for some $k$. This happens when
$\min_{k \in \BZ}(\varepsilon_k) = 2 \gamma$. For the short-range Kitaev chain,
the dispersion relation $\varepsilon_k$ takes its minimum value at
$k_{\mathrm{min}} = 0$ when $\mu < 0$, and at $k_{\mathrm{min}} = \pi$ when
$\mu > 0$; the jump of $k_{\mathrm{min}}$ from $0$ to $\pi$ occurs at $\mu = 0$
where the dispersion is flat. In contrast, in the long-range Kitaev chain with
$1 < \alpha < \infty$, there are two bifurcation points of the minimum of
$\varepsilon_k$ at $\mu_{b, \lessgtr}$: There is a single minimum at $k = 0$ for
$\mu < \mu_{b, <}$; then, in the range $\mu_{b, <} < \mu < \mu_{b, >}$, there
are two degenerate minima at momenta $\pm k_b$ with $k_b$ increasing
monotonically from 0 to $\pi$ for $\mu$ increasing from $\mu_{b, <}$ to
$\mu_{b, >}$; finally, for $\mu_{b, >} < \mu$, there is again a single minimum
at $k = \pi$.  This is illustrated for $\alpha = 3.3$ and $\alpha = 1.5$ in
Figs.~\ref{fig:long-range-phasediagram}(b) and~(d), respectively. To determine
the precise shape of the phase boundary, we note that $\Delta_k$ defined in
Eq.~\eqref{eq:J-k-Delta-k} vanishes at $k = 0, \pi$. Therefore, for
$\mu < \mu_{b, <}$ and $\mu_{b, >} < \mu$, the boundary of the PT-symmetric
phase is determined by $\varepsilon_0 = \abs{2 J_0 + \mu} = 2 \gamma$ and
$\varepsilon_{\pi} = \abs{2 J_{\pi} + \mu} = 2 \gamma$, respectively. These
conditions describe straight lines and are symmetric with respect to $\mu_m$
defined above. But for $\mu_{b, <} < \mu < \mu_{b, >}$, PT symmetry breaking
occurs for $\varepsilon_{k_b} = 2 \gamma$, yielding a smaller critical value of
$\gamma$ as compared to an extension of the straight boundaries over the entire
range of values of $\mu$.

The momenta $\pm k_b$ at which the dispersion relation takes on its minimum
value for $\mu_{b, <} < \mu < \mu_{b, >}$ can only be found
numerically. However, the bifurcation points $\mu_{b, \lessgtr}$ can be found
analytically by using the series expansions of the polylogarithm given in
Appendix~\ref{sec:polylog-zeta}: To determine the right bifurcation point
$\mu_{b, >}$, we employ the expansion of $\varepsilon_k$ around $k = \pi$ given
by $\varepsilon_k \sim \varepsilon_{\pi} + (k - \pi)^2/(2 m_{\pi})$, and solve
the equation $1/m_\pi = 0$ for $\mu$. Using Eq.~\eqref{eq:Li-k-to-pi}, for
$1 < \alpha < \infty$, we find
\begin{equation}
  \label{eq:mu-b-right}
  \mu_{b, >} = - 2 \left[ J_{\pi} + \left. \left( \Delta_{\pi}' \right)^2
      \middle/ J_{\pi}'' \right. \right],
\end{equation}
where primes denote derivatives with respect to $k$. The left bifurcation point
$\mu_{b, <}$ is determined by
$\varepsilon_k \sim \varepsilon_{0} + k^2/(2 m_{0})$ and $1/m_0=0$, which holds
for $3 < \alpha < \infty$, where the term $\sim k^{\alpha-1}$ in
Eq.~\eqref{eq:Li-k-to-0} can be neglected. We obtain
\begin{equation}
  \label{eq:mu-b-left-1}
  \mu_{b, <} =- 2 \left[ J_0 + \left. \left( \Delta_0' \right)^2 \middle/ J_0''
    \right. \right] \quad \text{for } 3 < \alpha < \infty.
\end{equation}
For $1 < \alpha < 3$, using the same idea as above but keeping only the relevant
terms in the expansion in Eq.~\eqref{eq:Li-k-to-0}, we are led to
\begin{equation}  
  \frac{\left(\Delta'_k \right)^2}{J^{\prime\prime}_k} \propto \frac{\left(
      k^{\alpha -2} + c_1 \right)^2}{k^{\alpha -3} +c_2} \xrightarrow{k \to 0}
  0,  
\end{equation}
where $c_1,c_2 \in \R$ are constants. Therefore,
\begin{equation}
  \label{eq:mu-b-left-2}
  \mu_{b, <} = -2 J_0 \qquad \text{for } 1 < \alpha < 3.
\end{equation}
In Fig.~\ref{fig:mbl-mbr}, our analytical expressions for the bifurcation points
are shown to agree with numerical solutions.

\subsection{Subsystem fermion parity}

\begin{figure}
  \centering
  \includegraphics[width=\linewidth]{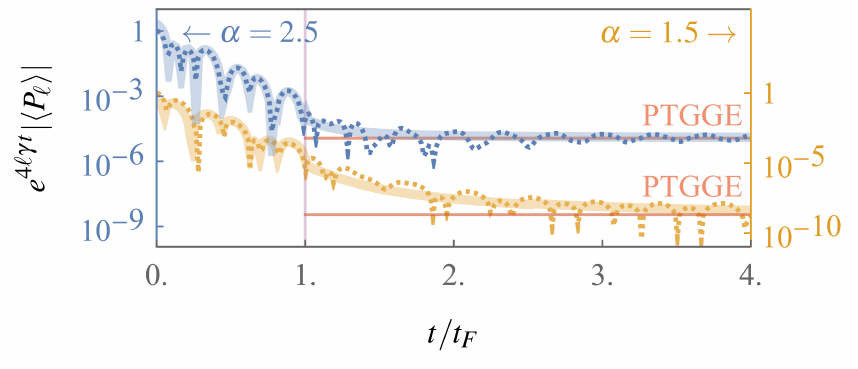}
  \caption{Relaxation of the subsystem parity after quenches in the
    driven-dissipative Kitaev chain with long-range couplings. Both quenches are
    to the topological phase, where the blue line corresponds to $\alpha=2.5$,
    $\mu = -J$, $\gamma = 0.3 J$, and $\ell = 20$, and the orange line to
    $\alpha=1.5$, $\mu = -0.5 J$, $\gamma = 0.15 J$, and $\ell = 80$. Dashed and
    solid lines show numerical data and the analytical conjecture in
    Eq.~\eqref{eq:subsystem-parity-space-time-scaling-topological},
    respectively. For smaller values of $\alpha$ corresponding to longer-range
    couplings, finite-size effects are more strongly pronounced and larger
    system sizes are required. The purple vertical line indicates the
    characteristic relaxation time scale $t_F$ and the red horizontal lines show
    the PTGGE predictions.}
  \label{fig:PTGGE-relax-long-range}
\end{figure}

The time evolution of the subsystem parity after quenches to the topological
PT-symmetric phase for $\alpha = 2.5$ and $\alpha = 1.5$ is shown in
Fig.~\ref{fig:PTGGE-relax-long-range}. In comparison to the short-range Kitaev
chain with Markovian drive and dissipation, there are more pronounced
oscillations after $t = t_F$. However, the stationary values of the rescaled
subsystem parity are still well described by the PTGGE.

\subsubsection{Topological zero crossings}

\begin{figure}
  \centering
  \includegraphics[width=\linewidth]{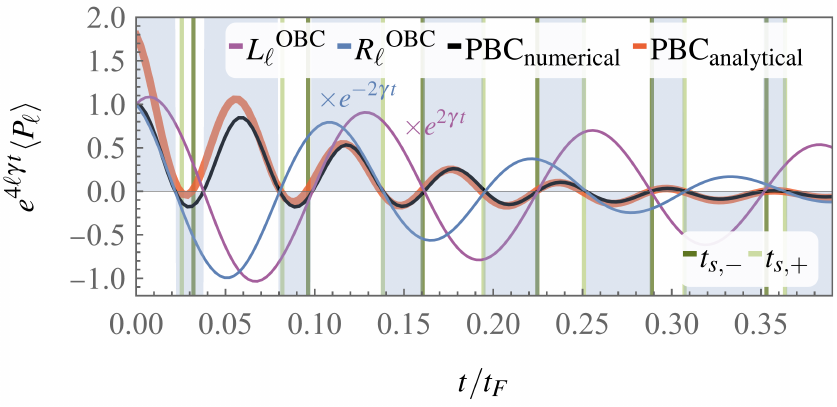}
  \caption{Directional pumping of subsystem parity for a quench to the
    topological PT-symmetric phase with $\alpha = 2.5$, $\mu=-0.5 J$,
    $\gamma=0.3J$, $\delta = 0$, and $\ell=20$. For PBC, the subsystem parity
    (black line) crosses zero at multiples of both $t_{s, +}$ (light green) and
    $t_{s, -}$ (dark green), with blue shading indicating the sign of the
    subsystem parity. In contrast, for OBC, zero crossings occur at multiples of
    $t_{s, -}$ or $t_{s, +}$ for a subsystem located at the left ($L_{\ell}$,
    violet line) and right end of the chain ($R_{\ell}$, blue line). Factors
    $\e^{\pm 2 \gamma t}$ compensate for additional exponential decay and growth
    due to edge modes. The analytical conjecture
    Eq.~\eqref{eq:subsystem-parity-space-time-scaling-topological} agrees well
    with the numerics also for the long-range model after stronger initial
    discrepancies (red line).}
  \label{fig:LR-corssings-25}
\end{figure}

The time scale of topological zero crossings of the subsystem parity is
determined by soft modes of the PTGGE. In the presence of long-range couplings,
the conditions Eq.~\eqref{eq:soft-mode-eqs-Kitaev}, which determine the soft
modes and associated frequency scales for the short-range Kitaev chain, are
generalized to
\begin{equation}
  \label{eq:soft-mode-eqs-Kitaev-lr}
  2 J_k + \mu = \sgn(\Delta_k) 2 \gamma, \quad \omega_k = 2 \abs{\Delta_k}.
\end{equation}
To solve these equations, we use the following properties of the function $J_k$
and $\Delta_k$ defined in Eq.~\eqref{eq:J-k-Delta-k}, which hold for
$1 < \alpha < \infty$: (i)~$J_k = J_{-k}$ and $\Delta_k = - \Delta_{-k}$ are
even and odd, respectively. (ii)~$J_k$ is monotonic for $k \in [0, \pi]$, and
can thus be inverted on that interval. We denote the inverse by
$k = J_{\mathrm{inv}}(J_k)$. (ii)~$\Delta_k \geq 0$ is nonnegative for
$k \in [0, \pi]$. The solutions to Eq.~\eqref{eq:soft-mode-eqs-Kitaev-lr} read
\begin{equation}
  k_{s, \pm} = \pm \sgn(\mu - \mu_s) J_{\mathrm{inv}} \! \left( - \left( \mu \mp
      \sgn(\mu - \mu_s) 2 \gamma \right) \middle/ \left( 2 J \right) \right),
\end{equation}
where we have introduced a value $\mu_s$ of the chemical potential that depends
on $\gamma$ and at which the designation of the solutions of
Eq.~\eqref{eq:soft-mode-eqs-Kitaev-lr} as $k_{s, \pm}$ is reversed. As in the
case of the Kitaev chain with short-range couplings, we have chosen $k_{s, -}$
and $k_{s, +}$ to describe the pumping of parity through the right and left ends
of a subsystem, respectively. The value of $\mu_s$ can be found numerically be
requiring the frequencies $\omega_{k_{s, \pm}}$ to be continuous functions of
$\mu$ and $\gamma$. As illustrated in Fig.~\ref{fig:LR-corssings-25} for the
long-range Kitaev chain with $\alpha = 2.5$, the soft mode periods
$t_{s, \pm} = \pi/\omega_{k_{s, \pm}}$ agree with the zero crossings of the
subsystem parity, both for PBC and OBC.

A unique property of the long-range model is the existence of a finite region in
the $\mu$-$\gamma$ plane which lies outside of the PT-symmetric phase but in
which the PTGGE has two soft modes as shown in
Figs.~\ref{fig:soft-mode-pd-lr}(a), (b) for $\alpha = 2.5$ and (c), (d) for
$\alpha =1.5$, where the boundary of the PT-symmetric phase is indicated by a
black line. This finding demonstrates that directional pumping phases are in
fact not bound to dynamical phases determined by the gap structure of the matrix
$z_k$ or PT symmetry. Moreover, this finding rules out a relation between
topological zero crossings and exceptional points, which are present in the
spectrum of the matrix $z_k$ in the gapless PT-mixed phase at the crossing of
the bands $\lambda_{\pm, k}$. For the SSH model, the band crossing is
illustrated in Fig.~\ref{fig:SSH-phasediagram}(c).

To understand the origin of the difference in phase boundaries, note that
solutions to Eq.~\eqref{eq:soft-mode-eqs-Kitaev-lr} exist for values of $\mu$
and $\gamma$ between the extrema of $J_k$ at $k = 0, \pi$ where $J_0 = J$ and
$J_{\pi} = - J \eta(\alpha)/\zeta(\alpha)$, leading to the conditions
$\abs{2 J_0 + \mu} = 2 \gamma$ and $\abs{2 J_{\pi} + \mu} = 2 \gamma$ that
determine the boundaries of directional pumping phases. In particular,
$\omega_{k_{s, -}}$ is nonzero for $\mu_{c, <} < \mu < \mu_{c, >}$ and
$\gamma < 2 \left( J_0 - J_{\pi} - \abs{\mu - \mu_m} \right)$. The PT-symmetric
phase is delimited by a smaller critical value of $\gamma$ for
$\mu_{b, <} < \mu < \mu_{b, >}$. Outside of that range, the phase boundaries
coincide.

\subsubsection{Dynamical criticality}
\label{sec:dynamical-criticality-LR}

\begin{figure}
  \centering
  \includegraphics{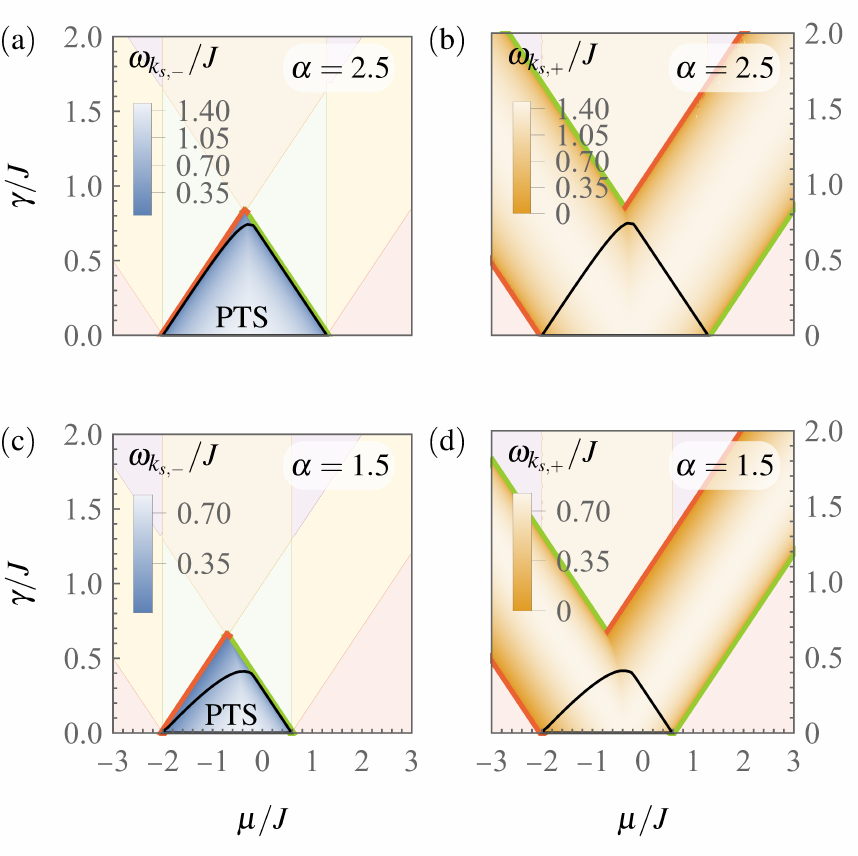}
  \caption{Directional parity pumping phase diagrams of the long-range Kitaev
    chain determined by the soft-mode frequencies $\omega_{k_{s, \pm}}$ for (a),
    (b) $\alpha = 2.5$ and (c), (d) $\alpha = 1.5$. Background colors indicate
    phases defined by PT symmetry as in
    Fig.~\ref{fig:long-range-phasediagram}. The boundary of the PT-symmetric
    topological phase is shown as a black line. Red and green lines indicate
    phase boundaries at which the critical exponents are and are not modified,
    respectively, in comparison to the short-range Kitaev chain.}
  \label{fig:soft-mode-pd-lr}
\end{figure}

The product $\nu' z'$ of critical exponents describes how the soft-mode time
scales $t_{s, \pm}$ diverge at the boundaries of directional pumping phases. For
the long-range Kitaev chain, this divergence is the same as in the short-range
Kitaev chain for the phase boundaries that are shown as green lines in
Fig.~\ref{fig:soft-mode-pd-lr}; but for the phase boundaries that are shown as
red lines, long-range hopping and pairing lead to modified critical behavior. To
derive the corresponding exponents, we note that according to
Eq.~\eqref{eq:soft-mode-eqs-Kitaev-lr}, at directional pumping phase boundaries,
$\abs{\Delta_{k_{s, \pm}}} = \omega_{k_{s, \pm}}/2 $ vanishes, which is the case
for $k_{s, \pm, c} = 0, \pm \pi$. We can then determine the scaling behavior of
$k_{s, \pm}$ as in the short-range Kitaev chain by inserting
$\mu = \mu_c + \delta_{\mu}$ and $\gamma = \gamma_c + \delta_{\gamma}$ in
Eq.~\eqref{eq:soft-mode-eqs-Kitaev-lr} and expanding in $k$ around
$k_{s, \pm, c} = 0, \pm \pi$. Modified scaling behavior of $k_{s, \pm}$ as
compared to the short-range model occurs if the expansions of $J_k$ and
$\Delta_k$ are modified as compared to $J \cos(k)$ and $\Delta \sin(k)$,
respectively. This is the case for the expansions around $k = 0$: Depending on
the value of $\alpha$, different exponents dominate in Eq.~\eqref{eq:Li-k-to-0},
leading to
\begin{equation}
  J_k - J_0 \sim
  \begin{cases}
    k^2 & \text{for } 3 < \alpha, \\
    k^{\alpha-1} & \text{for } 1 < \alpha < 3,
  \end{cases}
\end{equation}
and
\begin{equation}
  \Delta_k \sim \begin{cases}
    k  & \text{for }  2 < \alpha,\\
    k^{\alpha-1} & \text{for }  1 < \alpha < 2.
  \end{cases}
\end{equation}
For the phase boundaries with modified scaling behavior, which are indicated by
red lines in Fig.~\ref{fig:soft-mode-pd-lr}, we thus find
\begin{equation}
  \label{eq:t-s-long-range-critical-exponents}
  \begin{aligned}
    \nu' & = 1/2, & z' & = 1  & \text{for } & 3 < \alpha, \\
    \nu' & = 1/\left(\alpha -1\right), & z' & = 1 & \text{for } & 2 < \alpha < 3, \\
    \nu' & = 1/\left(\alpha -1\right), & z' & = (\alpha - 1) & \text{for } & 1 <
    \alpha < 2.
\end{aligned}
\end{equation}
Otherwise, when $k_{s, \pm, c} = \pm \pi$ which is the case for the phase
boundaries that are shown as green lines, the leading powers in the expansions
of $J_k$ and $\Delta_k$ and, therefore, the critical exponents $\nu'$ and $z'$
are the same as in the short-range Kitaev chain and the SSH model and given in
Eq.~\eqref{eq:critical-exponents-SSH}.

\begin{figure}
  \centering
  \includegraphics[width=\linewidth]{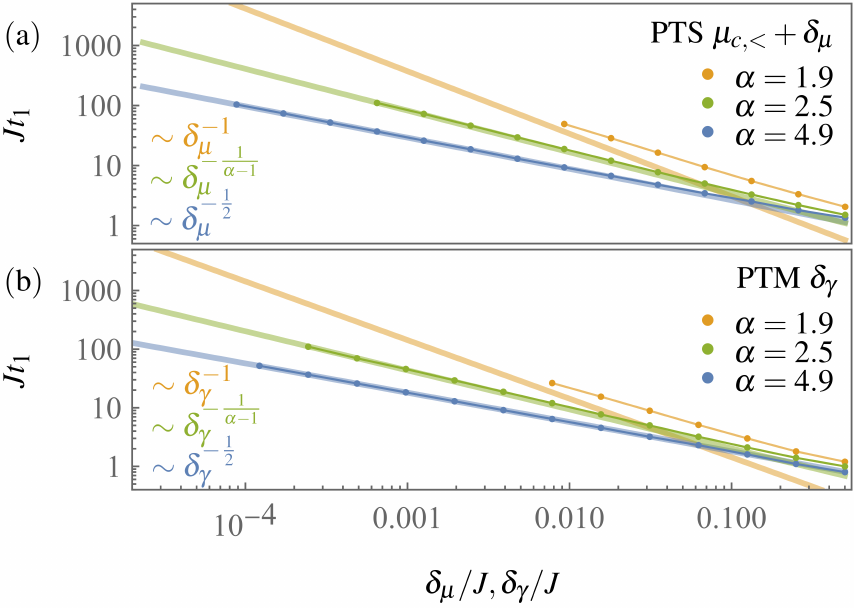}
  \caption{Scaling behavior of the first crossing time $t_1$ for the subsystem
    parity in the long-range Kitaev chain. The connected dots show numerical
    results for $\alpha = 4.9$ (blue), $\alpha = 2.5$ (green) and $\alpha = 1.9$
    (orange), and the expected scaling behavior according to
    Eq.~\eqref{eq:t-s-long-range-critical-exponents} is indicated by solid
    lines, with scaling exponents given in the left lower corner of each
    panel. We consider quenches to (a) the PT-symmetric (PTS) phase with
    $\mu = \mu_{c,<} + \delta_{\mu}$ and $\gamma = 0$, and (b) the PT-mixed
    (PTM) phase with $\mu = \mu_{c, <}$ and
    $\gamma =  \delta_\gamma$, for a chain with OBC and
    subsystems located at (a) the left ($L_{\ell}$ with $\ell = L/2 = 500$) and
    (b) the right end of the chain ($R_{\ell}$ with $\ell = L/2 = 300$).}
  \label{fig:LRK_t1}
\end{figure}

An efficient way to probe the modified exponents numerically is to perform
quenches for a range of values of $\mu$ and $\gamma$ close to a phase boundary
and measure the time $t_1$ at which the first zero crossing of the subsystem
fermion parity occurs. In Fig.~\ref{fig:LRK_t1}(a), we set
$\mu = \mu_{c,<} + \delta_{\mu}$ and $\gamma = 0$, and we consider the fermion
parity of the left half of the system $L_{L/2} = \{ 1, \dotsc, L/2 \}$ in a
chain with OBC and for different values of $\alpha$. By decreasing the value of
$\delta_{\mu}$, we expect to observe a scaling behavior
$t_1 \sim \delta_{\mu}^{- \nu' z'}$ with exponents given in
Eq.~\eqref{eq:t-s-long-range-critical-exponents}. Indeed, we find good agreement
between the predicted exponents and the numerical data for $\alpha > 2$. For
$\alpha < 2$, due to the limited system sizes in our simulations, we cannot
fully reach the scaling regime, we still observe a clear tendency toward the analytically
predicted scaling behavior. In Fig.~\ref{fig:LRK_t1}(b), we present an analogous
analysis for quenches to the PT-mixed phase, where we set $\mu = \mu_{c, <}$ and
$\gamma = \delta_{\gamma}$. Here, pumping of fermion parity occurs only through
the left end of a subsystem and, therefore, we consider a subsystem
$R_{L/2} = \{ L/2 + 1, \dotsc, L \}$ corresponding to the right half of the
chain. Even though numerics in the PT-mixed phase are restricted to smaller
system sizes, we again find compelling agreement with the analytical exponents
for $\alpha > 2$, and a clear trend toward the analytical prediction for
$\alpha < 2$.

\subsection{Connected density autocorrelation function}
\label{sec:conn-dens-dens}

For each momentum mode $k$, the Liouvillian single-particle eigenvalues
$\lambda_{\pm, k} = - \imag 2 \gamma \pm \omega_k$ are formally identical to the
eigenfrequencies of a damped harmonic oscillator with undamped natural frequency
$\varepsilon_k$ and damping rate $2 \gamma$. When the damping rate is increased,
the mode $k$ undergoes a transition from under- to overdamped oscillations---or,
equivalently, from PT-symmetric to PT-breaking---when $\varepsilon_k = 2 \gamma$
in Eq.~\eqref{eq:dispersion-open-Kitaev}. It is interesting to compare how
different observables are affected by this overdamping transition. The
oscillation frequencies of the topological disorder parameters of the SSH model
and the Kitaev chain are determined by the soft modes $k_{s, \pm}$, and
directional pumping transitions occur when either of these modes becomes
overdamped. According to our definition of these modes, $k_{s, -}$ is the first
to become overdamped, as shown in the phase diagrams in
Figs.~\ref{fig:SSH-soft-mode-phasediagram}, \ref{fig:soft-mode-phasediagram},
and~\ref{fig:soft-mode-pd-lr}. But even when $k_{s, -}$ is overdamped while
$k_{s, +}$ is not, for a system with PBC or a subsystem that is not on the left
end of a system with OBC, the topological disorder parameters exhibit
underdamped oscillatory decay as illustrated in
Fig.~\ref{fig:crossings-PT-mixed}. To observe this behavior, it seems to be
crucial that the disorder parameters act nontrivially on sufficiently large
subsystems of size $\ell$. Indeed, underdamped oscillatory decay persist until
leveling, caused by the finiteness of $\ell$, sets in. As an important example
of a local observable with support on only a small number of lattice sites, we
consider the density autocorrelation function in the steady state of the
driven-dissipative long-range Kitaev chain. As we show in the following, upon
increasing the strength of dissipation $\gamma$, the decay of the density
autocorrelation function becomes overdamped as soon as the first momentum mode
becomes overdamped, i.e., at the transition from the PT-symmetric to the
PT-broken phase. Concomitantly, one of the base oscillation periods of the
density autocorrelation function diverges. However, the exponent that governs
this divergence is fully determined by PT symmetry, and its value is not
modified by long-range couplings.

The density autocorrelation function is defined as
\begin{equation}
  \label{eq:density}
  A_l(t) = - 4 \left\langle \left( n_l(t) - 1/2 \right) \left( n_l(0) - 1/2 \right)
  \right\rangle_{\mathrm{SS}},  
\end{equation}
where $n_l = c_l^{\dagger} c_l$ and the normalization is chosen to obtain a
simple expression in terms of Majorana fermions. We evaluate the density
autocorrelation function in the steady state, which for $\gamma_l = \gamma_g$ is
at infinite temperature. In Appendix~\ref{sec:appendix-density}, we derive and
solve the equation of motion for the density autocorrelation function. The
solution can be expressed in terms of sums over contributions from momentum
modes, each containing a factor $\e^{- \imag \lambda_{\pm, k} t}$. This form
immediately implies overdamped late-time decay in PT-mixed and PT-broken phases:
In these phases, the dominant contribution stems from the mode with the smallest
decay rate, given by $\gamma_s$ in Eq.~\eqref{eq:gamma-s}. But this mode is
nonoscillatory. In contrast, in the PT-symmetric phase, the density
autocorrelation function shows oscillatory decay. As detailed in
Appendix~\ref{sec:appendix-density}, in the thermodynamic limit, the rescaled
density autocorrelation function $\e^{4 \gamma t} A_l(t)$ can be written as a
sum of squares of integrals of the form
\begin{equation}
  \label{eq:integral-stat-phase}
  I = \frac{1}{\pi} \Im \! \left( \int_0^{\pi} \diff k \, f_k \frac{\e^{\imag
        \omega_k t}}{\omega_k} \right),
\end{equation}
where $f_k = 2 J_k + \mu$ or $f_k = 1$. We obtain the asymptotic behavior of the
density autocorrelation function for $t \to \infty$ by evaluating these
integrals in a stationary phase approximation~\cite{Bender1999a}. To apply this
method, we need to identify the stationary points $k_0$ determined by
$\omega_k'|_{k = k_0} = \varepsilon_k'|_{k = k_0} = 0$, where primes denote
derivatives with respect to $k$. According to our discussion of the dispersion
relation in Sec.~\ref{sec:long-range-dispersion}, the stationary points in the
interval $0 \leq k_0 \leq \pi$ are given by $k_0 \in \{ 0, k_b, \pi \}$. The
asymptotic behavior of the integral in Eq.~\eqref{eq:integral-stat-phase} is
determined by momenta close to the stationary points, and we obtain
\begin{equation}
  \label{eq:I-asymptotic}
  I \sim \frac{1}{\sqrt{2 \pi t}} \left( I_s(0) + 2 I_s(k_b) + I_s(\pi) \right),
\end{equation}
where the terms $I_s(k_0)$ oscillate at the frequencies $\omega_{k_0}$,
\begin{equation}
  I_s(k_0) \sim \frac{f_{k_0} \sin \! \left( \omega_{k_0}
      t + \sgn \! \left(  \omega''_{k_0} \right) \pi/4
    \right)}{\omega_{k_0} \abs{\omega''_{k_0} }^{1/2}}.
\end{equation}
For $\alpha < 3$, the second derivative $\omega_k''$ is undefined at $k_0 = 0$,
and the corresponding contribution to $I$ has to be dropped. We omit the lengthy
explicit asymptotic expression for $A_l(t)$. But the asymptotic result for $I$
already shows that the density autocorrelation function exhibits oscillatory
decay with the base frequencies $\omega_0$, $\omega_{k_b}$, and
$\omega_{\pi}$.

As discussed in Sec.~\ref{sec:long-range-dispersion}, the minimum of
$\varepsilon_k$ is at $0$, $k_b$, and $\pi$, for $\mu < \mu_{b, <}$,
$\mu_{b, <} < \mu < \mu_{b, >}$, and $\mu_{b, >} < \mu$,
respectively. Therefore, upon increasing $\gamma$, gap closings with
$\omega_{k_0} \to 0$, which mark the transition to the PT-mixed phase, occur at
these momenta, similarly to $\omega_{k_{s, \pm}}$ going to zero at directional
pumping phase transitions. However, while the exponents that describe the
critical behavior of $\omega_{k_{s, \pm}}$ are modified in the presence of
long-range couplings as described by
Eq.~\eqref{eq:t-s-long-range-critical-exponents} and illustrated in
Fig.~\ref{fig:LRK_t1}, the base frequencies of the density autocorrelation
function always vanish as a square root of the deviation
$\delta_{\gamma} = \gamma - \gamma_c$ of the dissipation rate $\gamma$ from the
critical value for the transition to the PT-mixed phase,
$\omega_{k_0} \sim \abs{\delta_{\gamma}}^{1/2}$.

The value of $1/2$ of the exponent is, in fact, determined by PT symmetry. For
the Kitaev chain, the PT symmetry condition takes the same form as given in
Eq.~\eqref{eq:zk-IS-PTS} for the SSH model, but for the matrix
$x_k' = \mathbf{x}_k \cdot \boldsymbol{\sigma}$ corresponding to the traceless
part of the matrix $x_k$ defined in Appendix~\ref{sec:appendix-density}, and
given explicitly by
$\mathbf{x}_k = \left( 2 \Delta_k, - 2 J_k - \mu, - \imag 2 \gamma
\right)$~\cite{Starchl2022}.
Without resorting to the explicit form of $\mathbf{x}_k$ for the Kitaev chain,
PTS implies that $x_{x, k}, x_{y, k} \in \R$ whereas $x_{z, k} \in \imag \R$,
such that
$\omega_k = \sqrt{x_{x, k}^2 + x_{y, k}^2 - \left( \imag x_{z, k} \right)^2}$.
At the transition to the PT-mixed phase, the difference under the square root
vanishes for a particular momentum $k_0$. For the example of the Kitaev chain,
$\imag x_{z, k} = 2 \gamma$, and a gap closing occurs at $\gamma = \gamma_c$,
where $x_{x, k_0}^2 + x_{y, k_0}^2 - 4 \gamma_c^2 = 0$. If we now set
$\gamma = \gamma_c + \delta_{\gamma}$ and approach the gap closing by taking the
limit $\delta_{\gamma} \to 0$, we find
\begin{equation}
  \omega_{k_0} = \sqrt{x_{x, k_0}^2 + x_{y, k_0}^2 - 4 \left( \gamma_c +
      \delta_{\gamma} \right)^2} \sim \abs{\delta_{\gamma}}^{1/2},
\end{equation}
in agreement with the analysis of response functions in
Ref.~\cite{Sayyad2021}. We stress that this result follows directly from the
reality conditions imposed upon $x_{x, k}$, $x_{y, k}$, and $\imag x_{z, k}$ by
PT symmetry, and holds also for more general forms of $x_{z, k}$.

Finally, we note that for an isolated system with $x_{z, k} = 0$ such that
$\omega_k = \sqrt{x_{x, k}^2 + x_{y, k}^2}$, gap closings require that
$x_{x, k_0} = x_{y, k_0} = 0$. Considering again the Kitaev chain with
$x_{x, k} = 2 \Delta_k$ and $x_{y, k} = - 2 J_k - \mu$, gap closings occur at
$k_0 = 0, \pi$ where $x_{x, k_0} = 2 \Delta_{k_0} = 0$. Then, with
$\mu = \mu_{c, \lessgtr} + \delta_{\mu}$, we obtain
$\omega_{k_0} \sim \abs{\delta_{\mu}}$. In contrast, the soft-mode frequencies
vanish for $\gamma = 0$ as $\omega_{k_{s, \pm}} \sim \abs{\delta_{\mu}}^{1/2}$,
which provides further evidence for the independence of dynamical criticality at
the boundaries of directional pumping phases.

\section{Conclusions and outlook}
\label{sec:conclusion}

Through the study of quantum quenches in driven-dissipative many-body systems,
our work determines PT symmetry as the principal driver of local relaxation to a
PT-symmetric generalized Gibbs ensemble for two fundamental classes of quadratic
fermionic models. In this way, we substantially extend the field of quantum
quenches and relaxation of many-body systems by establishing local equilibration
also for open systems with finite coupling to external reservoirs.

We have presented the theoretical framework of the PTGGE for driven-dissipative
versions of the SSH model and the Kitaev chain, which can be regarded as natural
open-system generalizations of paradigmatic examples of one-dimensional
topological insulators and superconductors. These models differ by the presence
of a weak $\mathrm{U}(1)$ symmetry~\cite{Buca2012}, which leads to the vanishing
of anomalous correlations in the open SSH model, even though the coupling to
reservoirs breaks particle number conservation. Our analysis shows that
PT symmetry of the quadratic Liouvillian is the fundamental property leading to
coherent local relaxation in the presence of temporally uniform and spatially
global exponential decay.

After rescaling observables to compensate this exponential decay, key features
of the quench dynamics of isolated systems are revealed to persist in the
PT-symmetric phase. This includes the light cone spreading of correlations and
the linear-growth and volume-law saturation of the contribution to the subsystem
entropy due to the propagation of pairs of entangled quasiparticles---however,
with modified quasiparticle dynamics and statistics. Based on a dissipative
quasiparticle picture~\cite{Alba2022, Carollo2022, Alba2021}, we have proposed
an analytical conjecture for the time evolution of the quasiparticle-pair
contribution to the subsystem entropy in the space-time scaling limit, which is
in excellent agreement with our numerical results.

Furthermore, we have provided a detailed analysis of the dynamics of topological
disorder parameters---the \dsop{} and the subsystem fermion parity for the SSH
model and the Kitaev chain, respectively. In the isolated versions of these
models, the topological disorder parameters show oscillatory decay for quenches
from the trivial to the topological phase. This pumping phenomenon becomes
directional in driven-dissipative systems, i.e., the pumping of string order and
fermion parity happens at different rates through the left and right ends of a
subsystem. The pumping rates are determined by soft modes of the PTGGE, and
based on this insight, we have formulated analytical conjectures for the time
evolution of the topological disorder parameters, which we have found to match
numerical simulations very well. Interestingly, there are parameter regimes in
which there is pumping through only one end of a subsystem. This has led us to
introduce the notion of directional pumping phases. As we have demonstrated
using the example of a Kitaev chain with long-range hopping and pairing,
directional pumping phases do, in general, not coincide with the phases
determined by the breaking of PT symmetry. Moreover, we have identified a
distinct form of dynamical criticality at the transitions between directional
pumping phases, and we have shown that the critical exponents that govern the
divergences of pumping rates are modified when the effective long-wavelength
description is modified due to the presence of long-range couplings.

Our work opens up interesting prospects for future research. An important next
step is to establish the generality of our results, in particular, of relaxation
to the PTGGE. Our reasoning leading to the PTGGE applies to all fermionic
many-body systems described by a quadratic Liouvillian with a fully PT-symmetric
phase, and can be extended straightforwardly to more general quench protocols,
such as for systems prepared in excited or mixed states; and we have presented
first results for generalizations of the PTGGE to quadratic bosonic systems and
to noninteracting fermionic systems with quadratic Hermitian jump operators in
Ref.~\cite{Starchl2022}. However, it remains to be seen whether relaxation to a
suitably defined PTGGE occurs more generally in PT-symmetric integrable
driven-dissipative systems~\cite{Torres2014, Caspar2016, Medvedyeva2016,
  Foss-Feig2017a, Mesterhazy2017, Rowlands2018, Shibata2019a, Shibata2019,
  Essler2020, Buca2020, Ziolkowska2020, Nakagawa2021, DeLeeuw2021, Claeys2022}.

Furthermore, a possible topological origin of the zero crossings of the
topological disorder parameters that occur after quenches to the gapless
PT-mixed phase warrants further investigation. In Ref.~\cite{Sayyad2021}, zero
crossings of the subsystem parity were studied in a Kitaev chain with OBC and
for a subsystems given by the left half of the chain. Then, crossings occur only
within the gapped fully PT-symmetric phase, and have been attributed to the
nontrivial non-Hermitian topology of the postquench Liouvillian. Our finding,
that zero crossings occur for a subsystem at the right end of a chain with OBC
and or in a chain with PBC also within the PT-mixed phase, raises the question
whether the dynamical entanglement-spectrum bulk-boundary
correspondence~\cite{Gong2017a, Chang2018, Lu2019} can be extended to such phases
and how it can be refined to distinguish between left and right entanglement
cuts.

In addition to topological disorder parameter pumping or, equivalently,
entanglement spectrum crossings, nonanalyticities of the Loschmidt echo, which
have been dubbed dynamical quantum phase transitions~\cite{Heyl2018,
  Zvyagin2016, Naji2022}, have also been proposed as a dynamical signature of
topology~\cite{Vajna2015}. Indeed, for the isolated Kitaev chain, the soft-mode
period $t_s$ coincides with the period of singularities of the Loschmidt
echo. It will be interesting to see whether a suitable generalization of the
Loschmidt echo to open systems~\cite{Sedlmayr2018a, Lang2018, Poyhonen2021} can
capture the existence of two distinct soft-mode periods $t_{s, +} \neq t_{s, -}$
in the driven-dissipative Kitaev chain.

\begin{acknowledgements}
  We thank Jinlong Yu for helpful discussions and acknowledge support from the
  Austrian Science Fund (FWF) through the project P 33741-N.
\end{acknowledgements}  

\appendix

\section{Time evolution of the covariance matrix}
\label{sec:time-evolution-covariance-matrix}

In this appendix, we derive and solve the equation of motion of the covariance
matrix for time evolution generated by a quadratic Liouvillian. We consider
models with and without a weak $\mathrm{U}(1)$ symmetry, such that the state of
the system is described by a covariance matrix of the form given in
Eq.~\eqref{eq:G} for complex Dirac fermions and Eq.~\eqref{eq:gamma-kitaev} for
Majorana fermions, respectively. Representative examples are provided by the
driven-dissipative SSH model and the Kitaev chain considered in the main
text. In both cases, the covariance matrix takes the general form
\begin{equation}
  \label{eq:Ot}
  \langle O(t) \rangle = \tr \! \left( O \rho(t) \right) = \tr \! \left( O \e^{-
    \imag \mathcal{L} t} \rho_0 \right) = \tr \! \left( \e^{\imag
    \mathcal{L}^{\dagger} t}(O) \rho_0 \right),
\end{equation}
where $O$ is quadratic in fermionic operators, and Hermitian conjugation of the
Liouvillian $\mathcal{L}$ is defined with respect to the Hilbert-Schmidt inner
product of operators,
$\langle A, B \rangle_{\mathrm{HS}} = \tr \! \left( A^{\dagger} B \right)$. That
is, the defining relation for $\mathcal{L}^{\dagger}$ reads
$\langle A, \mathcal{L} B \rangle_{\mathrm{HS}} = \langle \mathcal{L}^{\dagger}
A, B \rangle_{\mathrm{HS}}$.
As stated in Eq.~\eqref{eq:app-L-adjoint}, we find
$\mathcal{L}^{\dagger} = \mathcal{H} - \imag \mathcal{D}^{\dagger}$, where we
have used that $\mathcal{H} = \mathcal{H}^{\dagger}$; the adjoint dissipator
$\mathcal{D}^{\dagger}$ is given in Eq.~\eqref{eq:D-adjoint}. By taking the
derivative of Eq.~\eqref{eq:Ot} with respect to time,
\begin{equation}
  \label{eq:dOt}
  \frac{\diff}{\diff t} \langle O(t) \rangle = \imag \tr \! \left( \e^{\imag
      \mathcal{L}^{\dagger} t} \left( \mathcal{L}^{\dagger}(O) \right) \rho_0
  \right) = \imag \langle (\mathcal{L}^{\dagger}O)(t) \rangle,
\end{equation}
we see that in order to obtain the evolution equations for the covariance
matrices in Eqs.~\eqref{eq:G} and~\eqref{eq:gamma-kitaev}, we have to identify
the relevant operator $O$ and apply $\mathcal{L}^{\dagger}$ to this operator.

\subsection{Complex Dirac fermions}
\label{sec:appendix-eom-dirac}

We first consider a quadratic particle-number conserving Hamiltonian, defined in
terms of $2 L$ complex fermionic annihilation and creation operators $c_l$ and
$c_l^{\dagger}$, respectively,
\begin{equation}
  \label{eq:H-H}
  H = \sum_{l, l' = 1}^{2 L} c_l^{\dagger} H_{l, l'} c_{l'}.
\end{equation}
The Hamiltonian of the SSH model in Eq.~\eqref{eq:H-SSH} is obtained as a
special case of this general form. Further, we consider linear jump operators
that describe particle loss and gain as in Eq.~\eqref{eq:L-l-L-g-Dirac-generic},
with bath matrices given in Eq.~\eqref{eq:bath-matrices}.

\subsubsection{Equation of motion of the covariance matrix}

To derive the equation of motion of the covariance matrix, it is convenient to
consider the single-particle density matrix defined by (note the order of
indices in the definition of $Q_{l, l'}$)
\begin{equation}
  \label{eq:single-particle-density-matrix}
  C_{l, l'}(t) = \langle Q_{l, l'}(t) \rangle, \qquad Q_{l, l'} =
  c_{l'}^{\dagger} c_l,
\end{equation}
as an auxiliary quantity, which is related to the covariance matrix by
\begin{equation}
	\label{eq:G-C}
	G = \id - 2 C.
\end{equation}
According to the discussion above, we can obtain the equation
of motion of the single-particle density matrix by applying
$\mathcal{L}^{\dagger}$ to $Q_{l, l'}$. To that end, it is convenient to
introduce the notation
\begin{equation}
  Q_{l, l'} = c_{l'}^{\dagger} c_l = \sum_{m, m' = 1}^{2 L} c_m^{\dagger} P^{l',
    l}_{m, m'} c_{m'}, \quad P^{l', l}_{m, m'} = \delta_{m, l'} \delta_{m', l}.
\end{equation}
The action of the Hamiltonian $\mathcal{H}$ is then given by
\begin{equation}
  \label{eq:H-Q}
  \mathcal{H}Q_{l,l'}=[H, Q_{l, l'}] = \sum_{m, m' = 1}^{2 L} [H, P^{l', l}]_{m,
    m'} Q_{m', m},
\end{equation}
and the contribution from the adjoint dissipator $\mathcal{D}^{\dagger}$ for
arbitrary loss and gain coefficients reads
\begin{equation}
  \label{eq:D-Q}
  \mathcal{D}^{\dagger} Q_{l, l'} = - \sum_{m,m' = 1}^{2 L} \{M_l + M_g,P^{l',l}\}_{m,m'} Q_{m',m} + 2 M_{g,l,l'}.
\end{equation}
Combining Eqs.~\eqref{eq:H-Q} and~\eqref{eq:D-Q}, we obtain
\begin{multline}
  \label{eq:app-dirac-adjoint-L-Q-action}
  \mathcal{L}^{\dagger} Q_{l, l'} = - \sum_{m, m' = 1}^{2 L} \left( P^{l', l} Z
    - Z^{\dagger} P^{l', l} \right)_{m, m'} Q_{m', m} - \imag 2 M_{g,l,l'},
\end{multline}
where the matrix $Z$ is defined in Eq.~\eqref{eq:Z-matrix-SSH}, and we have used
that the bath matrices defined in Eq.~\eqref{eq:bath-matrices} are Hermitian. By
inserting Eq.~\eqref{eq:app-dirac-adjoint-L-Q-action} in Eq.~\eqref{eq:dOt}, and
using relations such as
\begin{equation}
  \sum_{m, m' = 1}^{2 L} \left( P^{l', l} Z \right)_{m, m'} \tr(Q_{m', m}
  \rho) = \tr \! \left( P^{l', l} Z C \right) = \left( Z C \right)_{l, l'},
\end{equation}
where the trace on the left-hand side and after the first equality applies to
operators and matrices, respectively, we obtain
\begin{equation}
  \frac{\diff C}{\diff t} = - \imag \left( Z C - C Z^{\dagger} \right) + 2 M_g,
\end{equation}
which immediately leads to the equation of motion of the covariance matrix given
in Eq.~\eqref{eq:G-eom}.

\subsubsection{Formal solution of the equation of motion}

To find the solution of Eq.~\eqref{eq:G-eom}, we employ the following
ansatz~\cite{Prosen2011}:
\begin{equation}
  \label{eq:app-covariance-matrix-general-solution}
  G(t) = Q(t) G_0 Q(t)^{\dagger} - \imag P(t) Q(t)^{\dagger},
\end{equation}
where initial conditions are given by $G_0 = G(0)$, $Q(0) = \id$ and $P(0) =
0$. This ansatz satisfies Eq.~\eqref{eq:G-eom} if
\begin{equation}
  \frac{\diff Q}{\diff t} = - \imag Z Q, \quad \frac{\diff P}{\diff t} = - \imag
  Z P + \imag 2 \left( M_l - M_g \right) Q^{-\dagger}.
\end{equation}
The solutions to these equations read $Q(t) = \e^{- \imag Z t}$ and
\begin{equation}
  P(t) = \imag 2 Q(t) \int_0^t \diff t' \, Q(t')^{-1} \left( M_l - M_g \right)
  Q(t')^{-\dagger}.
\end{equation}
To perform the integration over time, we express the matrix $Z$ as
$Z = V \Lambda V^{-1}$, where $\Lambda$ is a diagonal matrix with entries
$\lambda_l$. For the last term in
Eq.~\eqref{eq:app-covariance-matrix-general-solution},
$R(t) = - \imag P(t) Q(t)^{\dagger}$, we find
\begin{equation}  
  R_{l, l'}(t) = - \imag 2 V \left[ \left( V^{-1} \left( M_l - M_g \right)
      V^{-\dagger} \right) \circ K(t) \right] V^{\dagger},
\end{equation}
where
\begin{equation}
  K_{l, l'}(t) = \frac{1 - \e^{-\imag \left( \lambda_l - \lambda_{l'}^{*}
      \right) t}}{\lambda_l - \lambda_{l'}^{*}},
\end{equation}
and where $A \circ B$ denotes the Hadamard product, i.e., the element-wise
product given by $\left( A \circ B \right)_{l, l'} = A_{l, l'} B_{l, l'}$. The
above form of $R(t)$ is particularly convenient to evaluate the covariance
matrix numerically.

\subsection{Majorana fermions}

Next, we consider a quadratic Hamiltonian for $2 L$ Majorana fermions,
\begin{equation}
  H = \frac{\imag}{4}  \sum_{l,l' = 1}^{2 L} w_l A_{l, l'} w_{l'},
\end{equation}
where without loss of generality we assume that $A$ is antisymmetric,
$A = - A^{\transpose}$. The representation of the Hamiltonian of the Kitaev
chain Eq.~\eqref{eq:H-Kitaev} in this form is provided in
Ref.~\cite{Starchl2022}. We further consider linear jump operators and define
the bath matrix $M$ through
\begin{equation}
  L_l = \sum_{l,l'=1}^{2L} B_{l,l'} w_{l'}, \qquad M = B^{\transpose} B^{*},
\end{equation}
which includes Eq.~\eqref{eq:L} as a special case.

\subsubsection{Equation of motion of the covariance matrix}

In analogy to Eq.~\eqref{eq:single-particle-density-matrix}, we define a
single-particle density matrix for Majorana fermions,
\begin{equation}
  \Xi_{l, l'}(t) = \imag \langle W_{l, l'}(t) \rangle, \qquad W_{l, l'} = w_l w_{l'},
\end{equation}
such that the covariance matrix in Eq.~\eqref{eq:gamma-kitaev} is given by
$\Gamma = \Xi - \imag \id$. To calculate the action of $\mathcal{L}^{\dagger}$
on $W_{l, l'}$ and thus obtain the equation of motion of the single-particle
density matrix, we write $W_{l, l'}$ as
\begin{equation}
  \label{eq:app-W}
  W_{l,l'} = w_l w_{l'} = \sum_{m, m' = 1}^{2 L} w_m P^{l, l'}_{m, m'} w_{m'},
  \quad P^{l, l'}_{m, m'} = \delta_{m, l} \delta_{m', l'}.
\end{equation}
The Hamiltonian contribution to $\mathcal{L}^{\dagger} W_{l, l'}$ is given by
\begin{equation}
  \label{eq:H-W}
  \mathcal{H} W_{l,l'} = [H,W_{l,l'}] = \imag \sum_{m, m' = 1}^{2 L} [A, P^{l',
    l}]_{m, m'} W_{m', m}.
\end{equation}
Next, we calculate the action of the dissipator on $W_{l,l'}$:
\begin{equation}
  \label{eq:dissipator-Q}
  \mathcal{D}^{\dagger} W_{l, l'} = - 4 \sum_{m, m' = 1}^{2 L} \{ M_R, P^{l', l}
  \}_{m, m'} W_{m', m} + 8 M_{l, l'}^{\transpose},
\end{equation}
where we have used that the real part $M_R = \Re(M)$ and the imaginary part
$M_I = \Im(M)$ of the bath matrix are symmetric and antisymmetric, respectively.
Combining Eqs.~\eqref{eq:H-W} and~\eqref{eq:dissipator-Q} we obtain
\begin{multline}
  \label{eq:app-L-dagger-W}
  \mathcal{L}^{\dagger} W_{l, l'} = \imag \sum_{m, m' = 1}^{2 L} \left(
    X^{\transpose} P^{l', l} + P^{l', l} X \right)_{m, m'} W_{m', m} - \imag 8
  M_{l, l'}^{\transpose}
\end{multline}
where
\begin{equation}
  \label{eq:X}
  X = - A + 4 M_R.
\end{equation}
We obtain the equation of motion for the single-particle density matrix by
inserting Eq.~\eqref{eq:app-L-dagger-W} in Eq.~\eqref{eq:dOt}, which then leads
to the equation of motion for the covariance matrix,
\begin{equation}
  \label{eq:Gamma-eom}
  \frac{d \Gamma}{d t} = - X \Gamma - \Gamma X^{\transpose} - Y,
\end{equation}
where $Y = - 8 M_I$.

\subsubsection{Formal solution of the equation of motion}

For the sake of completeness, we note that the formal solution of
Eq.~\eqref{eq:Gamma-eom} reads
\begin{equation}
  \Gamma(t) = \e^{- X t} \Gamma(0) \e^{- X^{\transpose} t} - \int_0^t \diff t'
  \, \e^{- X \left( t - t' \right)} Y \e^{- X^{\transpose} \left( t - t'
    \right)}.
\end{equation}
The integral over time can be performed as detailed in Ref.~\cite{Starchl2022}.

\section{Biorthogonal representation of quadratic operator evolution}
\label{sec:biorthogonal-representation}

In this appendix, we discuss how the biorthogonal eigenvectors of the matrix $Z$
defined in Eq.~\eqref{eq:Z-matrix-SSH} contribute to the expectation value
$\braket{O(t)}$. For concreteness, let us consider an observable that is
quadratic in fermionic operators, $O = \sum_{l,l'} c_l^\dagger O_{l,l'} c_{l'}$,
such that by using Eqs.~\eqref{eq:single-particle-density-matrix}
and~\eqref{eq:G-C} the expectation value can be expressed in terms of the
covariance matrix as
\begin{equation}
  \braket{O(t)} = \frac{1}{2} \sum_{l,l'} O_{l,l'} \left( \delta_{l,l'} -
    G_{l',l}(t) \right).
\end{equation}
Further, we focus on balanced loss and gain rates, such that $M_l - M_g = 0$ in
the evolution equation~\eqref{eq:G-eom}. Then, the solution to the evolution
equation that satisfies the initial condition $G(0) = G_0$ reads
\begin{equation}
  G(t) = \e^{-\imag Z t} G_0 \e^{\imag Z^\dagger t} = \sum_{l, l' = 1}^{2 L} \e^{-\imag \left(\lambda_l - \lambda_{l'}^\ast \right) t}
  \mathbf{v}_{R,l} \left( \mathbf{v}_{L,l}^\dagger G_0 \mathbf{v}_{L,l'} \right) \mathbf{v}_{R,l'}^\dagger,
\end{equation}
where we have employed the spectral decomposition
$Z = \sum_{l = 1}^{2 L} \lambda_l \mathbf{v}_{R,l} \mathbf{v}_{L,l}^\dagger$.
Here, $\mathbf{v}_{L, l}$ and $\mathbf{v}_{R, l}$ are left and right
eigenvectors of $Z$, respectively, corresponding to the eigenvalues
$\lambda_l$. The left and right eigenvectors obey the biorthogonality condition
$\mathbf{v}_{L, l}^{\dagger} \mathbf{v}_{R, l'} = \delta_{l, l'}$. We can thus
write the evolution of the expectation value of the observable $O$ as
\begin{equation}
  \braket{O(t)} = \frac{\tr\!\left(O\right)}{2} - \frac{1}{2} \sum_{l, l' =
    1}^{2 L} \e^{-\imag
    \left(\lambda_l - \lambda_{l'}^\ast \right) t} \left( \mathbf{v}_{L,l}^\dagger G_0 \mathbf{v}_{L,l'} \right)
  \left( \mathbf{v}_{R,l'}^{\dagger} O  \mathbf{v}_{R,l} \right).
\end{equation}
Therefore, the evolution of a quadratic operator can be expressed in terms of
matrix elements of the initial covariance matrix $G_0$ and the observable $O$
between left and right eigenvectors of $Z$, respectively. For Gaussian states,
expectation values of observables that involve products of more than two
fermionic fields can be reduced to expectation values of quadratic operators by
employing Wick's theorem.

\section{Diagonalization of $z_k$}
\label{sec:diagonalization-zk}

The non-Hermitian Bloch Hamiltonian of the driven-dissipative SSH model,
$z_k = z_\id \id + \mathbf{z}_k \cdot \boldsymbol{\sigma}$ defined in
Eq.~\eqref{eq:zk}, can be diagonalized in the PT-symmetric phase by applying two
consecutive rotations by a real and an imaginary angle, respectively, as
described by the nonunitary matrix
\begin{equation}
  \label{eq:Uk}
  U_k = U_{z,-\phi_k} U_{y,-\imag \theta_k + \pi/2} = \e^{-\imag
    \left(\phi_k/2\right)\sigma_z} \e^{\left[\left(\theta_k - \imag \pi/2\right)
      /2\right] \sigma_y},
\end{equation}
where the angles $\phi_k$ and $\theta_k$ are defined by the relations
\begin{align}
  \label{eq:phi-k}
  \varepsilon_k \e^{\imag \phi_k} & = J_1 + J_2 \cos(k) + \imag J_2 \sin(k), \\
  \label{eq:theta-k}
  \theta_k & = \acosh \! \left(\frac{\varepsilon_k}{\omega_k} \right)  =
             \asinh \! \left( \frac{2\Delta\gamma}{\omega_k} \right).
\end{align}
This leads to the following representation of $z_k$:
\begin{equation}
  \label{eq:zk-diag}
  z_k = z_\id \id + \omega_k U_k \sigma_z U_k^{-1},
\end{equation}
with the dispersion relation of the driven-dissipative SSH model $\omega_k$
given in Eq.~\eqref{eq:dispersion-open-SSH}. For the isolated SSH model with
$\Delta \gamma = 0$, the above relation reduces to Eq.~\eqref{eq:hk-diagonal}.

\section{Proof of Eq.~\eqref{eq:Gamma-l-G-l-SSH}}
\label{sec:appendix-majoran-cov-SSH}

As explained in Sec.~\ref{sec:SSH-driven-dissipative}, due to the weak
$\mathrm{U}(1)$ symmetry of the driven-dissipative SSH model, if there are no
anomalous correlations in the initial state, no such correlations will be
generated in the dynamics, and, therefore, the state of the system is at all
times fully determined by the covariance matrix defined in
Eq.~\eqref{eq:G}. Consequently, describing the state in terms of a covariance
matrix for Majorana fermions appears to be unnatural and inefficient. However,
as stated in Eq.~\eqref{eq:dual-SOP}, the \dsop{} has a simple representation in
terms of Majorana fermions, which, by applying Wick's theorem, immediately leads
to the expression Eq.~\eqref{eq:dual-SOP-expectation} for the expectation value
of the \dsop{} as the Pfaffian of a submatrix $\Gamma_{\ell}$ of the Majorana
covariance matrix. In this appendix, we provide a proof for
Eq.~\eqref{eq:Gamma-l-G-l-SSH} which relates $\Gamma_{\ell}$ to the reduced
covariance matrix $G_{1, \ell}$ defined in terms of complex
fermions---confirming that there is no information in $\Gamma_{\ell}$ that is
not already contained in $G_{1, \ell}$. For simplicity, we consider a
translationally invariant state of a system with PBC, but these assumptions are
not crucial.

We begin by defining a complex covariance matrix that includes anomalous
correlations $\braket{c_l c_{l'}}$ and
$\braket{c_l^{\dagger} c_{l'}^{\dagger}}$, even though these vanish for the SSH
model:
\begin{equation}
  F_{l,l'} = \left\langle \left[C_l, C_{l'}^{\dagger} \right] \right\rangle,
\end{equation}
with
$C = \left( c_1, c_1^{\dagger}, \dotsc, c_{2 L}, c_{2 L}^{\dagger}
\right)^{\transpose}$.
The Majorana covariance matrix for the SSH model, introduced in
Sec.\ref{sec:dual-string-order}, can then be written as
\begin{equation}
  \label{eq:app-Gamma-SSH}
  \Gamma = \imag R_{2 L}^{\dagger} F R_{2 L},
\end{equation}
where the matrix converting between Majorana and complex fermions is given by
\begin{equation}
  \label{eq:app-R}
  R_{\ell} = \bigoplus_{l=1}^{\ell} \frac{1}{\sqrt{2}}
  \begin{pmatrix}
    1 & -\imag \\ 1 & \imag
  \end{pmatrix}.
\end{equation}
We want to understand the structure of the submatrix $\Gamma_{\ell}$ introduced
in Eq.~\eqref{eq:Gamma-ell-SSH}. To that end, we first consider the structure of
$F$. For a translationally invariant state, $F$ can be decomposed into
$4 \times 4$ blocks $f_l$ which can be expressed in terms of the elements of the
covariance matrix Eq.~\eqref{eq:G} as
\begin{equation}
  f_{l - l'} =
  \begin{pmatrix}
    G_{2l-1,2l'-1} & 0 & G_{2l-1,2l'} & 0 \\
    0 & -G_{2l'-1,2l-1} & 0 & -G_{2l',2l-1} \\
    G_{2l, 2l'-1} & 0 & G_{2l,2l'} & 0 \\
    0 & -G_{2l'-1,2l} & 0 & -G_{2l',2l}
  \end{pmatrix},
\end{equation}
where the zero entries are due to the vanishing of anomalous correlations. Next,
we consider the transformation to Majorana fermions in
Eq.~\eqref{eq:app-Gamma-SSH}. For a $4 \times 4$ block $\gamma_l$ of $\Gamma$,
we obtain
\begin{equation}
  \gamma_l = \imag R_2^{\dagger} f_l R_2.
\end{equation}
To calculate the \dsop{} according to Eq.~\eqref{eq:dual-SOP-expectation}, we
require not the full matrix $\Gamma$ but rather the submatrix $\Gamma_{\ell}$
specified in Eq.~\eqref{eq:Gamma-ell-SSH}. This submatrix is composed of
$2 \times 2$ blocks $\left( \gamma_{\ell} \right)_l$ that are cut out from the
center of the $4 \times 4$ blocks $\gamma_l$ and given by
\begin{multline}
  \left(\gamma_{\ell}\right)_{l - l'} \\ = \frac{1}{2}
  \begin{pmatrix}
    \imag \left(G_{2l-1,2l'-1} - G_{2l'-1,2l-1}\right) & - \left( G_{2l-1,2l'} + G_{2l',2l-1} \right) \\
    G_{2l, 2l'-1} + G_{2l'-1, 2l} & \imag \left(G_{2l, 2l'} - G_{2l', 2l}\right)
  \end{pmatrix}.
\end{multline}
When we decompose the covariance matrix as $G = G_1 + G_2$ in analogy to
Eq.~\eqref{eq:gk(t)}, a drastic simplification of
$\left( \gamma_{\ell} \right)_l$ results from the symmetry properties of $G_1$
and $G_2$ stated in Eq.~\eqref{eq:PHS-g1l-g2l}. In particular, we find that the
contribution due to $G_2$ drops out,
\begin{equation}
  \left(\gamma_{\ell}\right)_{l - l'} =
  \begin{pmatrix}
    \imag G_{1,2l-1,2l'-1}  & - G_{1,2l-1,2l'} \\
    G_{1,2l, 2l'-1}  & \imag G_{1,2l, 2l'}
  \end{pmatrix}.
\end{equation}
Comparing this to the block structure of $G$ given by 
\begin{equation}
  g_{1,l - l'} =
  \begin{pmatrix}
    G_{1,2l-1,2l'-1}  & G_{1,2l-1,2l'} \\
    G_{1,2l, 2l'-1}  & G_{1,2l, 2l'}
  \end{pmatrix},
\end{equation}
we find that
\begin{equation}
  \left( \gamma_{\ell} \right)_l =  \imag P_1^{\dagger} g_{1,l} P_1,
\end{equation}
where $P_{\ell}$ is defined in Eq.~\eqref{eq:P-ell}. For the full matrices
$\Gamma_{\ell}$ and $G_{1, \ell}$, this relation leads to
Eq.~\eqref{eq:Gamma-l-G-l-SSH}.

\section{Polylogarithm and Riemann zeta function}
\label{sec:polylog-zeta}

The polylogarithm and the Riemann zeta function are defined as~\cite{Olver2010}
\begin{equation}
  \mathrm{Li}_{\alpha}(z)= \sum_{n=1}^{\infty} \frac{z^n}{n^{\alpha}}, \qquad
  \zeta(\alpha)=\sum_{n=1}^\infty \frac{1}{n^{\alpha}} = \mathrm{Li}_{\alpha}(1).
\end{equation}
To make analytical progress when working with the polylogarithm, we employ the
series expansions of $\mathrm{Li}_{\alpha} \! \left( \e^{\imag k} \right)$
around $k = 0$ and $k = \pi$. The expansion around $k = 0$
reads~\cite{Olver2010}
\begin{equation}
  \label{eq:Li-k-to-0}
  \mathrm{Li}_{\alpha} \! \left(\e^{\imag k}\right)= \Gamma \! \left(1-\alpha\right)\left(-\imag k\right)^{\alpha-1} +
  \sum_{n = 0}^{\infty}\frac{\zeta\!\left(\alpha - n\right)}{n!} \left(\imag k\right)^n,
\end{equation}
which holds for $\abs{k} < 2 \pi$ and $\alpha \notin \N$, and where
$\Gamma(\alpha)$ is the gamma function. For $\alpha \in \N$, the expansion is
given by~\cite{Gradshteyn2007}
\begin{equation}
  \label{eq:Li-k-to-0-int}
  \mathrm{Li}_\alpha \! \left(\e^{\imag k}\right) = \frac{\left(\imag
      k\right)^{\alpha - 1}}{\left(\alpha-1\right)!} \left(H_{\alpha-1} -
    \log\!\left(- \imag k\right)\right) + \sum_{\substack{n=0 \\ n\neq\alpha-1}}^{\infty}
  \frac{\zeta \! \left(\alpha-n \right)}{n!}\left(\imag k\right)^n,
\end{equation}
where $H_{\alpha} = \sum_{n=1}^{\alpha} \frac{1}{n}$ is the $\alpha$-th harmonic
number with $H_0 = 0$. The expansion of the polylogarithm around $k = \pi$ reads
\begin{equation}
  \label{eq:Li-k-to-pi}
  \mathrm{Li}_{\alpha} \! \left( \e^{\imag k} \right) = - \sum_{n=0}^\infty
  \frac{\eta \! \left(\alpha-n\right) }{n!} \left[\imag \left(k - \pi\right)
  \right]^n,
\end{equation}
where $\eta(\alpha) = \left( 1 - 2^{1 - \alpha} \right) \zeta(\alpha)$ is the
Drichlet eta function.

\section{Density autocorrelation function}
\label{sec:appendix-density}

The connected density autocorrelation function defined in
Eq.~\eqref{eq:density} can be regarded as a special case of the Majorana
four-point function
\begin{equation}
  \label{eq:B-l1-l2-l3-l4}
  B_{l_1, l_2, l_3}(t) = \langle W_{l_1, l_2}(t) W_{2 l_3 - 1, 2 l_3}(0)
  \rangle_{\mathrm{SS}},
\end{equation}
where $W_{l,l'}$ is given in Eq.~\eqref{eq:app-W} and, for
$\gamma_l = \gamma_g$, the expectation value in the steady state reduces to
taking the trace, $\braket{\dots}_{\mathrm{SS}} = \tr(\dotsb)/2^L$.  Note that
$l_1, l_2 \in \{ 1, \dotsc, 2 L \}$ label Majorana modes whereas
$l_3 \in \{ 1, \dotsc, L \}$ is a lattice-site index. The relation
$W_{2 l - 1, 2 l} = \imag 2 \left( n_l - 1/2 \right)$ leads to
\begin{equation}  
  A_l(t) = \langle W_{2 l - 1, 2 l}(t) W_{2 l - 1, 2 l}
  \rangle_{\mathrm{SS}} = B_{2 l - 1, 2 l, l}(t).
\end{equation}
As we show in the following, a closed equation of motion can be obtained for
four-point functions of the general type given in
Eq.~\eqref{eq:B-l1-l2-l3-l4}. Solving this equation yields the full time
dependence of the density autocorrelation function.

According to the quantum regression theorem, the two-time average in
Eq.~\eqref{eq:B-l1-l2-l3-l4} can be calculated as~\cite{Gardiner2014}
\begin{equation}
  \begin{split}
    B_{l_1, l_2, l_3}(t) & = \tr \! \left( W_{l_1, l_2} \e^{- \imag \mathcal{L}
        t} (W_{2 l_3 - 1, 2 l_3} \rho_{\mathrm{SS}}) \right) \\ & = \tr \!
    \left( \e^{\imag \mathcal{L}^{\dagger}}(W_{l_1, l_2}) W_{2 l_3 - 1, 2 l_3}
      \rho_{\mathrm{SS}} \right),
  \end{split}
\end{equation}
where the adjoint Liouvillian is defined in Eq.~\eqref{eq:app-L-adjoint}. By
taking the derivative with respect to time, we obtain
\begin{equation}
  \frac{dB_{l_1, l_2, l_3}}{dt} = \imag \tr \! \left( \mathcal{L}^{\dagger}
      (W_{l_1, l_2}) \e^{- \imag \mathcal{L} t} (W_{2 l_3 - 1, 2 l_3} \rho_{\mathrm{SS}}) \right),
\end{equation}
and inserting here Eq.~\eqref{eq:app-L-dagger-W} leads to
\begin{equation}
  \label{eq:density-eom-1}
  \frac{\diff B_{l_1, l_2, l_3}}{\diff t} = - \sum_{m = 1}^{2 L} \left(
    X_{l_1, m} B_{m, l_2, l_3} + B_{l_1, m, l_3} X^{\transpose}_{m, l_2}
  \right).
\end{equation}
In the last term, we have used
$\braket{W_{2 l_3 - 1, 2 l_3}}_{\mathrm{SS}} = \imag 2 \langle n_{l_3} - 1/2
\rangle_{\mathrm{SS}} = 0$.
Let us now consider a system with PBC and exploit translational invariance to
represent $B_{l_1, l_2, l_3}(t)$ in terms of $2 \times 2$ blocks,
\begin{equation}
  b_{l_1 - l_3, l_2 - l_3} =
  \begin{pmatrix}
    B_{2l_1-1,2l_2-1, l_3} & B_{2 l_1-1,2l_2, l_3} \\
    B_{2l_1,2l_2-1, l_3} & B_{2l_1,2l_2, l_3}
  \end{pmatrix},
\end{equation}
that depend only on relative coordinates. Indeed, note that here all three of
$l_1, l_2, l_3 \in \{ 1, \dotsc, L \}$ are lattice-site indices. Similarly, we
introduce $2 \times 2$ blocks of the matrix $X$. Rewritten in terms of block
matrices, the equation of motion reads
\begin{equation}
  \label{eq:b-l-l'-eom}
  \frac{\diff b_{l, l'}}{\diff t} = - \sum_{m = 1}^L \left(
    x_{l - m} b_{m, l'} + b_{l, m} x_{l' - m}^{\transpose}  \right).
\end{equation}
This equation can be solved straightforwardly in momentum space. To this end, we
introduce the discrete Fourier transforms of the block matrices,
\begin{equation}
  b_{k,k'} = - \imag \sum_{l,l' = 1}^L \e^{-\imag\left(k l + k' l' \right)}
  b_{l,l'}, \quad x_k = - \imag \sum_{l = 1}^L \e^{-\imag k l} x_l,
\end{equation}
in terms of which Eq.~\eqref{eq:b-l-l'-eom} can be recast as
\begin{equation}
  \frac{d b_{k,k'}}{dt} = - \imag \left(x_k b_{k,k'} + b_{k,k'} x_{k'}^\transpose \right).
\end{equation} 
This equation of motion is solved by
\begin{equation}
  b_{k,k'}(t) = \e^{-\imag x_k t} b_{k,k'}(0) \e^{-\imag x_{k'}^\transpose t},
\end{equation}
where the initial condition $b_{k,k'}(0)$ in momentum space can be obtained by
taking the discrete Fourier transform of
$b_{l,l'}(0) = - \imag \sigma_y \delta_{l,0} \delta_{l',0}$, leading to
$b_{k,k'}(0) = - \sigma_y$. In the PT-symmetric phase, where
$\omega_k \in \R_{>0}$ for all $k \in \BZ$, we find
\begin{multline}
  \label{eq:b-k-k-prime}
  b_{k,k'}^{1,2}(t) = - \imag \e^{- 4 \gamma t} \left[ a_k(t) a_{k'}(t)^{*}
    + \left( \cos\!\left(\omega_k t\right) - \frac{2 \gamma}{\omega_k} \sin\!\left(\omega_k t\right)\right) \right. \\
  \left. \times \left( \cos\!\left(\omega_{k'} t \right) + \frac{2
        \gamma}{\omega_{k'}} \sin\!\left(\omega_{k'} t\right)\right) \right],
\end{multline}
where
$a_k(t) = \left( 2 J_k + \mu - \imag 2 \Delta_k \right) \sin(\omega_k
t)/\omega_k$. Finally, we obtain the density autocorrelation function as
\begin{equation}
  \label{eq:app-auto-correlation-final}
  \begin{split}
    A_l(t) & = b_{0,0}^{1,2}(t) = \frac{\imag}{L^2} \sum_{k,k'\in\BZ}
    b_{k,k'}^{1,2}(t) \\ & = \e^{-4 \gamma t} \left[ \abs{A_{1, l}(t)}^2 - 4
      \gamma^2 A_{2, l}(t)^2 + \left( \frac{\diff A_{2, l}(t)}{\diff t}
      \right)^2 \right],
  \end{split}
\end{equation}
where
\begin{equation}
  A_{1, l}(t) = \frac{1}{L} \sum_{k \in \BZ} a_k(t), \quad A_{2, l}(t) =
  \frac{1}{L} \sum_{k \in \BZ} \frac{\sin(\omega_k t)}{\omega_k}.
\end{equation}
In the thermodynamic limit, sums over momenta are replaced by integrals over
$k \in [- \pi, \pi]$. Since the range of integration is symmetric, the
contribution to $a_k(t)$ that contains $\Delta_k$ and is odd in $k$ can be
dropped. For $\omega_k \in \R_{> 0}$, we may thus write
\begin{equation}
  \begin{split}
    A_{1, l}(t) & = \frac{1}{\pi} \Im \! \left( \int_0^{\pi} \diff k \, \left( 2
        J_k + \mu \right) \frac{\e^{\imag \omega_k t}}{\omega_k} \right), \\
    A_{2, l}(t) & = \frac{1}{\pi} \Im \! \left( \int_0^{\pi} \diff k \,
      \frac{\e^{\imag \omega_k t}}{\omega_k} \right).
  \end{split}
\end{equation}
The asymptotic behavior of these integrals for $t \to \infty$ can be obtained by
means of a stationary phase approximation as described in
Sec.~\ref{sec:conn-dens-dens}.

\bibliography{bibliography.bib}

\end{document}